\documentclass[12pt]{article}

\RequirePackage[authoryear]{natbib}
\RequirePackage[colorlinks,citecolor=blue,urlcolor=blue]{hyperref}
\RequirePackage{graphicx}
\usepackage[export]{adjustbox}

\usepackage[utf8]{inputenc}
\usepackage[T1]{fontenc}
\usepackage{comment}
\usepackage[english]{babel}

\usepackage{amscd,amsfonts,amsmath,amssymb,amsthm,bbm,bm,latexsym,mathrsfs}
\allowdisplaybreaks
\usepackage{bookmark}
\usepackage{enumitem}
\usepackage{adjustbox}
\usepackage{textcomp, gensymb}
\usepackage{natbib}
\usepackage{float,rotating}
 \usepackage{vcell}
\hypersetup{hidelinks}

\usepackage{tabu}

\usepackage[usenames]{color}
\usepackage{epsfig,epstopdf,graphics,graphicx,longtable}
\graphicspath{{./figures/}}
\usepackage{tikz}
\usepackage{tkz-berge}
\usetikzlibrary{shapes,arrows,bayesnet}	

\usepackage{color}
\definecolor{France}{RGB}{32,221,60}
\definecolor{Germany}{RGB}{255,77,20}
\definecolor{Italy}{RGB}{50,197,225}
\definecolor{Spain}{RGB}{255,133,20}

\usepackage[toc,page]{appendix}
\usepackage[normalem]{ulem}
\usepackage{cancel}

\makeatother

\usepackage{enumitem}

\usepackage[section]{placeins}
\usepackage{afterpage}

\newtheorem{corollary}{Corollary}[section]
\newtheorem{assumption}{Assumption}[section]
\newtheorem{proposition}{Proposition}[section]

\theoremstyle{remark}

\theoremstyle{plain}

\newcommand\sdot[1][.5]{\mathbin{\vcenter{\hbox{\scalebox{#1}{$\bullet$}}}}}

\theoremstyle{plain}

\theoremstyle{remark}

\addto\captionsenglish{}

\usepackage{caption}
\usepackage{subcaption}
\usepackage{graphicx}

\newcommand{\blind}{0}

\renewcommand{\arraystretch}{0.8}

\begin{document}

\def\spacingset#1{\renewcommand{\baselinestretch}%
{#1}\small\normalsize} \spacingset{1}


\if0\blind
{
\title{\bf Media Bias and Polarization through the Lens of a Markov Switching Latent Space Network Model}
  \author{Roberto Casarin\thanks{Department of Economics, Ca' Foscari University of Venice and VERA Centre, \texttt{r.casarin@unive.it}} \and
          Antonio Peruzzi\thanks{Department of Economics, Ca' Foscari University of Venice, \texttt{antonio.peruzzi@unive.it}} \and
          Mark F.J. Steel\thanks{Department of Statistics, University of Warwick, \texttt{m.steel@warwick.ac.uk}}}
  \maketitle
} \fi

\if1\blind
{
  \bigskip
  \bigskip
  \bigskip
  \begin{center}
    {\LARGE\bf Media Bias and Polarization  through the Lens of a Markov Switching Latent Space Network Model}
\end{center}
  \medskip
} \fi

\bigskip
\begin{abstract}
News outlets are now more than ever incentivized to provide their audience with slanted news, while the intrinsic homophilic nature of online social media may exacerbate polarized opinions. Here, we propose a new dynamic latent space model for time-varying online audience-duplication networks, which exploits social media content to conduct inference on media bias and polarization of news outlets. We contribute to the literature in several directions: 1) Our model provides a novel measure of media bias that combines information from both network data and text-based indicators; 2) we endow our model with Markov-Switching dynamics to capture polarization regimes while maintaining a parsimonious specification; 3) we contribute to the literature on the statistical properties of latent space network models. The proposed model is applied to a set of data on the online activity of national and local news outlets from four European countries in the years 2015 and 2016. We find evidence of a strong positive correlation between our media slant measure and a well-grounded external source of media bias. In addition, we provide insight into the polarization regimes across the four countries considered.
\end{abstract}

\noindent%
\vfill

\spacingset{1.2} 
\section{Introduction}
\label{sec:intro}

We propose a new statistical model able to offer meaningful insights into the perceived media bias and regime changes in polarization within online social media. The risk of being unintentionally exposed to biased news and polarized opinions has gained awareness both in the public debate (\citealt{wef2022}) and in the academic sphere (see \citealt{puglisi2015empirical, gentzkow2015media,cinelli2021echo}) due to the rapid changes in the news consumption landscape (\citealt{newman2017reuters}). Luckily, the current availability of social-media data provides a privileged perspective on phenomena related to people's preferences and homophilous behavior (\citealp{zhang2018discovering, chen2022monitoring,yu2022collaborative}).
\begin{figure}[p]
  \centering
  \begin{tabular}[t]{p{2cm}c}\vspace{-30pt}
        \emph{\small Year 2015} & \includegraphics[trim={1cm 2.5cm 1cm 2cm},clip,valign=t,scale=0.3]{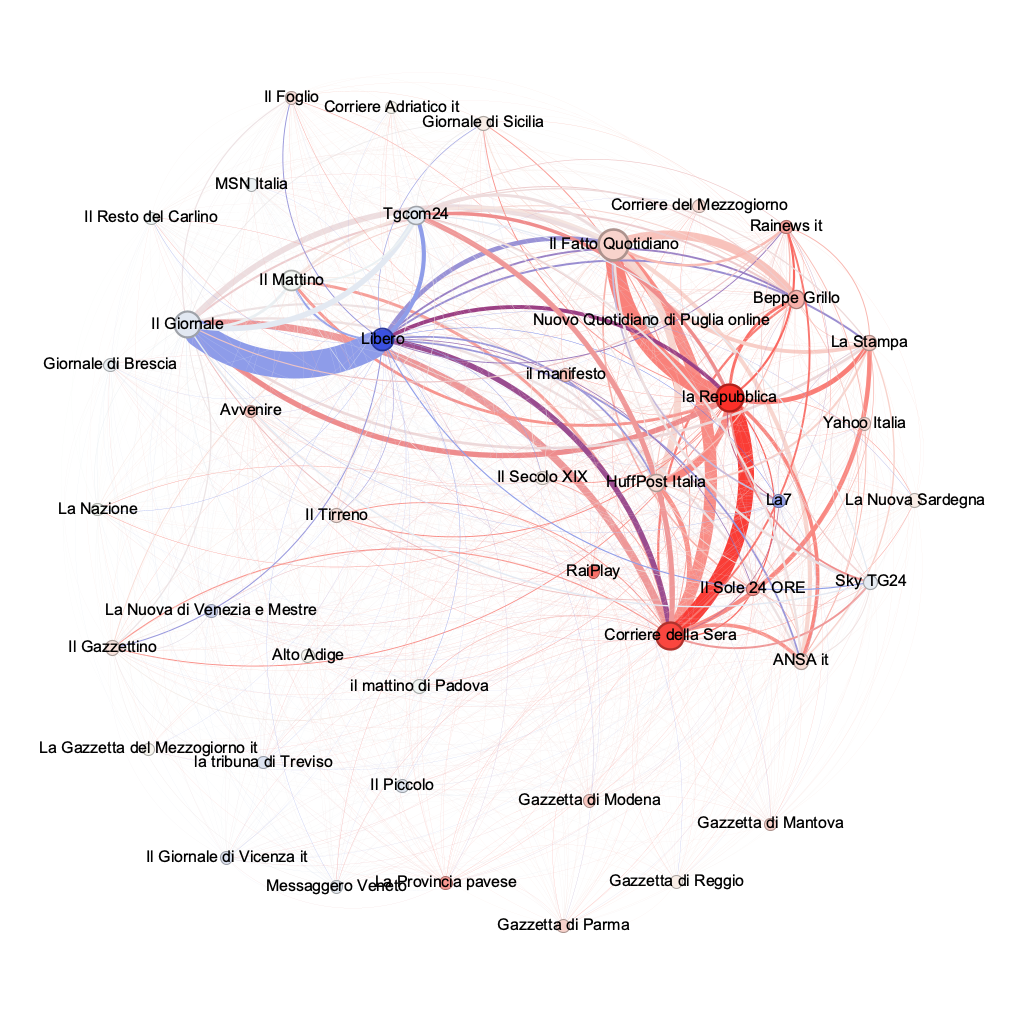}\\\vspace{-20pt}
       \emph{\small Year 2016} &\includegraphics[trim={1cm 2.5cm 1cm 2cm},clip,valign=t,scale=0.3]{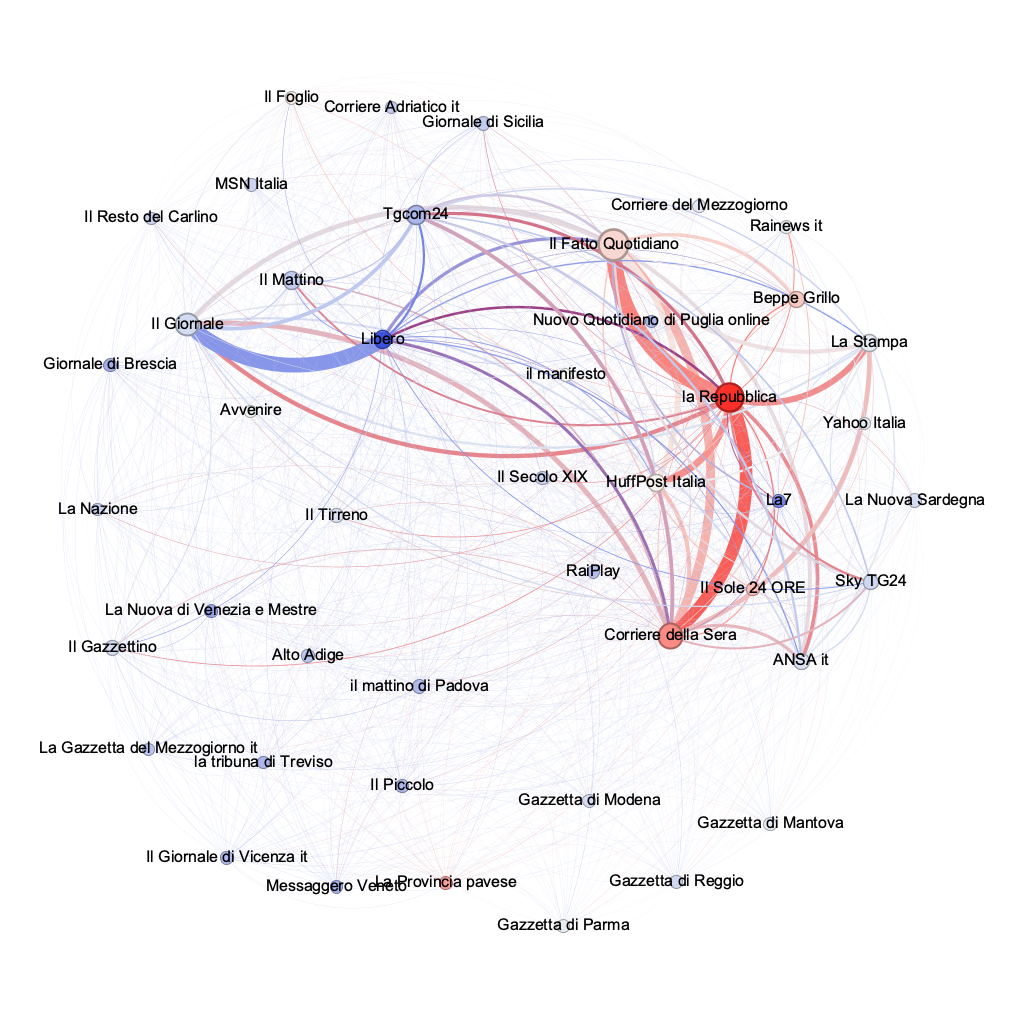}
   \end{tabular}
 \caption{\textbf{Italian Audience-Duplication Networks} obtained from the bipartite network of Italian news outlets and their Facebook commenters in 2015 (top) and 2016 (bottom). Node size is proportional to the news outlets' engagement in terms of comments. Nodes are labeled with the name of the outlet and colored from red (left) to blue (right) according to the text-analysis political-leaning score computed following \citet{gentzkow2010drives} and \citet{garz2020partisan}. Edge thickness is proportional to the number of commenters in common between any two news outlets.} \label{fig:frenchnet}
\end{figure}

\begin{figure}[t]
  \centering
\begin{tabular}{c}
    \includegraphics[scale=0.44]{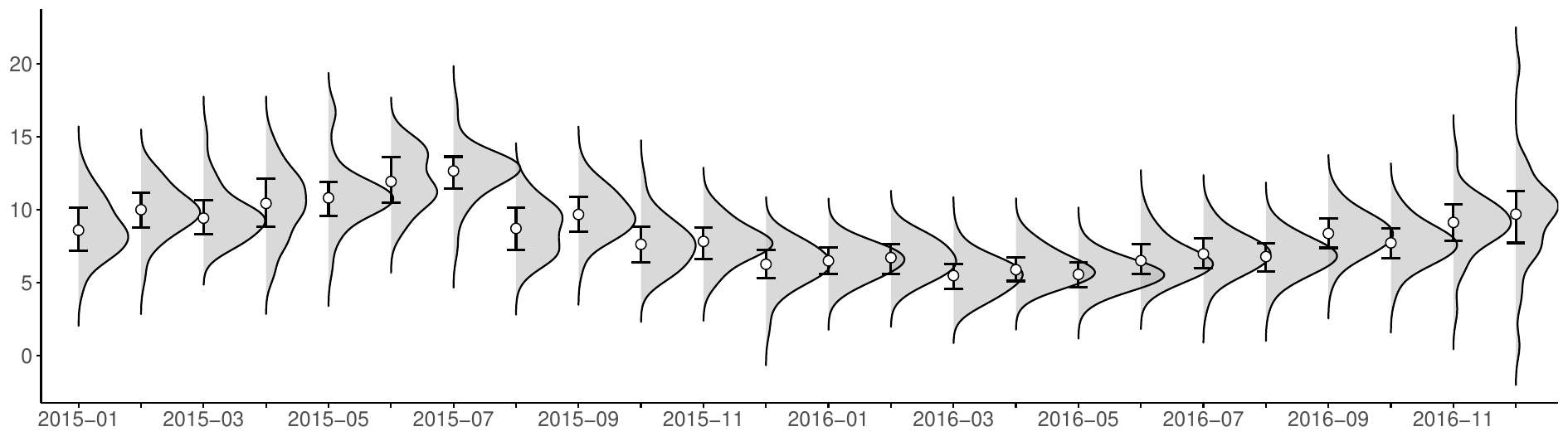} 
\end{tabular}
 \caption{\textbf{Monthly network density} of the average number of commenters in common between any two Italian outlets. The gray areas indicate the kernel density estimates, the bars display the interquartile range, and the dots denote the median.}
\label{fig:frenchnet1}
\end{figure}

Figure \ref{fig:frenchnet} provides an illustrative example of both media bias and polarization starting from a preliminary analysis of the dataset described in Section \ref{sec:application}. The figure displays a network of Italian news outlets in which the  thickness of the edges is proportional to the number of Facebook commenters in common between any two outlets in the years 2015 (top) and 2016 (bottom). While media bias, in terms of political leaning, can be inferred indirectly from the network's structure or directly by analyzing news outlets' content production, an increase in polarization can be detected when the average number of users interacting with various sources decreases (see Figure \ref{fig:frenchnet1}).

Media bias refers to the dissemination of biased news pieces, often with the aim of supporting the interests of certain individuals or groups, such as political parties. The phenomenon is considered detrimental to consumer welfare (\citealt{gentzkow2015media}) as it entails a reduction in the informativeness of news pieces, while some argue that biased news outlets driven by both ideological interests and profits could even affect political outcomes (\citealt{anderson2012media}). Recent advancements in measuring media bias include the implementation of text-analysis techniques to account for the similarity between news articles and political content (see \citealt{gentzkow2010drives, garz2020partisan}), as well as the use of both Stochastic Block Models (SBMs; \citealt{lee2019review}) and Latent-Space (LS) models  (\citealt{hoff2002latent, friel2016interlocking, sewell2016latent}). SBMs are employed to capture discrete group structures, such as clusters of ideologically aligned actors or political communities (see \citealt{Peixoto19}). LS models, on the other hand, allow for infinitely many levels of political leaning and induce a node ordering by representing social-media relational data in a continuous latent space (\citealt{barbera2015birds}; \citealt{aoas_irish}).

\begin{figure}[t]
  \centering
  \begin{tabular}{c}
   \includegraphics[trim={0cm 6.3cm 0cm 0cm},clip,width= 0.85\textwidth]{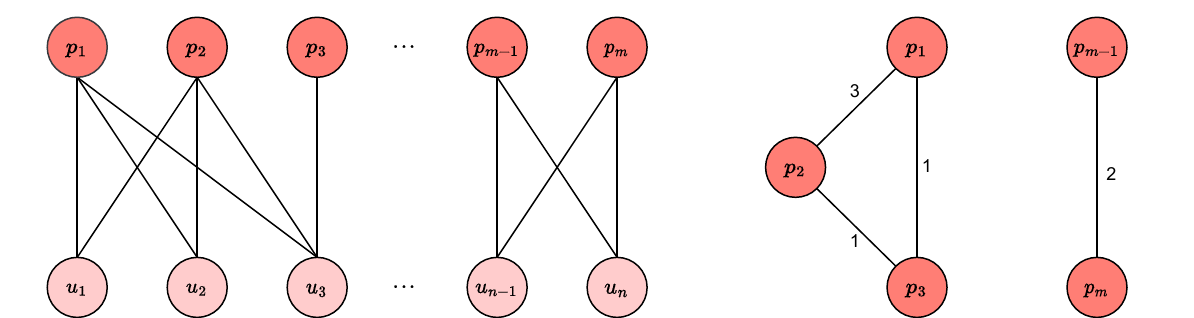}\\
    \includegraphics[trim={1cm 1.1cm 0cm 7.3cm},clip,width= 0.9\textwidth]{Figures/Bip_A.drawio_4.pdf}
   \end{tabular}
   \caption{\textbf{Media Networks:} An example of a bipartite network of $n$ readers, $u_1, \ldots, u_n$ and $m$ news outlets, $p_1, \ldots, p_m$ (top left) and the corresponding audience duplication network obtained using one-mode projection (top right). At the bottom, the adjacency-matrix representation of both the bipartite network, $B_{n\times m}$, and the audience-duplication network $A_{m \times m}$. The matrix $A$ is obtained as $A = B'B$.  }
 \label{fig:bptproj}
\end{figure}

Polarization refers to the radicalization of people's opinions in the sense that they are further apart from one another. Some fear that this collective change in attitudes may be reflected in more partisan positions of people's representatives even though there is no obvious evidence of this (\citealt{prior2013media}). Others claim that online social media exacerbate polarization by offering incentives for homophilous behavior, i.e.~the tendency to interact with similar individuals (\citealt{dandekar2013biased}). However, while a predisposition toward homophily has been observed on several social platforms (see \citealt{hanusch2019journalistic, cinelli2021echo}), evidence of an exacerbation of polarization in social media environments is mixed (\citealt{kubin2021role}).
Several different methodologies have been adopted for measuring polarization  (see \citealt{esteban1994measurement,yarchi2021political}), including in the field of network science (see \citealt{garimella2018quantifying, cinelli2021echo}).
Two common objects of investigation are bipartite networks, which relate social media users to online pages, and audience duplication networks, where nodes represent pages and weighted edges denote the number of users in common between any pair of pages (a one-mode projection of the bipartite network). Figure \ref{fig:bptproj} illustrates the two concepts. 

In this paper, we present a novel dataset of time-varying media networks. We construct reader-user bipartite networks and audience duplication networks using tick-by-tick information on the Facebook activity of national and local news outlets, as collected by \citet{schmidt2018polarization}.  The dataset comprises all posts published by the news outlets from four European countries (France, Germany, Italy, and Spain) along with the corresponding users' interactions for the years 2015 and 2016. Our bipartite network explicitly captures the presence or absence of user–outlet interactions in terms of comments, while the derived audience duplication network is constructed by assigning edge weights between pairs of news outlets proportional to the number of users who comment on posts from both outlets. Following common practice in the literature \citep{peng2022anatomy}, we aggregate user comments at the outlet level to obtain a more stable representation of each outlet’s overall editorial line, avoiding the noise inherent in post-level heterogeneity. The resulting dataset is openly available. See Appendix~\ref{J:repository}, in the Supplementary Materials (\citealp{Casarin2025Supplement}) and the online repository (\citealp{Casarin2025Repository}).

Previous studies about media polarization use heuristics to detect communities and informal sequential analysis for time variation, while we propose a formal statistical framework for dynamic polarization analysis, treating polarization as a collective feature of the actors involved. 
In particular, we introduce a novel dynamic LS network model that leverages both time-varying online audience-duplication network data and textual content to characterize a set of news outlets in terms of both a dynamic latent political leaning dimension and popularity via individual effects.

LS models project the nodes of a network on a lower $d$-dimensional latent space \citep{hoff2002latent}. Extensions of the original model include, e.g. a dynamic component for the latent coordinates \citep{friel2016interlocking,sewell2016latent, kim2018review}, mixtures of latent coordinates (see \citealp{handcock2007model}), and common or multiple latent spaces for multi-layer networks (see \citealp{gollini2016joint,AOASDangelo,sosa2022latent}). 
The statistical properties of LS models have been addressed by \citet{rastelli2016properties} for binary networks, while \citet{barbera2015birds} presents an early application of LS modeling for the estimation of latent ideology on social media and \citet{denicola23} provides a static assessment of media polarization using LS models.

Our paper adds to the methodological literature in several respects: \begin{itemize}

\item The latent coordinates of our LS model provide a novel measure of media bias which combines information from network data (e.g. \citealp{barbera2015birds}) and from a state-of-the-art text-based indicator (e.g. \citealp{gentzkow2015media,garz2020partisan});

\item We endow the latent coordinates with Markov-Switching (MS) dynamics which allows for capturing polarization regimes over time. A similar approach has been adopted for change-point detection in the context of LS projection models (\citealp{Park2020}). The choice of modeling regimes rather than continuously varying levels of polarization is coherent with the literature on opinion formation (see \citealp{iyengar2012affect, tornberg2021modeling}). In addition, the MS model is parsimonious and more easily scalable than other dynamic specifications.

\item We extend the results of \citet{rastelli2016properties} on statistical properties of LS models to weighted temporal networks with MS dynamics (MS-LS models) by obtaining closed-form expressions for the first and second moment of the strength distribution. Besides the theoretical relevance of the results, we show how such closed-form expressions can be used to assess the adequacy of the proposed model against other competing models (See Section \ref{subsec:dynamic_analysis}).
\end{itemize}

Our model is applied to our novel temporal network dataset to provide estimates of media bias that
are coherent with the PEW Research survey \citep{mitchell2018western}. We also shed light on the in-platform (i.e.~within Facebook) polarization regimes across the four countries. We find evidence of cross-country heterogeneity in the shifts from low to high polarization regimes, in line with the sociological argument of \citet{prior2013media}.

Section 2 provides an overview of the model within a Bayesian framework and discusses its statistical properties.  In Section 3, we discuss posterior inference along with the constraints used in our model and present a simulation exercise. Finally, Section 4 describes the dataset of European news outlets and applies our model in both static and dynamic setups.
    
\section{The Longitudinal Markov-Switching Latent Space Model}

\subsection{The Model}\label{sec:Model}

Let $\mathcal{G}=\{\mathcal{G}_{t},\, t=1,2,\ldots, T\}$ be an undirected and weighted temporal (in our application audience-duplication) network with $\mathcal{G}_{t}=(V_{t},E_{t},\mathbf{Y}_{t})$. For each time instance $t$, the vertex set (collection of news outlets) is constant, i.e. $V_{t}=V \subset \mathbb{N}$, and the edge set $E_{t}\subset V\times V$ (common commenters between any two outlets) is time-varying. For each edge $(i,j)\in E_{t}$ we assume the $(i,j)$-th element of the weighted adjacency matrix $\mathbf{Y}_{t}$, i.e. $Y_{ijt}$, is observed and denotes the number of connections (commenters in common) between news outlets $i$ and  $j$ at time $t$. We adopt a Poisson model for the connections:
\begin{equation}
{Y_{ijt}|\lambda_{ijt} }\stackrel{ind}{\sim} \mathcal{P}oi(\lambda_{ijt}),
\label{eq:pois}
\end{equation}
for $i,j=1,\ldots,N$ $i \neq j$, $t=1,\ldots,T$, where $N=Card(V)$ is the number of nodes in the network and $\mathcal{P}oi(\lambda)$ denotes a Poisson distribution with intensity parameter $\lambda>0$. In our LS model, the intensity is driven by static ($\alpha_{i}$) and $d$-dimensional dynamic node-specific latent features ($\mathbf{x}_{it}\in\mathcal{X}\subset\mathbb{R}^{d}$):
\begin{equation}
\log \lambda_{ijt} = \alpha_{i} + \alpha_{j}  - \beta||\mathbf{x}_{it}- \mathbf{x}_{jt}||^2.
\label{eq:intensity} 
\end{equation} 

The parameters $\alpha_{i}$, $i=1,\ldots, N$ have the natural interpretation of individual effects which are news-outlet specific and capture the popularity of the outlet (the engagement of the audience with the newspaper). 
 The latent features $\mathbf{x}_{it}$ 
 are realizations of the latent variables $\mathbf{X}_{it}$, $i=1,\ldots, N$, and enter the log intensity through the squared Euclidean distance as suggested in \citet{gollini2016joint} and in \citet{AOASDangelo}. This accounts for a clearer representation of the proximity of news outlets on a latent manifold and leads to faster computational convergence compared to the standard  Euclidean distance. Assuming $\beta>0$ lends an interpretation of similarity features to the latent variables. The more similar the news outlets (the closer the nodes), the higher the number of commenters they tend to have in common. We also employ an observed political leaning proxy, {$L_{it}$}, to provide additional information on the location of news outlets within the latent space. This political leaning measure takes values between zero, denoting extreme left, and one, denoting extreme right (e.g., see \citealp{gentzkow2010drives, garz2020partisan}). Our modeling choice is not to include $L_{it}$ in the log-intensity equation to preserve the tractability of the random graph model properties. Rather, we assume that the political-leaning proxy {$L_{it}$} is driven by the same latent variables $\mathbf{X}_{it}$ as appear in the network log-intensity. Since the leaning measure $L_{it}$ is continuous in $(0,1)$, a Beta-logistic regression model is assumed, which is a convenient and interpretable specification, although other choices can be used (e.g., see \citealp{branscum2007bayesian, Roberto2012}):
\begin{eqnarray}\label{eq:betaLean}
{L_{it}}|\mu_{it}, \phi &\stackrel{ind}{\sim}& \mathcal{B}e(\mu_{it}\phi, (1-\mu_{it})\phi),
\end{eqnarray}
where $\mathcal{B}e$ denotes the Beta distribution with location and precision parameters $\mu_{it} \in (0,1)$ and $\phi > 0$, respectively.  To endow the latent features $\mathbf{X}_{it}$ with a media-bias interpretation while ensuring $Y_{ijt}$ is independent of $L_{it}$ conditional on $\mathbf{X}_{it}=\mathbf{x}_{it}$, we assume:
\begin{eqnarray}\label{eq:betaLean2}
\mu_{it} =  \varphi(\gamma_{0} +\boldsymbol{\gamma}_{1}^{\prime} \mathbf{x}_{it}),
\end{eqnarray}
where $\varphi(x)=1/(1+\exp(-x))$ is the logistic function. With this modeling choice, we can interpret the latent space as the political spectrum of news outlets.

In the media environment, news outlets are subject to time-varying polarization regimes (e.g., \cite{macy2021polarization, leonard2021nonlinear} on polarization dynamics). One can think about a state of low polarization, where news outlets are perceived, on average, closer within the political spectrum, and a state of high polarization, in which news outlets are perceived politically further apart. For this reason, we specify a Markov-Switching (MS) regime process $\{S_t,t=1,\ldots, T\}$ that drives the dynamic latent features. We assume a Hidden Markov Chain with $K < \infty$ possible states of the world. The MS process allows our latent features $\mathbf{x}_{it}$ to vary jointly through time across the different polarization states. We can reparameterize our dynamic latent features $\mathbf{x}_{it}$ to account for MS dynamics:
\begin{equation}\label{eq:ms}
\mathbf{x}_{it}=\sum_{k=1}^{K}\mathbb{I}(s_{t}=k)\boldsymbol{\zeta}_{ik},
\end{equation}
where $\boldsymbol{\zeta}_{ik}$ denotes the latent position of news outlet $i$ in state $k$, while $\mathbb{I}(s_{t}=k)$ is an indicator function which is 1 if the realization of the hidden variable $S_{t}$ is $s_t=k$ at time $t$, and 0 otherwise. 
Moreover, we characterize the transition between states through:
\begin{equation}
\mathbb{P}({S_{t}}=k|{S_{t-1}}=l)=q_{lk},\, l,k=1,\ldots,K,
\end{equation}
 which can be grouped in the transition probability matrix $\mathbf{Q} = \{\boldsymbol{q}_{1}, \ldots, \boldsymbol{q}_{k}, \ldots,  \boldsymbol{q}_{K}  \}$, where $\boldsymbol{q}_{k} = \left(q_{k1}, \ldots, q_{kK}\right)$ denotes a column vector such that $\boldsymbol{q}'_{k}\mathbf{1} = 1$ for each state $k$. Since the number of latent variables increases as $\mathcal{O}(dKN + T)$, our model is more parsimonious than a continuously varying LS model \citep[e.g., see][]{sewell2016latent}, where the number of latent variables increases as $\mathcal{O}(dTN)$. Thus, our approach is computationally much less demanding for dynamic network applications with large $T$. Our approach also differs substantially from SBMs. While SBMs would cluster the news outlets into discrete groups, our MS-LS ranks news outlets on a continuum interpretable as a political spectrum. More importantly, the specification in (\ref{eq:ms}) allows identifying polarization regimes (periods in time characterized by common behavior in terms of political leaning and polarization). Appendix \ref{alternative_specs} in \citet{Casarin2025Supplement} provides more detail on alternative dynamic specifications.
 
We take a Bayesian approach to inference and choose the following prior structure (as explained in Section \ref{subsec:identification} we fix $\beta$ in our empirical application):
\begin{eqnarray}
&&\pi(\alpha_{1},\dots,\alpha_{N}, \{\boldsymbol{\zeta}_{1k}, \dots,\boldsymbol{\zeta}_{Nk}, \sigma_{k}^2\}_{k=1,\dots,K},\gamma_{0},\boldsymbol{\gamma}_{1}, \phi, \mathbf{Q}) =\nonumber\\
&&\pi(\gamma_{0})
\pi(\boldsymbol{\gamma}_{1}) \pi(\phi)\left(\prod_{i=1}^{N} \pi(\alpha_{i})\right) 
\prod_{k=1}^{K}\pi(\sigma_{k}^2)\pi(\boldsymbol{q}_{k})\prod_{i=1}^{N} \pi(\boldsymbol{\zeta}_{ik}|\sigma_{k}^2) , \label{prior}\end{eqnarray}
where we find it useful to augment the parameter space with $\sigma_{k}^2, k=1,\dots, K$ and the specific prior distributions assumed are detailed in Appendix \ref{sec:AppPrior} (\citealp{Casarin2025Supplement}). 
In our implementation of the model, we have little prior information at our disposal and we opt for the use of relatively vague priors to let the data speak. Our results are robust to substantial changes in these priors. The directed acyclic graph in Figure \ref{fig:DAG} summarizes our MS-LS model. 

\begin{figure}[t]
\centering

\resizebox{0.7\textwidth}{!}{  

\tikzstyle{hyper}= [circle, fill=white, inner sep=1pt, dashed, minimum size=40pt, font=\fontsize{15}{10}\selectfont, draw=black]
\tikzstyle{latent}= [circle, fill=lightgray, inner sep=1pt, solid, minimum size=40pt, font=\fontsize{15}{10}\selectfont, draw=black]
\tikzstyle{par}= [circle, fill=lightgray, inner sep=1pt, solid, minimum size=40pt, font=\fontsize{15}{10}\selectfont, draw=black]
\tikzstyle{obs}= [circle, fill=white, inner sep=1pt, solid, minimum size=50pt, font=\fontsize{15}{10}\selectfont, draw=black]
\tikzstyle{obse}= [circle, fill=white, inner sep=1pt, solid, minimum size=40pt, font=\fontsize{15}{10}\selectfont, draw=white]
\tikzstyle{vertex} = [draw, black, circle, minimum size=30pt, inner sep=0pt]
\tikzstyle{plate} = [draw, rectangle, rounded corners, minimum width=40pt, minimum height=40pt, fit=#1]
\begin{tikzpicture}[x=1.8cm,y=2.0cm]
	\node [latent] (xt) at (-6.8,2) {{$\mathbf{X}_{t}$}};	
	\node [latent] (st) at (-6.8,3) {{$S_{t}$}};
 \node [latent] (sigt) at (-9.5, 2) {$\sigma^2_{k}$};
  \node [latent] (zigt) at (-8.5, 2) {{$\mathbf{Z}_{\sdot k}$}};	
 \node [latent] (g0) at (-9.5,  -0.75) {$\gamma_{0}$};
 \node [latent] (g1) at (-8.5,   -0.75){$\boldsymbol{\gamma}_{1}$};
\node [latent] (phi) at (-10.5,  -0.75){$\phi$};

    \node [latent] (alpha) at (-9.5, 0.6) {$\boldsymbol{\alpha}$};
\node [latent] (beta) at (-8.5, 0.6){$\beta$};

 \node [obs] (Yt) at (-6.8, 0.40) {$Y_{ijt}$};
	\node [obs] (Zt) at (-6.8,-1) {{$L_{it}$}};
	\node [latent] (xt1) at (-4.5,2) {{$\mathbf{X}_{t+1}$}};
 	\node [latent] (st1) at (-4.5,3) {{$S_{t+1}$}};

    \node [obs] (Yt1) at (-4.5, 0.40) {$Y_{ij,t+1}$};
    \node [obs] (Zt1) at (-4.5,-1) {{$L_{i,t+1}$}};
     
    \node [obse] (empx) at (-10.5, 3) {$\ldots$};
    \node [obse] (empx1) at (-2, 3) {$\ldots$};   
   \plate [] {plate_l}{(g0)  (g1)  (phi) }{};
    \plate [] {plate_y}{ (alpha) (beta) }{};
    \plate [] {plate_zi}{ (sigt) (zigt) }{$k = 1, \ldots, K$};
\tikzstyle{EdgeStyle}=[post]

\tikzstyle{EdgeStyle}=[post,bend right=20]

\tikzstyle{EdgeStyle}=[post]
   
    \Edge[](st1)(empx1)
    \Edge[](empx)(st)
    \Edge[](st)(st1)
    \Edge[](st)(xt)
    \Edge[](sigt)(zigt)
    \Edge[](xt)(Yt)
    \Edge[](st1)(xt1)

    \Edge[](xt1)(Yt1)

\tikzstyle{EdgeStyle}=[post,bend right=40]

  \Edge[](plate_y)(Yt)
  \Edge[](plate_y)(Yt1)
  \Edge[](plate_zi)(xt)
  \Edge[](plate_zi)(xt1)

  \tikzstyle{EdgeStyle}=[post,bend right = 40]
  \Edge[](plate_l)(Zt)
  \Edge[](plate_l)(Zt1)

  \tikzstyle{EdgeStyle}=[post,bend left=40]
    \Edge[](xt)(Zt)
    \Edge[](xt1)(Zt1)

\end{tikzpicture}}
\caption{\textbf{Directed Acyclic Graph} of the  Markov-switching latent-space model. The graph exhibits the conditional independence structure of the observation model for $Y_{ijt}$ and {$L_{it}$} within white circles with parameters, $\boldsymbol{\alpha} = \{\alpha_1,\ldots,\alpha_{N}\}$, $\beta$, $\gamma_{0}$, $\boldsymbol{\gamma}_1$, $\phi$, latent coordinates ${\mathbf{X}_{t} = \{\mathbf{X}_{1t},\ldots,\mathbf{X}_{Nt}\}}$, their state-dependent counterparts {$\mathbf{Z}_{\sdot k} =\{\mathbf{Z}_{1k},\ldots,\mathbf{Z}_{Nk}\}$}  with variance $\sigma^2_{k}$, and latent states, {$S_{t}$}, within gray circles.}\label{fig:DAG}
\end{figure}

\subsection{Model Properties}\label{sec:properties}

We now present some of the properties of the MS-LS Model in (\ref{eq:pois}),  (\ref{eq:intensity}) and (\ref{eq:ms}). With Assumption \ref{Ass:Indep}, we extend the scope of the Latent Variable Model provided in \citet{rastelli2016properties} to weighted temporal networks. In this section, we do not consider the political-leaning equation (\ref{eq:betaLean}) since this is quite specific to our application and we aim to present the properties of a general network model. 

\begin{assumption}\label{Ass:Indep}
Given an undirected temporal network, $\mathcal{G}_t=(V,E_t, \mathbf{Y}_t)$, for $t=1,2,\ldots$ having vertex set $V \subset \mathbb{N} $ and weighted edge sets $E_t \subset V\times V$ with characteristic weight $Y_{ijt}$, we assume a sequence of latent coordinates $\{\boldsymbol{X}_{1t}, \ldots,\boldsymbol{X}_{Nt}\}$ for $t=1,2,\ldots$ with $\boldsymbol{X}_{it}\in\mathcal{X}\subset\mathbb{R}^{d}$ for each node $i\in V$ and time index $t\in\mathbb{N}$. 
\end{assumption}

With Assumption \ref{Ass:HMM}, we introduce the Markov-Switching dynamics  and the independence between latent variables across states.

 \begin{assumption}\label{Ass:HMM}
Given a $K$-state latent Markov-chain process ${S_t} \in \{1,2,\ldots, K\}$ for $K< \infty$ and $t=1,2,\ldots$ with transition probabilities $q_{lk}={\mathbb{P}(S_{t}= k |S_{t-1}= l)}$, we assume the latent variables $\boldsymbol{X}_{it}=\sum_{k= 1}^K\mathbb{I}(S_{t}=k)\mathbf{Z}_{ik}$. We also define the set $\boldsymbol{\zeta}_{\sdot k} = \{\boldsymbol{\zeta}_{1k} ,\ldots,  \boldsymbol{\zeta}_{Nk} \}$ consisting of the i.i.d.~realizations of the latent random variables $\{\mathbf{Z}_{1k},\ldots,\mathbf{Z}_{Nk}\}$ with $k\in\{1, \ldots, K\}$, where each $\mathbf{Z}_{ik}$ is distributed according to $\pi_k( \cdot )$, a given probability measure. 
\end{assumption}

Assumption \ref{Ass:Poisson} introduces the conditional independence between any two edges given the latent variables and the current state of the world.

\begin{assumption}\label{Ass:Poisson}
We assume conditional independence between any two edges given the latent variables applicable to the current state {$S_t$}. Given the intensity parameter $\lambda_{ijt}$, $\forall (j, i) \in E_t$, $Y_{ijt}|\boldsymbol{X}_{it} = \mathbf{x}_{it} , \boldsymbol{X}_{jt} = \mathbf{x}_{jt}  \stackrel{ind}{\sim} \mathcal{P}oi\left(\lambda_{ijt}\right)$ is a Poisson random variable.
\end{assumption}

Moreover, we assume that our latent variables are normally distributed and conditionally independent given their variance.

\begin{assumption}\label{Ass:Normal}
The latent variables are normally distributed: $\mathbf{Z}_{ik}|\sigma_k^2\sim \mathcal{N}(\boldsymbol{0}, \sigma_k^2 I_d)$ and take values in $\mathbb{R}^{d}$, for a fixed $d$. 
\end{assumption}

Finally, we specify the form of the intensity parameter $\lambda_{ijt}$.

\begin{assumption}\label{Ass:IntensityPoiss}
Given the individual effects $\alpha_{i}$ and $\alpha_{j}$ and the latent variables, we assume the Poisson rate parameter:
\begin{equation*}
\lambda_{ijt} =\exp\left\{\alpha_{i} + \alpha_{j} - \sum_{k =1}^K\mathbb{I}(s_t = k)\beta ||\boldsymbol{\zeta}_{ik} - \boldsymbol{\zeta}_{jk}||^2 \right\}.\end{equation*}
\end{assumption}

Under Assumptions \ref{Ass:Indep}-\ref{Ass:IntensityPoiss} our model is a time-varying MS-LS model with Poisson weights and normally distributed latent variables. 

The nodal strength, defined as ${Y}_{it} = \sum_{j \neq i}Y_{ijt}$, is a quantity of particular interest when dealing with weighted networks as it provides information on how strongly connected a node is with its neighbors. We can derive the following properties of the probability generating function (pgf) of the nodal strength, where in the sequel we will focus on the strength of a random node (and,  thus, omit the index $i$): 
\begin{proposition}
The $m$-th derivative of the conditional pgf $G_l$ of the nodal strength evaluated in $x=1$  given ${S_{t-1}}=l$ for the MS-LS model can be written as:
\begin{equation*}
    \footnotesize
    \left.\frac{\partial^{m} G_l(x)}{\partial x^{m}}\right|_{x=1} = \sum_{k =1}^Kq_{lk}   \left.\frac{\partial^{m} \widetilde{G}_{k}(x)}{\partial x^{m}}\right|_{x=1}=\sum_{k =1}^K \sum_{\underline{h}_i \in \mathcal{H}_i}  \binom{m}{\underline{h}_i} e^{\sum_{j \neq i}(\alpha_{i}+\alpha_{j})h_j}(\sigma^2_k)^{-\frac{d}{2}} b_{\underline{h},k}q_{lk},
\end{equation*} 
where $\widetilde{G}_{k}(x)$ is the conditional pgf given ${S_t=k}$ and ${S_{t-1}}=l$ and 
\begin{equation*}
b_{\underline{h},k}=\left(\frac{1}{\sigma^2_k}+ \sum_{j \in \mathcal{J}_i}\frac{2\beta h_j}{2\beta h_j\sigma^2_k + 1}\right)^{-\frac{d}{2}}  \prod_{j \in \mathcal{J}_i}\left(2\beta h_j\sigma^2_k+1\right)^{\frac{d}{2}}
\end{equation*} 
with multi-index
$\underline{h}_i = \{h_1,\ldots,h_{i-1}, h_{i+1},\ldots, h_{N} \} $ and index set $\mathcal{H}_i = \{h_j\in\{0,\ldots,m\}, j \neq i| \sum_{j \neq i}h_j = m\}$, $\mathcal{J}_{i} = \{ j| j \neq i, h_j > 0\}$ and $\beta > 0$.
\end{proposition}
 The first derivative of the pgf returns the conditional expectation of the strength for a random node, $\mathbb{E}(Y_t|{S_{t-1}= l})$.

\begin{corollary}
Defining $\alpha =  \alpha_{i} + \alpha_{j} $ for each $i$ and each $j$, the expected nodal strength of the underlying network $\mathcal{G}_t$ can be expressed as 
$$\mathbb{E}(Y_t| {S_{t-1}} = l) = \left.G_l^{\prime}(x)\right|_{x=1} = \sum_{k =1}^Kq_{lk}\left.\widetilde{G}_{k}^{\prime}(x)\right|_{x=1} = (N-1) e^{\alpha}\sum_{k =1}^Kq_{lk}\left(4 \sigma_k^{2}\beta+1\right)^{-\frac{d}{2}}.$$
\end{corollary}

Note that  $\mathbb{E}(Y_t|{S_{t-1}}= l)$ turns out to be a weighted sum of the expected nodal strength obtained by conditioning on each possible state of the world. The result in \citet{rastelli2016properties} can be obtained as a special case imposing all but one of the conditional probabilities $q_{lk}$ for $k \in \{1, \ldots, K\}$ to be zero. This is due to the fact that the expected conditional strength in their unweighted network is the same as in our setup, only with the restriction that $\alpha$ has to be negative. 
The expected strength for each regime increases linearly with the number of nodes, $N$, and exponentially with the intercept parameter $\alpha$. The lower $\sigma^2_k\beta$, the larger the similarity between the nodes in that state and, in turn, the higher the expected strength.

\begin{corollary}
The analytical expression of the variance of the strength distribution uses the first and the second factorial moment of the pgf, resulting in
{\small
\begin{align*}
  \mathbb{V}ar(Y_t|{S_{t-1}}=l) =\sum_{k =1}^Kq_{lk}\mathbb{V}ar(Y_t|{S_{t}} =k)+\sum_{k =1}^Kq_{lk}\left( \left.\widetilde{G}_k^{\prime}(x)\right|_{x=1}- \left.G_l^{\prime}(x)\right|_{x=1} \right)^2,
\end{align*}}
where $\mathbb{V}ar(Y_t|{S_{t}} =k) = \left.\widetilde{G}_k^{\prime\prime}(x)\right|_{x=1} + \left.\widetilde{G}_k^{\prime}(x)\right|_{x=1} -  \left.\widetilde{G}_k^{\prime 2}(x)\right|_{x=1}$.

Similarly, an analytical expression for the dispersion index can be obtained:
\begin{align*}
  \mathfrak{D}(Y_t|{S_{t-1}}= l)&=\sum_{k =1}^Kq_{lk}\mathfrak{D}(Y_t|{S_t}= k)  + v,
\end{align*}
where 
\begin{align*}
\mathfrak{D}(Y_t|{S_t}= k) &= 1+ v_{k} - \left.\widetilde{G}_{k}^{\prime}(x)\right|_{x=1},\quad
v =   \frac{  \sum_{k =1}^Kq_{lk}\left.\widetilde{G}_{k}^{\prime\prime}(x)\right|_{x=1}}{ \sum_{k =1}^Kq_{lk}\left.\widetilde{G}_{k}^{\prime}(x)\right|_{x=1}} - \sum_{k =1}^Kq_{lk}v_{k},
\end{align*}
with $v_{k}=\left.\widetilde{G}_{k}^{\prime\prime}(x)/\widetilde{G}_{k}^{\prime}(x)\right|_{x=1}$. The expressions of the derivatives $\widetilde{G}_{k}^{\prime}(x)$ and $\widetilde{G}_{k}^{\prime\prime}(x)$ are given in (\ref{eq:Gprime}) and (\ref{eq:Gprimeprime}) in Appendix \ref{A:properties} (\citealp{Casarin2025Supplement}).
\end{corollary}

The result for the second factorial moment differs from the results in \citet{rastelli2016properties} as our derivation reflects the existing heterogeneity in weighted edges. Moreover, our MS-LS model allows modeling overdispersion in the strength distribution; 
in particular, we show in Appendix \ref{Sec:AppDisp} (\citealp{Casarin2025Supplement}) that $\mathfrak{D}(Y_t|{S_{t-1}}= l)>1$.

We derive the above results and present a sensitivity analysis of the moments of the strength distribution to different combinations of the model parameters in Appendix \ref{A:properties} (\citealp{Casarin2025Supplement}). To help clarify these results,  Figure \ref{fig:lvm_prop} presents a sensitivity analysis of these network metrics to the parameters of the generative network model: it displays contour plots of moments of the strength distribution of an MS-LS model with $d=1$ and $K=2$. Labels "L" and "H" denote low and high polarization states.
The mean, variance and dispersion index of the strength distribution increase with $\alpha$ and $N$ (see also Figure \ref{fig:di_nodes}, in the Appendix, \citealp{Casarin2025Supplement}), and the mean decreases with  {$\sigma^2_{\mbox{\footnotesize L}}\beta$}. For {$q_{\mbox{\footnotesize LL}}=1$}, when the network stays in low polarization, both the variance and the dispersion initially increase with {$\sigma^2_{\mbox{\footnotesize L}}\beta$}  up to a maximum and then decrease. When a state of high polarization kicks in with {$\sigma^2_{\mbox{\footnotesize H}}\beta > \sigma^2_{\mbox{\footnotesize L}}\beta$}, {\it i.e.}~{$q_{\mbox{\footnotesize LL}}< 1$}, both variance and dispersion  decrease with {$\sigma^2_{\mbox{\footnotesize L}}\beta$}. The smaller  {$q_{\mbox{\footnotesize LL}}$}, the sharper the variance and especially the dispersion index decrease with {$\sigma^2_{\mbox{\footnotesize L}}\beta$}.

\newpage
\section{Inference}

\subsection{Posterior sampling algorithm}
In this section, we will go back to the setup in (\ref{eq:pois})-(\ref{prior}), so including (\ref{eq:betaLean}).
Let $\mathbf{Y}=(\mathbf{Y}_{1},\ldots,\mathbf{Y}_{T})$ be the collection of observed network weights with characteristic element of $\mathbf{Y}_{t}$ given by $Y_{ijt}$, { $\mathbf{L}=(\mathbf{L}_{1},\ldots,\mathbf{L}_{T})$} are the observed  political-leaning proxies with characteristic element  $L_{it}$ and $\mathbf{S} =(S_{1},\ldots,S_{T})$. Consider the parameters $\boldsymbol{\theta} =  \left(\boldsymbol{\alpha}, {\mathbf{Z}}, \boldsymbol{\sigma}^2, \gamma_{0} , \boldsymbol{\gamma}_{1}, \phi, \boldsymbol{Q} \right)$, while $\beta$ is fixed in our empirical implementation (see Section \ref{subsec:identification}). Here {$\mathbf{Z}=(\mathbf{Z}_{\sdot 1},\ldots,\mathbf{Z}_{\sdot K})$ denotes the latent coordinate parameters, where $\mathbf{Z}_{\sdot k}$ is a $d \times N$ matrix}, while $\boldsymbol{\sigma}^2 = (\sigma^2_{1}, \ldots, \sigma^2_{K} )$  is a vector of state-specific variables. The joint posterior $p(\boldsymbol{\theta}| \mathbf{Y}, {\mathbf{L}})$ $\propto f( \mathbf{Y}, {\mathbf{L}}| \boldsymbol{\theta})\pi(\boldsymbol{\theta})$ is not tractable. Thus, we follow a data augmentation approach and apply a Gibbs sampler for posterior inference (see Appendix \ref{C:mcmc}, \citealp{Casarin2025Supplement}). Let us denote with $\boldsymbol{\xi}=(\boldsymbol{\xi}_{1},\ldots,\boldsymbol{\xi}_{T})$ the collection of state indicator variables, where $\boldsymbol{\xi}_{t}=(\xi_{1t},\ldots,\xi_{Kt})'$ and $\xi_{kt}=\mathbb{I}({S_t}=k)$. The complete-data likelihood function is the product  of the following:
\begin{equation}
f(\mathbf{Y},{\mathbf{L}},\boldsymbol{\xi}\vert \boldsymbol{\theta})=\prod_{t=1}^{T}\prod_{i=1}^{N}\left(f_{B}({L_{it}}|{S_t}, \boldsymbol{\theta})\prod_{j = i +1}^{N}f_{P}(Y_{ijt}|{S_t}, \boldsymbol{\theta})\right)\prod_{l=1}^{K}\prod_{k=1}^{K} q_{lk}^{\xi_{lt-1}\xi_{kt}},
\end{equation}
where $f_{P}(Y_{ijt}|{S_t}, \boldsymbol{\theta})$ is the probability mass function of the Poisson in (\ref{eq:pois})  with dynamic intensity given in (\ref{eq:intensity}), $f_{B}({L_{it}}|{S_t}, \boldsymbol{\theta})$ the probability density function of the Beta given in (\ref{eq:betaLean}). With this notation, $\mathbf{x}_{it}$ can be written as ${\boldsymbol{\zeta}_{i\sdot}}\boldsymbol{\xi}_{t}$, in (\ref{eq:intensity}) and (\ref{eq:betaLean2}) where ${\boldsymbol{\zeta}_{i\sdot}}=(\boldsymbol{\zeta}_{i1},\ldots,\boldsymbol{\zeta}_{iK})$ is a $d\times K$ matrix.

We approximate the joint posterior distribution by Markov-chain Monte Carlo (MCMC) sampling. Our Gibbs sampling algorithm iterates the following steps:
\begin{enumerate}
\item Draw $\alpha_{i}$ from $p(\alpha| \ldots)$, $i=1,\ldots,N$ via Adaptive Metropolis-Hastings (MH);
\item Draw $\phi$ from $p(\phi| \ldots )$ via MH with truncated normal proposal;
\item Draw $\gamma_{0}$ and $\boldsymbol{\gamma}_{1}$ from $p(\gamma_{0}, \boldsymbol{\gamma}_{1}| \ldots )$ via MH;
\item Draw $\boldsymbol{\zeta}_{ik}$ from $p(\boldsymbol{\zeta}_{ik}| \ldots )$, $i=1,\ldots,N$ and for $k = 1, \ldots, K$ via Adaptive MH;
\item Draw $\sigma^2_{k}$ from $p(\sigma^2_{k}| {\boldsymbol{\zeta}_{\sdot k}})$ for $k = 1,\ldots,K$;
\item Draw $\boldsymbol{q}_{l}$ from $p(\boldsymbol{q}_{l}|\boldsymbol{\xi})$ for $l = 1,\ldots,K$;
\item Draw $\mathbf{s}$ via the forward-filtering and backward-sampling algorithm  \citep[see][]{fruhwirth2006finite}. 

Further details on the algorithmic design can be found in Appendix \ref{C:mcmc} (\citealp{Casarin2025Supplement}).
\end{enumerate}

\subsection{Identifying Restrictions}\label{subsec:identification}
The model presents several identification challenges. The first issue is related to the multiplication of the squared Euclidean distance $||{\boldsymbol{\zeta}_{i\sdot}}\boldsymbol{\xi}_{t}-{\boldsymbol{\zeta}_{j\sdot}}\boldsymbol{\xi}_{t}||^2$ by the parameter $\beta$. As there is a clear scale indeterminacy between  $\beta$ and the variance of the latent variables in terms of $\sigma_k^2$, we choose to set $\beta = 1$. In addition, latent coordinates enter the parameter $\lambda_{ijt}$ only through the squared distance. This makes -- in principle -- positions that differ just by means of reflection, translation, and rotation equally likely (see \citealp{hoff2002latent} and  \citealp{friel2016interlocking}). Nonetheless, the introduction of  (\ref{eq:betaLean}) helps prevent the emergence of many equivalent latent-space representations. To overcome translation issues, we center the latent coordinates to the origin of the axes at each Gibbs-sampling iteration. For $d=1$, a reflection of the latent coordinate around the origin is possible. To counteract  reflection, we assume that the position of a single outlet is known in terms of left and right political leaning (e.g. $\zeta_{i^*k} < 0$ for a left-leaning outlet $i^*$ and for each state $k$), and we apply a reflection transformation to the latent leaning coordinates every time the latent leaning of  $i^*$ is in the wrong orthant, similarly to \citet{barbera2015birds}.
For $d > 1$, not only reflection remains possible but also rotation. This causes indeterminacy of the parameter $\mathbf{\gamma}_1$. To prevent rotation of the latent coordinate with media bias interpretation, we impose the following restrictions to (\ref{eq:betaLean}): $\boldsymbol{\gamma}_1 = (\gamma_{11},0, \ldots, 0)$ and $\gamma_{11} = 1$. Such a strategy is similar in spirit to the loading restrictions in factor analysis (see \citealp{fruhwirth2024sparse}). For the other coordinates, we apply Procrustes transformation to solve the identification issue, as commonly done for standard LS models (see \citealp{hoff2002latent}).

Finally, as pointed out in \cite{fruhwirth2006finite}, the joint posterior in Markov-Switching models is invariant with respect to a re-labeling of the hidden states. We tackle this issue by imposing an ordering restriction on the latent regimes across states. In particular, we label latent regimes in increasing order of median distance, $\Tilde{D}_{k} = med_{j>i}\left(||\zeta_{ik}-\zeta_{jk}||\right)$. Alternative ordering restrictions may be considered, but we leave these for future research.

\subsection{Analysis of Simulated Data} 
We assess the performance of our inference method by running the algorithm on simulated data from an MS-LS model with $d=1$ and $K=2$. Our simulation consists of 20 fictitious news outlets observed for 100 periods. Each period may belong to one of two polarization states $k \in \{\mbox{L, H}\}$, where State L is characterized by a lower average distance in the political-leaning dimension across news outlets, and State H has a higher average distance. So news outlets jointly undergo periods of high polarization and low polarization.
The simulation setting is inspired by previous studies (\citealp{tornberg2021modeling}) and aligns with our empirical findings. In periods of high polarization, news outlets exhibit a stronger and more radicalized identity, corresponding to a higher average distance in the latent space. This manifests itself with news content that is more aligned with the language of the reference political parties and with an audience that is more homogeneous in terms of news diet, partially reflected by well-separated clusters along the political leaning direction of the latent space. Vice versa, we expect a less pronounced and more heterogeneous identity of news outlets in periods of low polarization with lower outlet separation in the latent space.

The latent leaning positions in State H,  {$Z_{i\mbox{\footnotesize H}}$}, are drawn from a normal distribution centered at -$0.75$ for $i \in \{1,\dots,10\}$ and at  $0.75$ for $i \in \{11,\dots,20\}$ with a standard deviation equal to $0.15$, while the latent leaning positions in State L,  {$Z_{i\mbox{\footnotesize L}}$}, are drawn from a normal centered around the positions -$0.25$ for $i \in \{1,\dots,10\}$ and  $0.25$ for $i \in \{11,\dots,20\}$ 
also with a standard deviation of $0.15$;
the individual effect parameters $\alpha_i$ are randomly drawn from a normal distribution with $\mu_\alpha = 0$ and $\sigma_\alpha = 2$; the transition probability matrix $\boldsymbol{Q}$ has 0.95 on the diagonal and 0.05 as off-diagonal elements, and the sequence of states is randomly drawn from the Markov Chain initialized at State L; finally, we set $\phi = 200$, $\gamma_0 = -0.1$, and $\gamma_1 = 0.5 $. We sample $Y_{ijt}$ and {$L_{it}$}  from the data generating process (\ref{eq:pois})-(\ref{eq:ms}). Our simulation represents a situation in which news outlets diverge in magnitude -- via $\alpha_i$ -- and in terms of political leaning -- via $Z_{ik}$. 

We run our MCMC algorithm for 50,000 iterations and we discard the first 30,000 iterations as burn-in and thin by a factor of 10 to reduce auto-correlation in the draws. To correctly identify left and right-leaning, we consider news outlet 3 as known to be left.

\begin{figure}[!htb]
    \centering
        \includegraphics[width= 0.9\textwidth]{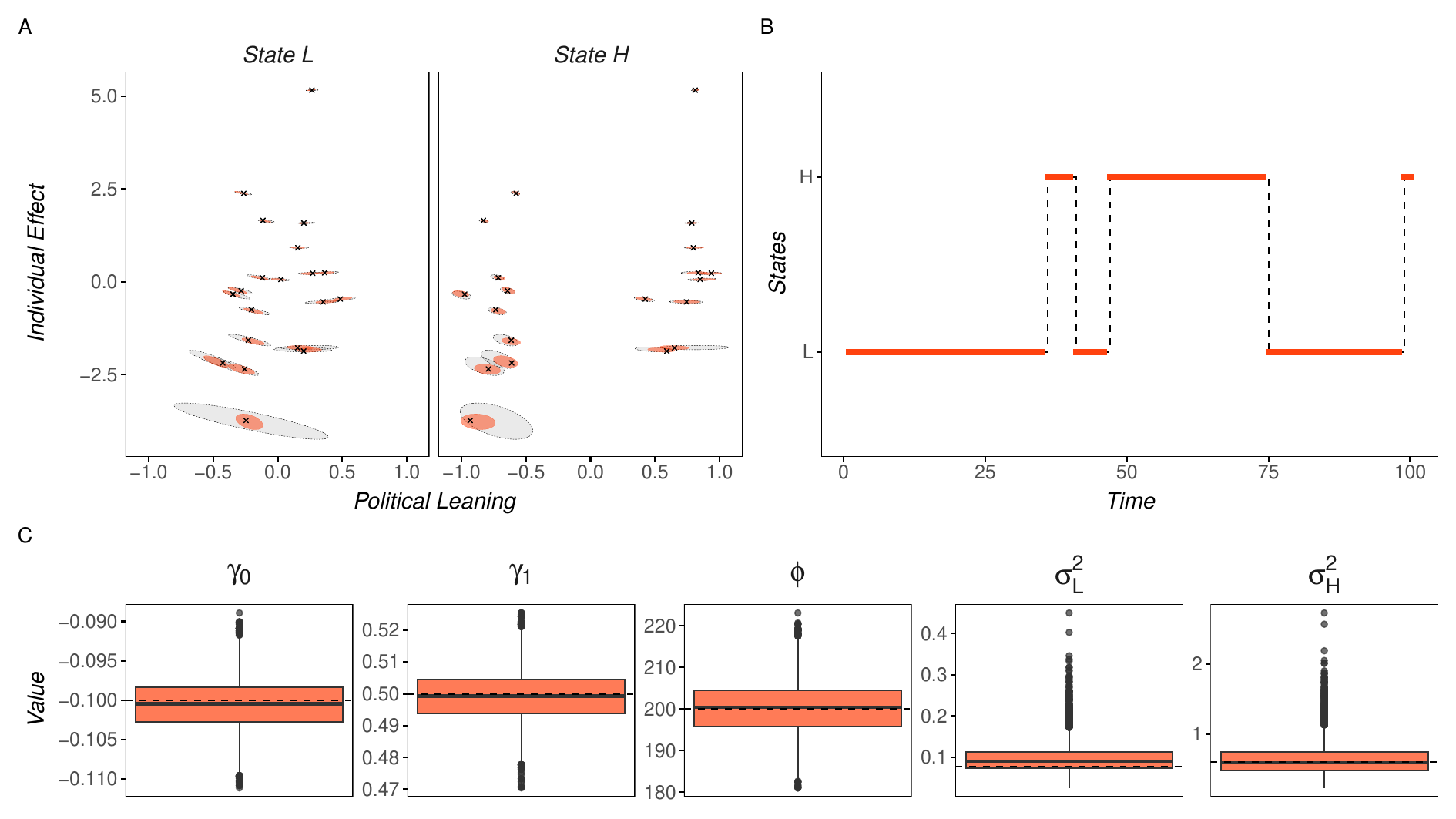} 
    
        \caption{\textbf{Simulated data:} Panel A: Estimated Coordinates in the Latent Leaning - Individual Effect Plane with 99\% credible ellipses for the mode with (\ref{eq:betaLean}) (solid {inner ellipses}), without  (\ref{eq:betaLean}) (dashed {outer ellipses}) and true values (crosses). Panel B: estimated latent states ({squared} dots) and true states (dashed {step line}). Panel C: boxplots of the marginal posteriors for the parameters $\gamma_0$, $\gamma_1$, $\phi$, $\sigma^2_L$ and $\sigma^2_H$, with dashed lines indicating the true values.}
 
  \label{fig:simul_res}
\end{figure}

Figure \ref{fig:simul_res} reports a summary of the simulation results. From a comparison between the true values and the marginal  posterior distributions, the model performs very well in terms of identification of the individual effects and latent variables (Panel A), the latent states (Panel B), as well as the other parameters in the simulation (Panel C). Credible regions for the pairs $(\alpha_{i}, Z_{ik})$ estimated with our model (solid ellipses in Panel A) are narrower than those obtained disregarding the  political-leaning proxy ${L_{it}}$, i.e. dropping  (\ref{eq:betaLean}) from the model (dashed ellipses). This suggests that the information-borrowing strategy is effective in improving the estimation accuracy of the latent variables. As expected, the detection of the regimes deteriorates when the regimes are not well separated in the latent space (see Appendix \ref{sens_analysis}, \citealp{Casarin2025Supplement}). Properties of the MCMC chains are reported in the Supplementary Material (Appendix \ref{E:traceplots}, \citealp{Casarin2025Supplement}). Our MCMC algorithm is implemented in \emph{R} and \emph{C++} and we have made the scripts freely available (see Appendix \ref{J2:repository_script}, \citealp{Casarin2025Supplement}).

\section{Political Leaning and Polarization of News Outlets}
\label{sec:application}
We apply our model to a dataset of daily Facebook activities related to 225 national and local news outlets in France, Germany, Italy and Spain. We provide both static and dynamic analyses to assess the media slant in these outlets as well as the polarization regimes across countries. 

\subsection{Dataset Description and Construction}
Our novel media temporal network dataset, \emph{network dataset} in what follows, comprises the news outlets in the list reported in the Reuters Digital News Report (2017) \citep{newman2017reuters}. 
The tick-by-tick data in the \emph{source dataset} of \citet{schmidt2018polarization} contains all posts published, along with their associated metadata, as well as all data on anonymized user interactions with these posts, in the form of comments. Table \ref{tab:db_desc} reports a summary description of the source dataset. We aggregate comment interactions to daily data at the outlet level by constructing the set of daily bipartite networks between news outlets and \emph{commenters} - those Facebook users commenting on news outlets' posts.  For each country, i.e. \emph{France, Germany, Italy} and \emph{Spain}, at time $t$, we obtain the set of audience-duplication networks $\mathcal{G}_{t}$ presented in Figure~\ref{fig:sample2} by performing the one-mode projection on the side of news outlets (as in  Figure~\ref{fig:bptproj}).

\begin{table}[!htp]
\renewcommand{\arraystretch}{1.1}

\caption{\textbf{Source Dataset Description:} Description of the Facebook dataset on national and local news outlets of the four European countries (France, Germany, Italy and Spain) gathered by \citet{schmidt2018polarization}. The dataset entirely covers the years 2015 and 2016. The news outlet list is  reported in the Reuters Digital News Report (2017).} 
 \label{tab:db_desc}
 
\centering
\begin{tabular}{lcccc}
\hline\hline

\multicolumn{1}{c}{Country}  & Pages & Posts          & Comments     & Commenters  \\ \hline
\multicolumn{1}{l|}{France}  & 65    & 1,008,018  & 47,225,675 & 5,755,268 \\
\multicolumn{1}{l|}{Germany} & 49    & 749,805   & 31,881,407 & 5,338,195 \\
\multicolumn{1}{l|}{Italy}   & 54    & 1,554,817 & 51,515,121 & 4,086,351 \\
\multicolumn{1}{l|}{Spain}   & 57    & 1,372,805  & 34,336,356 & 6,494,725 \\ \hline\hline

\end{tabular}
\end{table}
We complement our network dataset with data from Crowdtangle \citep{team2020crowdtangle} and Chapel Hill Expert Survey (CHES) data \citep{polk2017explaining}. 
Crowdtangle allows for retrieving Facebook posts for public pages and provides additional metadata for each post. In particular, the fields \emph{Link Text} and \emph{Description} contain information on the text of linked pages, such as the texts of news articles published on the Facebook walls of the news outlets. There is not a perfect match between all the pages available in the source dataset of \citet{schmidt2018polarization} and those available in Crowdtangle. Some news outlets may have changed account or ceased to exist. In this case,  information about these news outlets may no longer be available on Facebook at the time of writing. The CHES questions political scientists on different aspects related to politics and European integration. The \emph{CHES dataset} contains all the information at an aggregate level about scientists' opinions on the ideological position of political parties in Europe. Here we will make use of the \emph{lrgen} variable, which provides the ideological stance of a political party from 0 (extreme left) to 10 (extreme right). The information retrieved from Crowdtangle and CHES allows us to construct a text-analysis proxy for media slant. In particular, we obtain our observed proxy for daily media slant  ${L_{it}}$ by computing the index proposed by \citet{gentzkow2010drives} and adapted to online media outlets by \citet{garz2020partisan}. Such a media slant index relies on text analysis techniques to assess the similarity between pieces by news outlets and texts published by politicians. We then associate a political leaning to each news outlet as a function of this similarity and the parties' political leaning. Further information on the adopted methodology can be found in Appendix \ref{F:MediaSlantProxy} (\citealp{Casarin2025Supplement}). 

The network dataset and the media slant index are publicly available as described in Appendix \ref{J1:repository_data} (\citealp{Casarin2025Supplement}).

\begin{figure}[t]
  \centering
   \renewcommand{\arraystretch}{1.3}
\setlength{\tabcolsep}{10pt}
\resizebox{0.75\textwidth}{!}{ 
\begin{tabular}{cc}
   {\scriptsize France} &  {\scriptsize Germany}  \vspace{-2pt}\\\vspace{-10pt}
     \includegraphics[trim={0cm 0cm 0cm 0cm},clip,scale = 0.055]{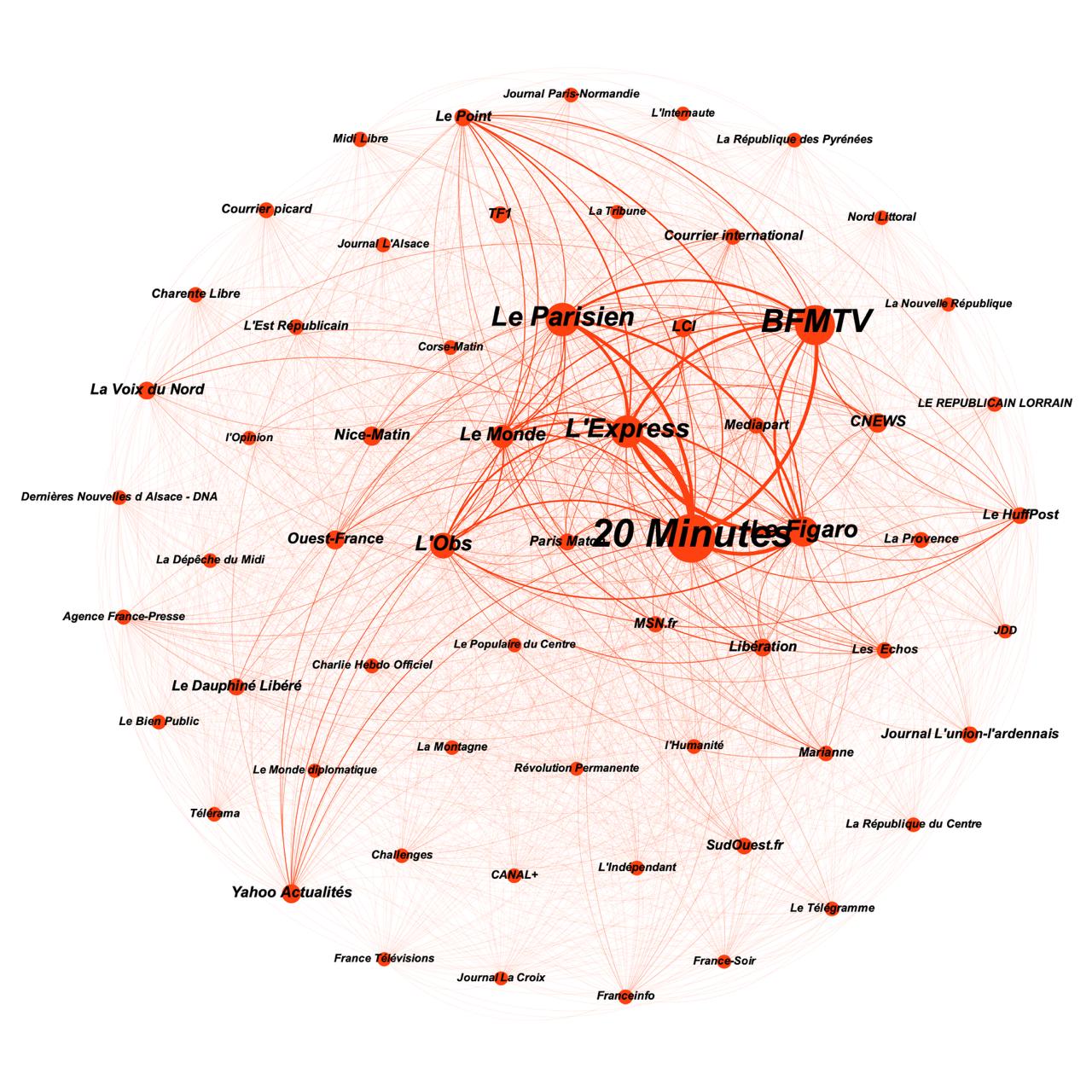} & \includegraphics[scale = 0.055]{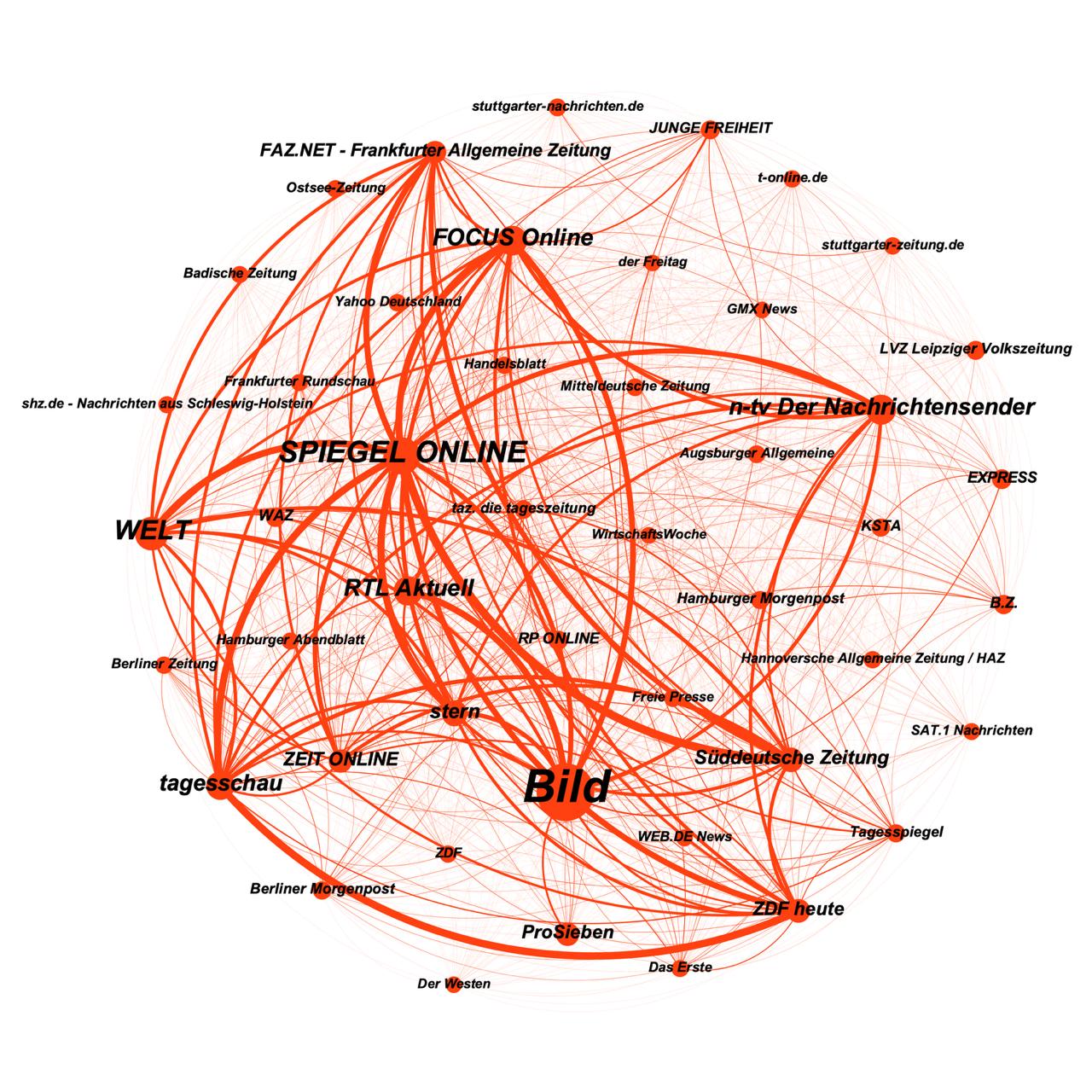}\\
  {\scriptsize Italy} & {\scriptsize Spain}\vspace{-2pt}\\
     \includegraphics[scale = 0.055]{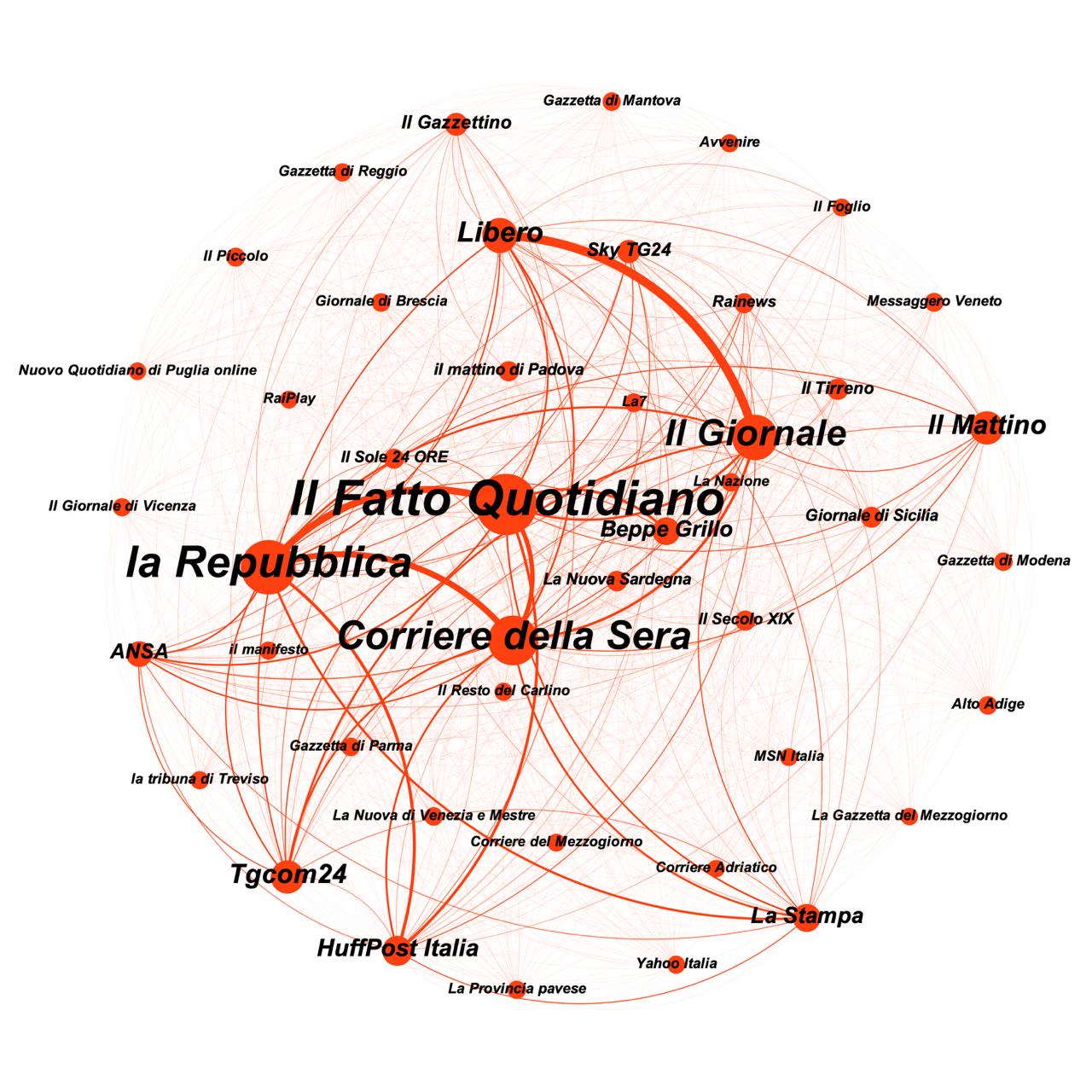}  & \includegraphics[scale = 0.055]{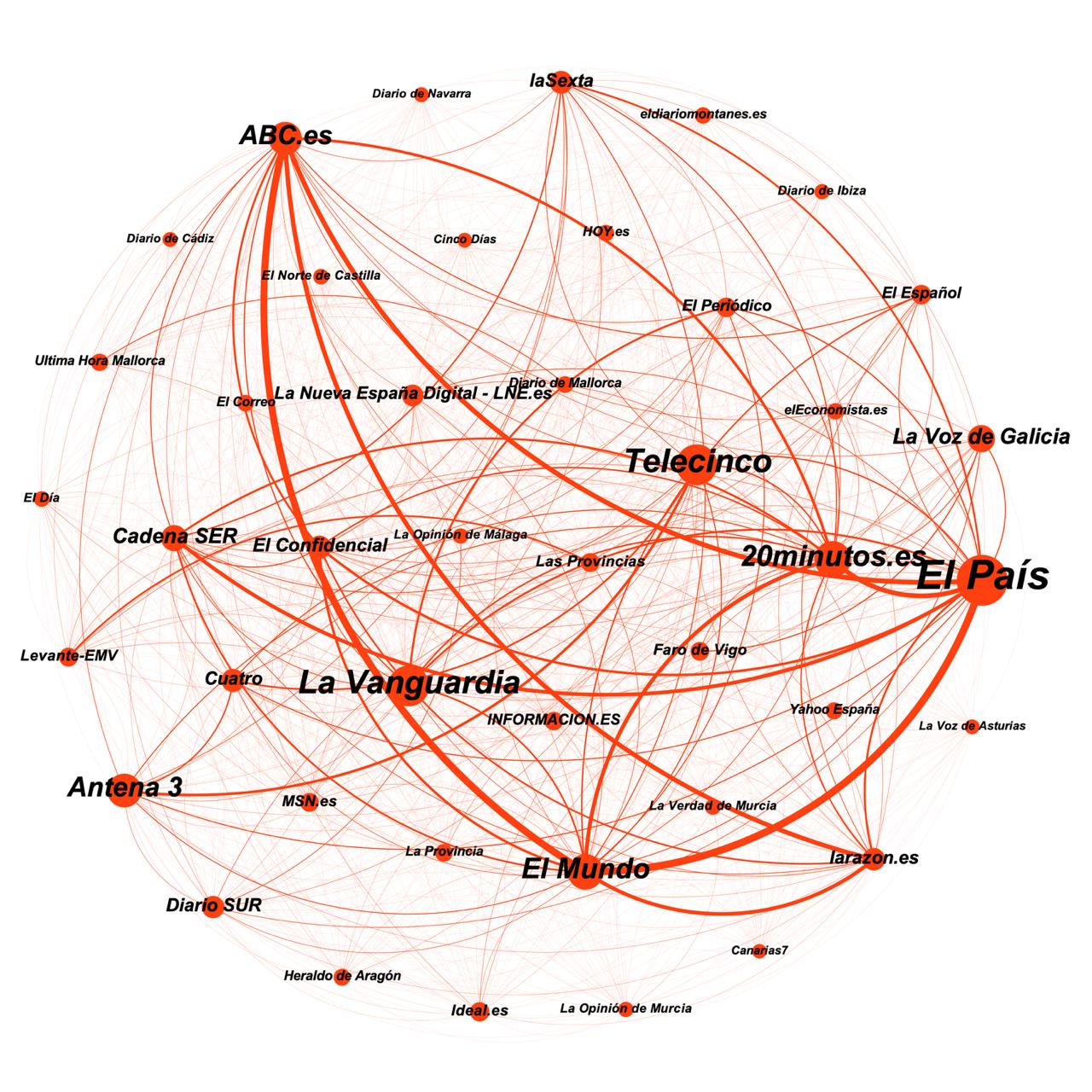}
 \end{tabular}}
  \caption{{\textbf{Audience-Duplication Networks:}} Cumulative networks for France (62 outlets), Germany (47 outlets), Italy (45 outlets) and Spain (43 outlets) from January $1^{st}$ 2015 to December $31^{st}$ 2016. Node size is proportional to the cumulative number of comments received in the time interval. Edge thickness is proportional to the number of common commenters between each pair of outlets. Quantities are normalized for each country.}
   \label{fig:sample2}
 \end{figure}

\subsection{Results from a Static Analysis}

\begin{figure}[t]
  \centering
   \includegraphics[width= 0.85\textwidth]{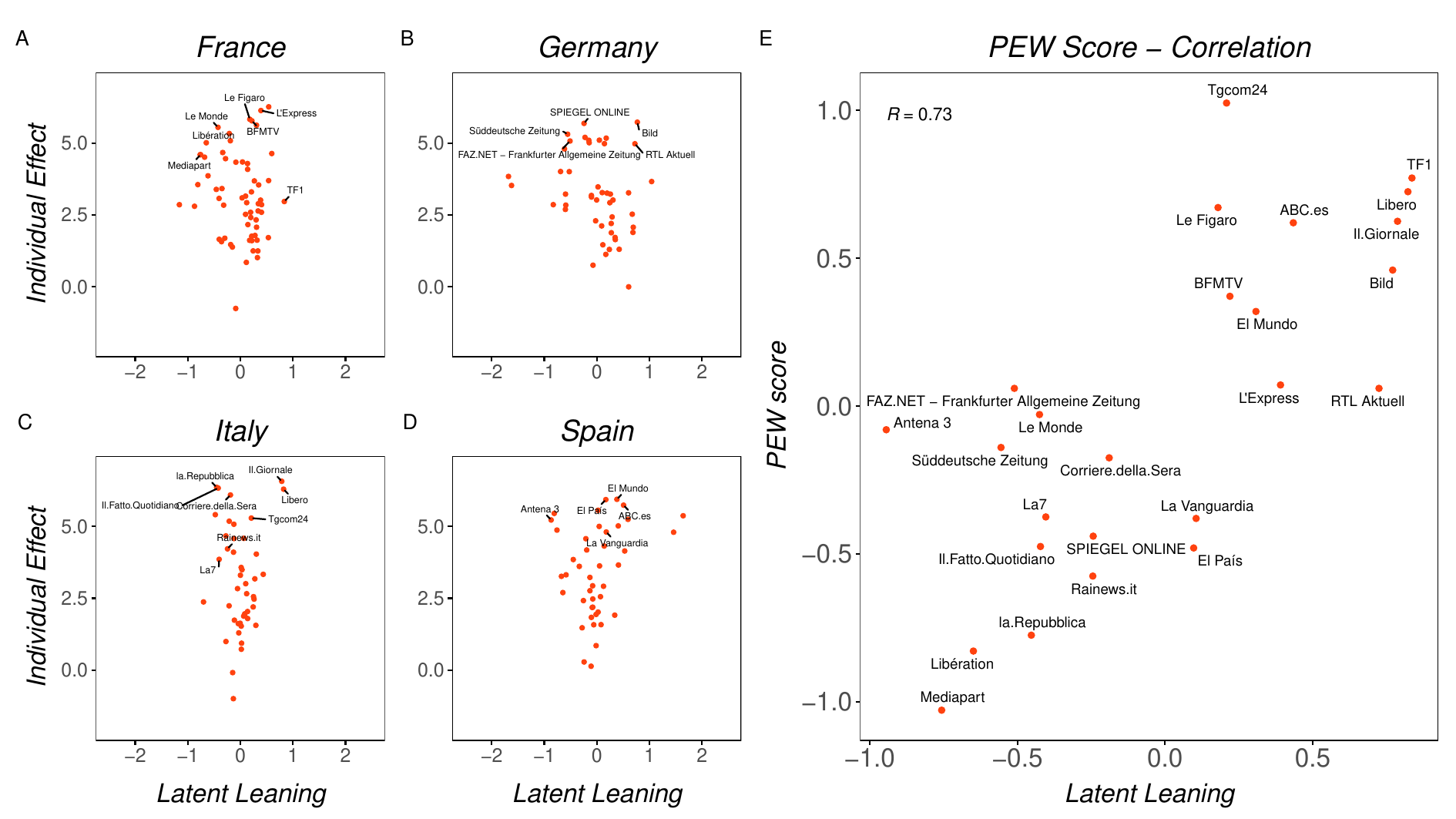}
    \caption{\textbf{PEW Index - Latent Leaning Comparison for the Static Model:} Panels A, B, C, and D present the posterior mean of the latent coordinates of our news outlets, while Panel E displays the scatter plot comparing the PEW survey results, available for 25 major national news outlets, with our estimated latent leaning variable (country-specific means have been subtracted from the PEW scores to improve readability).} 
 \label{fig:rob}
 \end{figure}

First, we implement a static version of our model  with $d=1$ on the whole 2-year time period, without the MS dynamic component. For this, we use an overall audience duplication network $\tilde{\mathcal{G}}_{t}$ for each country, where the weighted edge for each pair of outlets is $\tilde{Y}_{ij} = \sum_{t =1}^TY_{ijt}$, and the overall observed leaning-feature is constructed as  ${\tilde{L}_{i} = T^{-1}\sum_{t =1}^T L_{it}}$. Panels A, B, C, and D in Figure \ref{fig:rob} report the estimated (posterior mean) latent coordinates in the latent leaning-individual effect space. We notice how the individual effect parameter $\alpha_{i}$ associated with each news outlet and country may be interpreted in terms of news outlet's engagement, as major national news outlets are concentrated at the top in each graph. 

We correlate our posterior mean media slant with the results obtained by the PEW Research Survey \citep{mitchell2018western}. In this survey, participants were asked to assess the left-right leaning of major national news outlets (25 of these also appear in our dataset) on a 0-6 scale with 0 indicating far left and 6 indicating far right.  To the best of our knowledge, a survey assessing the left-right leaning of all the outlets in our dataset is not available. We will refer to the left-right ranking obtained by PEW Research as the PEW Research index. We find that the PEW Research index has a $0.73$ correlation with our estimated latent leaning, see Panel E in Figure \ref{fig:rob}. Moreover, we notice the presence of both a left-leaning cluster (bottom-left) and a right-leaning cluster (top-right). As a further validation, Figure \ref{fastgreedy} in Appendix \ref{H:furtherplots} (\citealp{Casarin2025Supplement}) provides a comparison between the estimated leaning and the partition obtained via traditional cluster analysis, using the Fast Greedy algorithm (\citealp{clauset2004finding}). This algorithm provides an optimal number of clusters larger than two and in itself is not able to discriminate between left and right-leaning outlets or rank news outlets on the political spectrum. Nevertheless,
a graphical inspection shows that both the PEW score and our leaning measure can be used to separate the clusters (different symbols) into two large groups: left-wing outlets (bottom-left quadrant) and right-wing outlets (top-right quadrant), coherently with the cluster analysis partition.

Figure \ref{fig:boxplots_static} presents the marginal posterior distribution of the parameters $\gamma_{0}$, $\gamma_{1}$ and $\phi$. The parameter $\gamma_{1}$ conveys information on the relationship between the latent variable $x_{i}$ and the observed leaning proxy {$L_{i}$}.  Latent leaning appears to be a strong driver for the observed proxy only in the case of Italy, as the posterior mass of $\gamma_{1}$ is located far away from zero, while it seems a weak driver for France and mostly irrelevant for both Germany and Spain. Nonetheless, the strong correlation with the PEW Research index suggests that having information on online users' interactions with news outlets may still be sufficient to provide an effective classification on the political spectrum.

\subsection{Results from a Dynamic Analysis}
\label{subsec:dynamic_analysis}
\begin{figure}[t]
  \centering
   \includegraphics[width= 0.85\textwidth]{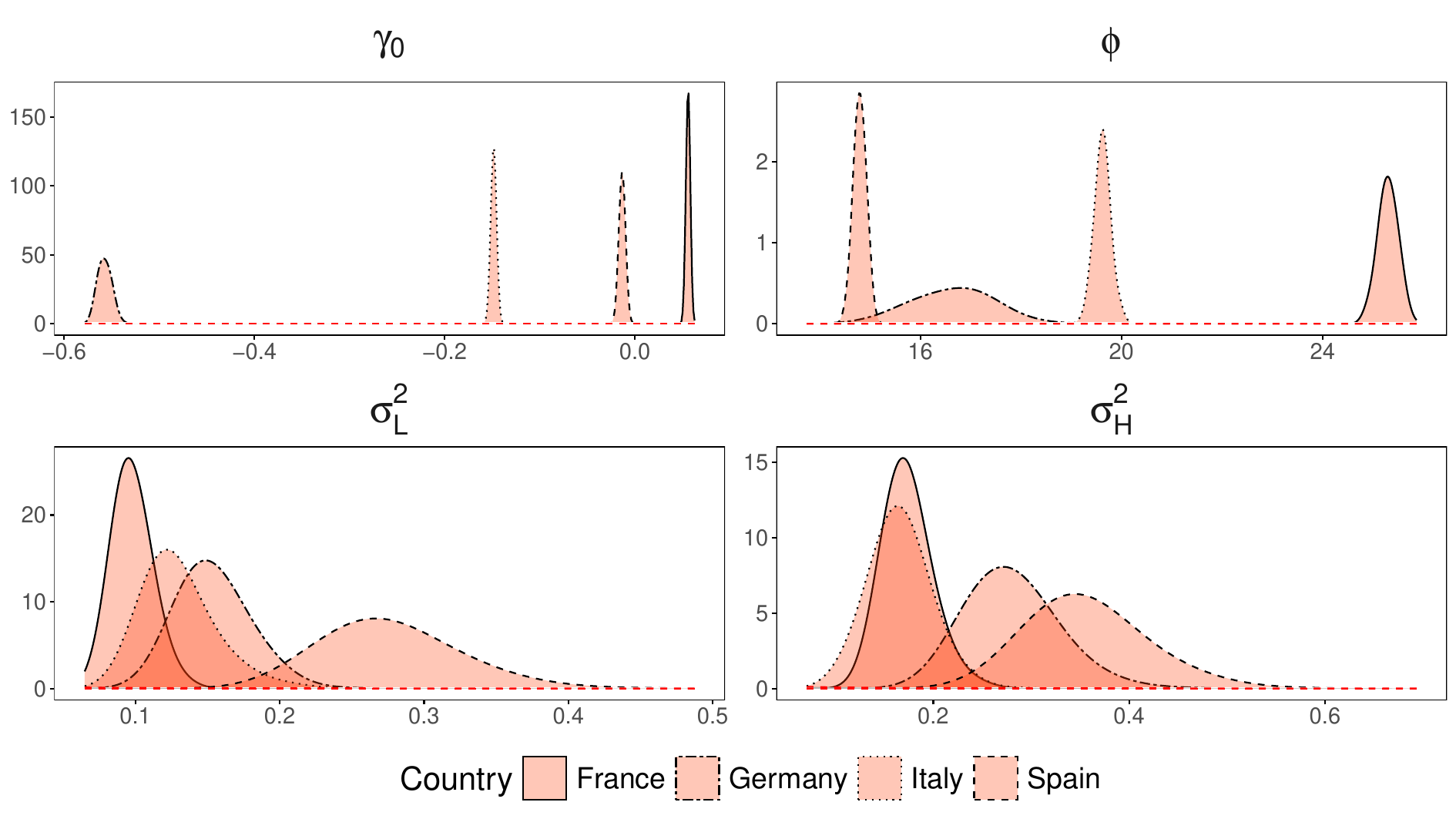}
 
\caption{\textbf{Marginal Posteriors for the Dynamic Model  $\mathcal{M}_4$:} Kernel posterior density estimates for the parameters $\gamma_{0}$,  $\phi$, $\sigma^2_{L}$ and $\sigma^2_{H}$ for France, Germany, Italy and Spain. The corresponding prior distributions (dashed lines) are flat and nearly indistinguishable from the horizontal axis.}

 \label{fig:postDyn}
 \end{figure}

\begin{figure}[t]
  \centering
   \includegraphics[height=450pt, width= 0.8\textwidth]{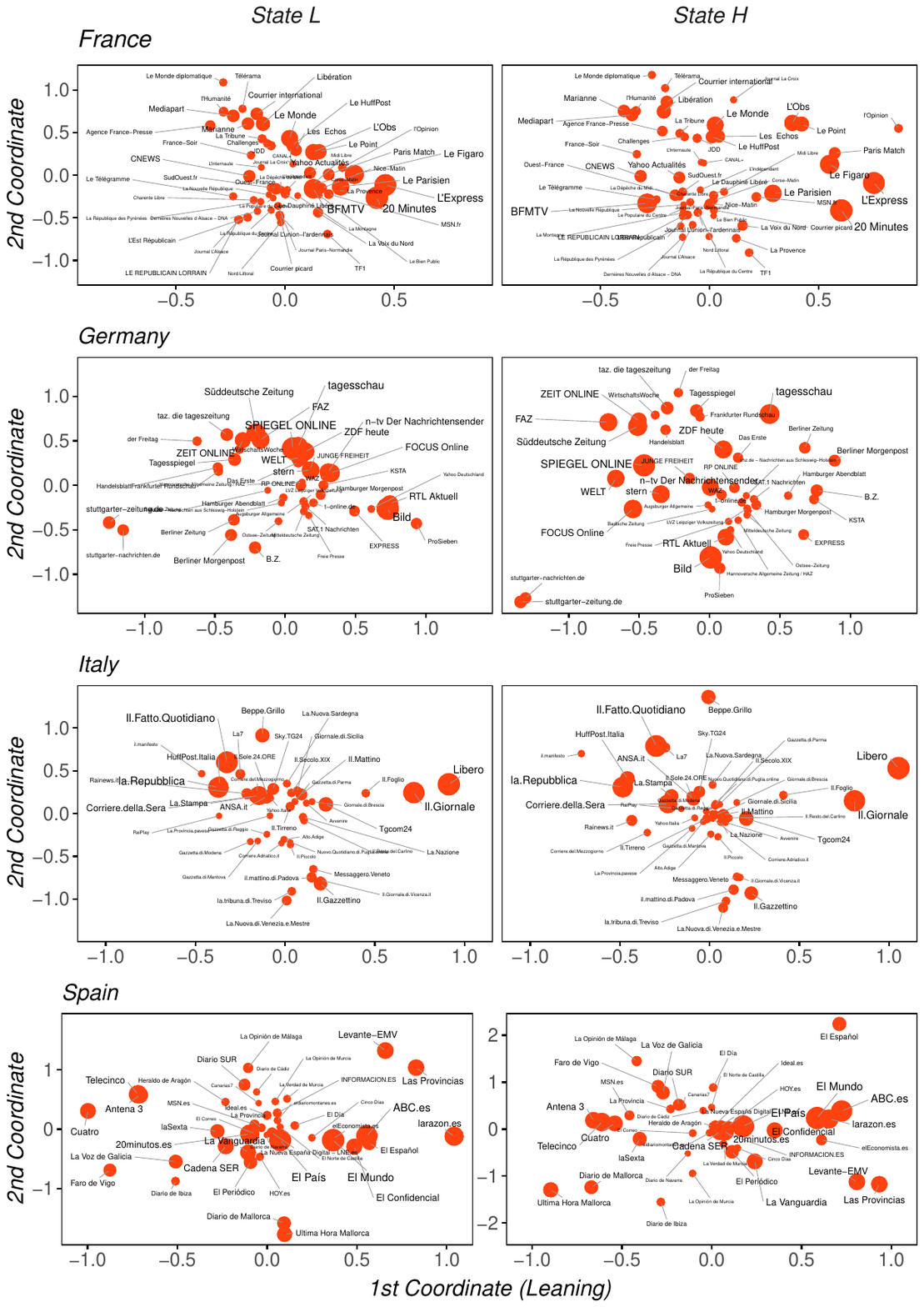}
 \caption{\textbf{Latent Positions for the Dynamic Model  $\mathcal{M}_4$:} Posterior means of latent coordinates of the news outlets for France, Germany, Italy and Spain in State L and in State H. The node size is proportional to the posterior mean of the individual effects.}
\label{fig:dyn_latpos}
\end{figure}

\begin{figure}[!htb]
  \centering
   \includegraphics[width= 0.80\textwidth]{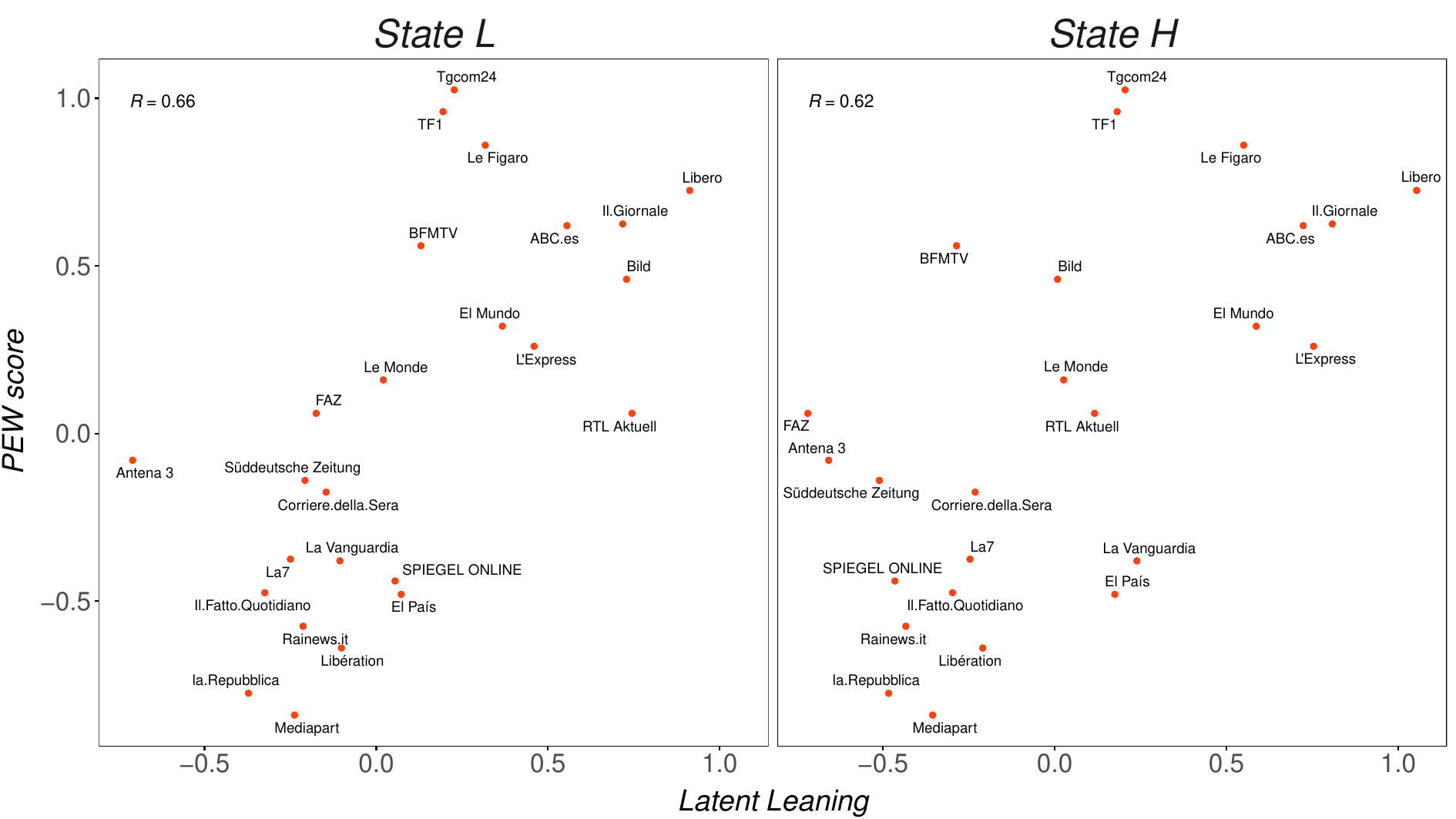}
 \caption{\textbf{PEW Index - Latent Leaning  Comparison for Dynamic Model $\mathcal{M}_4$:} Scatter plot comparing the PEW survey results with the estimated latent leaning variable in State L and H. The country-specific mean has been subtracted from the PEW Score to improve readability.}
\label{fig:rob_dyn_lat}
 \end{figure}

We now estimate the model described in Subsection \ref{sec:Model} in its dynamic specification with $d = 2$ and $K = 2$ (called $\mathcal{M}_4$ in the next subsection) and with $d = 2$ and $K = 5$ ($\mathcal{M}_6$). The former specification is instrumental for a clear  representation of the main characteristics of the latent space, while the latter is the preferred model (see the next section) which also provides an improved fit of the dynamic network features. Model selection is discussed in some detail in the next section. The choice of a different model for each country is also legitimate.
The dynamic analysis uses daily data, and we deleted from our dataset those outlets that remained inactive -- i.e.~did not receive any comment -- for more than 15 consecutive days. Overall, we removed 13 news outlets (DE: 4 outlets, FR: 5, IT: 2, SP: 2, see Appendix \ref{G:Identifiers}, \citealp{Casarin2025Supplement}), which displayed unusual behavior. Posterior results for the parameters of the MS-LS model with $d = 2$ and $K = 2$ are presented in Figure \ref{fig:postDyn}. As expected, values of {$\sigma_{\mbox{\footnotesize H}}^2$} tend to be larger than those for {$\sigma_{\mbox{\footnotesize L}}^2$}. In Figure \ref{fig:dyn_latpos}, we report the posterior means of the latent positions for the four countries in states L and H. The individual-effect values are coherent with the engagement interpretation in both states: well-known national newspapers display larger individual effects (larger point size), while local newspapers display smaller individual effects (smaller point size). Moreover, our latent variable dimension with media bias interpretation shows a positive correlation with the PEW Research Survey Index in this setting. Figure \ref{fig:rob_dyn_lat} illustrates the correlation of 0.66 in the lower polarisation state and 0.62 in the state of higher polarisation. Finally, the second latent coordinate captures news outlets' similarities in other distinctive characteristics, such as geographic vicinity or editorial ownership.

\begin{figure}[htb!]
  \centering
    \includegraphics[height=190pt, width=0.95\textwidth]{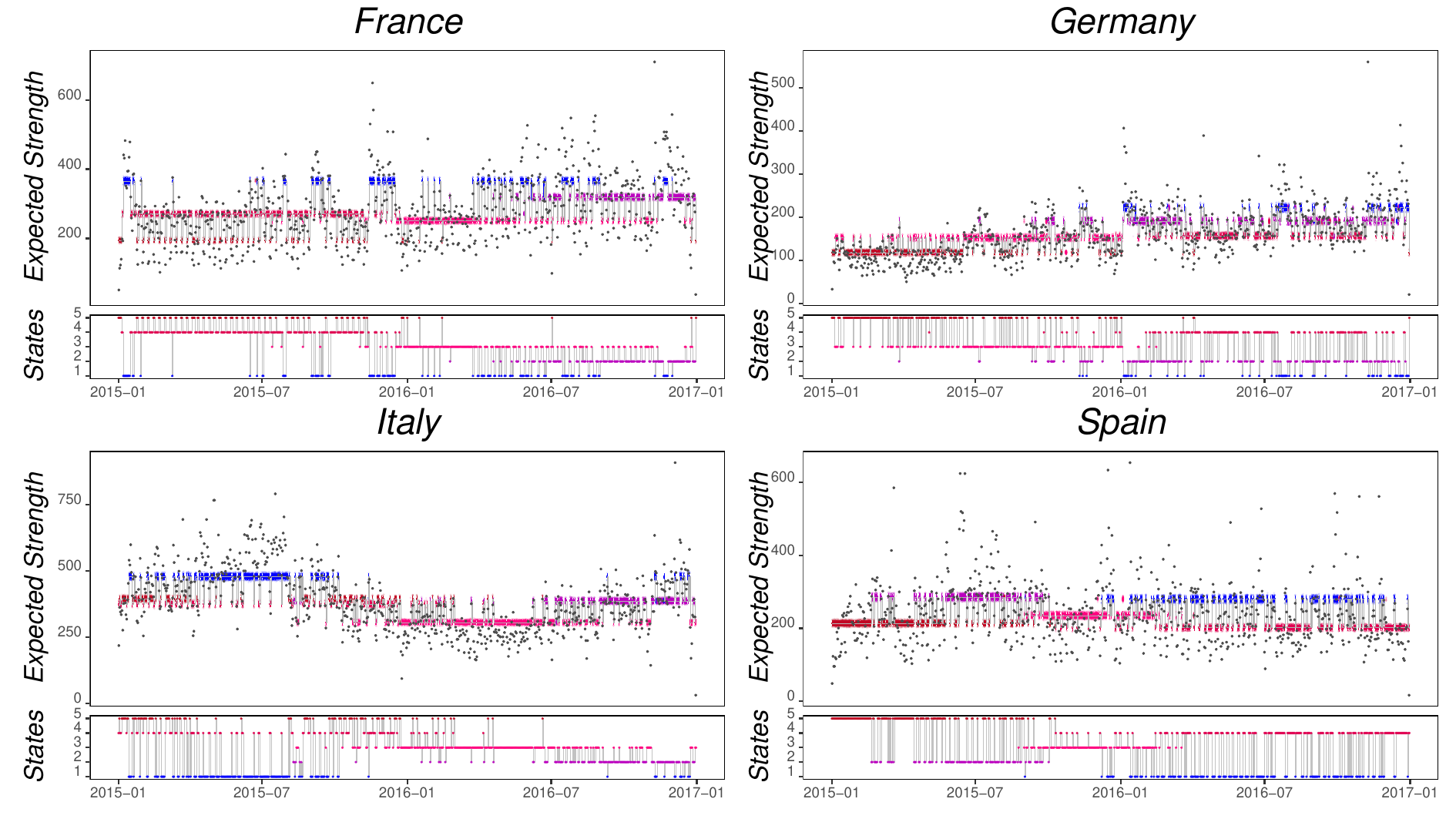}

 \caption{{\textbf{Latent States for Dynamic MS-LS Model $\mathcal{M}_6$:} Top panels show the posterior distribution boxplots for the average expected strength in each regime, with the underlying daily observations for information in the background (gray dots); Bottom panels show the polarization states through time, where 1 denotes the lowest polarization state and 5 the highest polarization state. In the online version, colors in both panels relate to the polarization level, with red corresponding to high, purple to intermediate, and blue to low polarization.}}
 \label{fig:latent_states_res} 
\end{figure}

Latent states signal the presence of lower or higher in-platform polarization regimes. In state L, the average distance between outlets is lower in terms of political leaning than in state H, making Facebook users more prone to interact with news outlets with different political tendencies. Figure~\ref{fig:latent_states_res} reports the estimated polarization regimes (bottom panels) through time for the MS-LS model with $d = 2$ and $K = 5$ estimated for the four countries. State 1 denotes the lowest polarization state, while State 5 indicates the highest polarization state.
We notice a tendency to move from a state of high in-platform polarization to a lower polarization for France, Germany and Spain. Italy and, to some extent, Spain follow an oscillating pattern, with Spain exhibiting frequent switches between very-high and very-low polarization. Overall, our findings contradict the hypothesis of a global shift toward a high polarization regime on social media in this time frame.
In line with the opinion dynamics literature in media contexts (e.g., see \citealp{hu2024opinion, brooks2020model}), polarization within our model is strongly interconnected with the degree of interactions within a network (expected strength). In the opinion dynamics literature, weak network connectivity and fragmented structures (e.g. echo chambers), can favor the emergence of polarization driven by ideological differences.  Overall, our model agrees with these media theories, since periods of higher polarization (high values in the bottom panel of Figure~\ref{fig:latent_states_res}) correspond to periods of lower expected strength in the observed networks (low values in the top panel). Despite not being the primary purpose of our analysis,  polarization changes can be used to effectively track the medium-term dynamics of the expected strength (boxplots in the top panels) by borrowing information across periods classified within the same polarization regime. We stress that our main aim is not to replicate the short-term features of the daily observations, but rather to conduct inference on polarization and media bias regimes.  See Appendix \ref{alternative_specs} (\citealp{Casarin2025Supplement}) for a discussion on alternative MS-LS dynamic specifications. 

\subsection{Model Selection} 
We perform model selection considering eight alternative models. $\mathcal{M}_{1}$ is an MS-LS model with $d =1$ and $K = 2$ omitting the text-analysis interpretation in (\ref{eq:betaLean}) (i.e.~imposing $\gamma_{1} = 0$), $\mathcal{M}_{2}$ is an LS model with $d =1$ and $K = 1$ 
omitting the Markov-switching dynamics described in (\ref{eq:ms}), $\mathcal{M}_{3}$ is an unrestricted MS-LS model with $d =1$ and $K = 2$, $\mathcal{M}_{4}$ is an unrestricted MS-LS model with $d =2$ and $K = 2$, $\mathcal{M}_{5}$ is an unrestricted MS-LS model with $d =2$ and $K = 3$, and $\mathcal{M}_{6}$ is an unrestricted MS-LS model with $d =2$ and $K = 5$. In addition, we consider the standard Poisson random graph model $\text{RG}_1$ with intensity $\lambda_{ijt} = \exp\{\alpha\}$, and the Poisson random graph model with individual effects and observed leaning distances $\text{RG}_2$ with intensity $\lambda_{ijt} = \exp\{\alpha_i + \alpha_j - \beta||L_{it} - L_{jt}||^2\}$.

Model selection is carried out via two popular predictive measures, the Deviance Information Criterion (DIC) \citep{spiegelhalter2002bayesian} and the log pointwise predictive density (lppd) of \cite{gelman2014understanding}. Table \ref{tab:DIC} reports both criteria (see Appendix \ref{subsec:model_sel}, \citealp{Casarin2025Supplement}, for further details). A first comparison conducted using the DIC and lppd of model $\mathcal{M}_1$, $\mathcal{M}_2$ and $\mathcal{M}_3$ highlights the contributions of the observed  leaning \emph{a-là} \citet{gentzkow2015media} and the dynamic component. The model without the dynamic component ($\mathcal{M}_{2}$) is dominated by the other two specifications for each country. Except for Spain and to some extent Italy, very similar scores are obtained for $\mathcal{M}_{1}$ and $\mathcal{M}_{3}$. This is to be expected as $\mathcal{M}_{3}$  aims to offer more interpretable latent coordinates rather than an improved fit for the network. Table \ref{tab:DIC} also includes results for the random graph models $\text{RG}_1$ and $\text{RG}_2$ which are clearly performing worse than our models.
A comparison between models $\mathcal{M}_3$, $\mathcal{M}_4$, $\mathcal{M}_5$ and $\mathcal{M}_6$  highlights the effect of increasing the number of dimensions. The best-performing model on the basis of DIC and lppd is $\mathcal{M}_6$ ($d = 2$ and $K = 5$). Graphical inspection reveals that the latent space results for $\mathcal{M}_6$ are in line with those of the other models (see Appendix \ref{G:high-order}, \citealp{Casarin2025Supplement}, for additional results).

\begin{table}[t]
\renewcommand{\arraystretch}{1}
\caption{\textbf{Network Model Selection:} (a) DIC scores and (b) Log pointwise predictive densities (lppd) as in \cite{gelman2014understanding} for the models introduced here and two random graph models $\text{RG}_1$ and $\text{RG}_2$ with intensity $\lambda_{ijt} = \exp\{\alpha\}$ and $\lambda_{ijt} = \exp\{\alpha_i + \alpha_j - \beta||L_{it} - L_{jt}||^2\}$, respectively. Best performance in bold.} 
\label{tab:DIC}
\centering
\resizebox{0.9\linewidth}{!}{  
\begin{tabular}{c cccc | cccc}
\hline\hline
\multicolumn{1}{l}{} & \multicolumn{4}{c|}{(a) DIC$\times 10^{-6}$} & \multicolumn{4}{c}{(b) lppd $\times 10^{-6}$} \\ \hline

Model & France & Germany & Italy & Spain & France & Germany & Italy & Spain \\ \hline

$\mathcal{M}_1$ & 4.4698 & 2.3669 & 3.3066 & 4.6390 & -2.2784 & -1.2190 & -1.6771 & -2.3471 \\
$\mathcal{M}_2$ & 4.6434 & 2.4766 & 3.4825 & 4.9049 & -2.3597 & -1.2702 & -1.7657 & -2.4511 \\
$\mathcal{M}_3$ & 4.4696 & 2.3669 & 3.3049 & 4.6139 & -2.2784 & -1.2191 & -1.6776 & -2.3347 \\
{$\mathcal{M}_4$} & 4.2654 & 2.2582 & 2.9797 & 4.2796 & -2.1697 & -1.1500 & -1.5171 & -2.1576 \\
{$\mathcal{M}_5$} & 4.2533 & 2.2143 & 2.9766 & 4.1819 & -2.1644 & -1.1290 & -1.5127 & -2.1075 \\
{$\mathcal{M}_6$} & \textbf{4.0949} & \textbf{2.1570} & \textbf{2.6588} & \textbf{3.9590} & -\textbf{2.0827} & -\textbf{1.1107} & -\textbf{1.4641} & -\textbf{2.0633} \\
\begin{tabular}[c]{@{}c@{}}${\text{RG}_1}$\end{tabular} & 24.9489 & 9.6074 & 26.1551 & 15.1828 & -12.5047 & -4.8341 & -13.1011 & -7.6049 \\
\begin{tabular}[c]{@{}c@{}}${\text{RG}_2}$\end{tabular} & 5.1913 & 2.8212 & 4.4554 & 5.4713 & -2.6153 & -1.4121 & -2.2424 & -2.7350 \\ \hline \hline
\end{tabular}
}
\end{table}

In addition, we compare the empirical and model-implied network metrics averaged over the entire sample. Table \ref{tab:net_char2} reports a posterior predictive check (see \citealp{gelman2014understanding}) in which the posterior predictive expected nodal strength, nodal strength's standard deviation and dispersion index -- derived in Section \ref{sec:properties} -- are compared with the empirical values, for the random graph models  $\text{RG}_1$ and $\text{RG}_2$, and {models $\mathcal{M}_3$ to $\mathcal{M}_6$}. All models are able to mimic the first moment of the strength distribution, but, as expected, the simple Poisson random graph model can not accommodate the large amount of overdispersion in the data. The MS-LS models are able to capture the observed dispersion in the strength distribution, leading to a rather similar fit. Appendix \ref{G:high-order} in the Supplementary Material (\citealp{Casarin2025Supplement}) presents time-varying diagnostic figures for these metrics. Appendix \ref{sec:sbm} (\citealp{Casarin2025Supplement}) provides a comparison with SBMs, which generally offer a good fit and provide an alternative modeling solution for applications where media clustering is more relevant than ranking media on the political spectrum. 

\section{Conclusion}
We propose a dynamic Markov-Switching Latent Space model to extract insightful information concerning media ideology and in-platform polarization. The model projects the audience duplication network of news outlets on a $d$-dimensional Euclidean space where the latent positions can be interpreted in terms of political leaning through a suitable proxy. Inference is carried out within a Bayesian framework, allowing reliable results with relatively standard MCMC methods. We derive the theoretical model properties and assess the effectiveness of the proposed methodology on simulated data.  Our model is applied to a Facebook dataset of news outlets in four European countries.  We find that the inferred latent leaning strongly correlates with the independent PEW Research Survey Index and correctly ranks news outlets in terms of left and right leaning. Moreover, inference on the latent states does not support the hypothesis of a unidirectional shift toward high polarization on Facebook. Finally, model selection suggests that the dynamic specification should be preferred over a static one, and that a text-analysis index may aid the leaning identification. Our novel modeling technique is not free from limitations. First, few longitudinal studies have focused on polarization in general and more specifically on media polarization in Europe. This makes it difficult to find a ground truth for validating polarization changes. Moreover, latent factor and state identification is not trivial in this context. Finally, the current model does not benefit from cross-country information-sharing which could be obtained, {\it e.g.}~with a suitable hierarchical prior choice. Future research may be directed toward the testing of more advanced text-analysis indicators for media bias, a more fine-grained analysis,  {\it e.g.}~at post level, the adoption of alternative latent space embedding specifications, such as circular and hyperbolic, and the exploration of different identifying restrictions.

\newpage

\medskip 

\textbf{Acknowledgments}
 The authors are grateful to the Editors and the anonymous Referee for the insightful comments and to A.L.~Schmidt, F.~Zollo, A.~Scala, and W.~Quattrociocchi for providing access to the Facebook \emph{source} dataset used in this work. We are also grateful to the conference participants for helpful discussions at the ISBA World Meeting (June 2022, Montreal, Canada), Bayes Comp (March 2023, Levi, Finland) and the 10th Italian Congress of Econometrics (May 2023, Cagliari, Italy). This research used the HPC system of the ``Venice Centre in Economic and Risk Analytics for Public Policies'' (VERA). MFJS gratefully acknowledges the hospitality of Ca' Foscari University of Venice, where he was a visiting scholar during the latter stages of this research.


\medskip 

\textbf{Funding}
    RC and AP acknowledge support from the Italian Ministry of University -- PRIN project ‘Discrete random structures for Bayesian learning and prediction’ under g.a. n. 2022CLTYP4; the European Union (EU) --  NextGenerationEU, Growing Resilient, INclusive and Sustainable project (GRINS PE00000018 - CUP H73C22000930001), National Recovery and Resilience Plan (NRRP) -- PE9. The views and opinions expressed are solely those of the authors and do not necessarily reflect those of the EU. 


\medskip 

\textbf{Supplement}
\begin{itemize}
    \item \textbf{Supplement to “Media Bias and Polarization through the Lens of an MS-LS Model”:} the online Supplementary Materials include derivations of the results and further analyses.
    \item \textbf{Repository for “Media Bias and Polarization through the Lens of an MS-LS Model”:} This online repository contains the data and code used to replicate the results in the paper. The online repository is also available at \url{https://github.com/BayesianEcon/Dyn-MS-LS-Media} and at \url{https://codeocean.com/capsule/9380600/tree/v1}.
\end{itemize}

\newpage



\bibliographystyle{apalike} 
\bibliography{biblio}       


\newpage
\appendix

\begin{center}
{\Large Supplement to\\\vspace{12pt}
Media Bias and Polarization  through the Lens of a Markov Switching Latent Space Network Model\\
\vspace{5pt}\large
Roberto Casarin, Antonio Peruzzi, Mark F.J. Steel}
\vspace{12pt}
\end{center}

This document contains the supplementary material for the MS-LS model. Theoretical properties and numerical illustrations are presented in Appendices A and B. Appendices C to F include the derivation of the full conditional distributions used in the Gibbs sampler, a detailed description of the sampling algorithm, simulation results, and model selection procedures. Additional preliminary analyses and empirical findings related to the media bias application are provided in Appendices G to J.  Comparisons with alternative specifications are discussed in Appendices K and L. Finally, Appendix M contains descriptions of the data and code used in the analysis.

\appendix

\renewcommand\thefigure{A.\arabic{figure}}
\setcounter{figure}{0}
\renewcommand\theequation{A.\arabic{equation}}
\setcounter{equation}{0}
\renewcommand\thetable{A.\arabic{table}}
\setcounter{table}{0}

\section{Derivation of the Markov--Switching Latent Space Model Properties}
\label{A:properties}
In this section, we derive the properties reported in Subsection \ref{sec:properties} for an MS-LS model applied to a weighted network. 

\subsection{Relevant Results}

Here we report some background material that will turn out to be useful in the derivation of the main results.

\begin{proposition}\label{prop:Multinomial}

\textbf{Multinomial Theorem}
Let $x_j\in\mathbb{R}$~ for $j=1,\ldots,m$ and $n\in\mathbb{N}$, then
\begin{equation*}
    (x_1 + x_2 + \cdots + x_m)^n = \sum_{\underline{k} \in \mathcal{K}}  \binom{n}{\underline{k}}\prod_{i = 1}^{m}x_i^{k_i},
\end{equation*}
where $\underline{k} = \{k_1,\ldots, k_m \} $, $\mathcal{K} = \{k_i \in \mathbb{N}| \sum_{i}k_i = n\}$.

\end{proposition}

\begin{proposition} \label{prop:normalint}
\textbf{Integral of the Product of $N$ zero-mean independent MVNs}
\begin{equation*}
    \left.\int_{\mathbb{R}^{d}} \prod_{i = 1}^{N}f_N(\mathbf{x}; \boldsymbol{0}, \sigma_{i}^2I_d)d\mathbf{x} \right. = (2\pi)^{-\frac{(N-1)d}{2}}\left(\sum_{i =1}^N \frac{1}{\sigma^2_i}\right)^{-\frac{d}{2}}\prod_{i= 1}^{N}(\sigma_i^2)^{-\frac{d}{2}},
\end{equation*}
where $f_N(\mathbf{x}; \boldsymbol{0}, \sigma_{i}^2I_d)$ is the pdf of a $d$-dimensional multivariate normal distribution with mean ${\mathbf{0}}$ and variance $\sigma_{i}^2I_d$ for each $i = 1, \ldots, N$.
\end{proposition}
The proofs are straightforward; hence, they are omitted.

\begin{proposition}\label{prop:normalconv}
\textbf{Convolution of Normal Distributions}
\begin{align*}
&f_N^{(1)} * f_N^{(2)}(z) =\int f_N^{(1)}(\mathbf{x}; \mathbf{a}_1, A_1) \cdot f_N^{(2)}(\mathbf{z}- \mathbf{x} ; \mathbf{a}_2, A_2) d \mathbf{x} \\
&=\int \frac{1}{(2 \pi)^{p / 2}|A_1|^{1 / 2}} e^{-\frac{1}{2}(\mathbf{x}-\mathbf{a}_1)^{\prime} A_1^{-1}(\mathbf{x}-\mathbf{a}_1)} \frac{1}{(2 \pi)^{p / 2}|A_2|^{1 / 2}} e^{-\frac{1}{2}(\mathbf{z}-\mathbf{x}-\mathbf{a}_2)^{\prime} A_2^{-1}(\mathbf{z}-\mathbf{x}-\mathbf{a}_2)} d \mathbf{x} \\
&=\frac{1}{(2 \pi)^{p / 2}|A_1+A_2|^{1 / 2}} e^{-\frac{1}{2}(\mathbf{z}-(\mathbf{a}_1+\mathbf{a}_2))^{\prime}(A_1+A_2)^{-1}(\mathbf{z}-(\mathbf{a}_1+\mathbf{a}_2))} \\
&=f_N(\mathbf{z} ; \mathbf{a}_1+\mathbf{a}_2, A_1+A_2),
\end{align*}
where $\mathbf{a}_1$ and $\mathbf{a}_2$ are $p$-dimensional mean vectors and $A_1$ and $A_2$ are $p \times p$ variance-covariance matrices.
\end{proposition}

\subsection{Probability Generating Function}

For a general LS model, following the conditional independence and HMM assumptions  (Assumptions \ref{Ass:Indep} and \ref{Ass:HMM}) and from the law of iterated expectation, the probability generating function (pgf) for the weighted degree can be written as:
\begin{align*}
G_l(x) &= \sum_{v=0}^{\infty} x^{v} p_{v}
=\sum_{k =1}^K\widetilde{G}_k(x)q_{lk},
\end{align*}
where the state-specific pgf $\widetilde{G}_k(x)$ is
\begin{align*}
&\widetilde{G}_k(x)=\sum_{v=0}^{\infty} x^{v} \int_{\mathcal{Z}^{N}_k} \mathbb{P}\left(Y_{it}=v \mid S_t=k, {\boldsymbol{\zeta}_{\sdot k}}\right) \left(\prod_{j=1}^{N}f_N\left(\boldsymbol{\zeta}_{jk}\right) \right) d \boldsymbol{\zeta}_{1k} \cdots d\boldsymbol{\zeta}_{Nk} \\
&=\int_{\mathcal{Z}^{N}_k} \left(\sum_{v=0}^{\infty} x^{v} \mathbb{P}\left(Y_{it}=v \mid s_t=k,{\boldsymbol{\zeta}_{\sdot k}}\right)\right)\left(\prod_{j=1}^{N} f_N\left(\boldsymbol{\zeta}_{jk}\right) \right) d\boldsymbol{\zeta}_{1k} \cdots d\boldsymbol{\zeta}_{Nk} \\
&=\int_{\mathcal{Z}_k}f_N(\boldsymbol{\zeta}_{ik})\left(\int_{\mathcal{Z}^{(N-1 )}_k} \left(\prod_{j \neq i} \varphi_{ijk}\left(x ; \boldsymbol{\zeta}_{ik}, \boldsymbol{\zeta}_{jk}\right)\right)\left(\prod_{j \neq i} f_N\left(\boldsymbol{\zeta}_{jk}\right)\right) d \boldsymbol{\zeta}_{jk}\right)d\boldsymbol{\zeta}_{ik} \\
&=\int_{\mathcal{Z}_k} f_N\left(\boldsymbol{\zeta}_{ik}\right)\left( \prod_{j \neq i}\int_{\mathcal{Z}^{(N-1)}_k}\varphi_{ijk}\left(x ; \boldsymbol{\zeta}_{ik}, \boldsymbol{\zeta}_{jk}\right)f_N(\boldsymbol{\zeta}_{jk})d\boldsymbol{\zeta}_{jk} \right)d\boldsymbol{\zeta}_{ik}\\
&=\int_{\mathcal{Z}_k} f_N\left(\boldsymbol{\zeta}_{ik}\right) \prod_{j \neq i}\theta_{ijk}\left(x ; \boldsymbol{\zeta}_{ik}\right) d \boldsymbol{\zeta}_{ik},
\end{align*}
with latent coordinates ${\boldsymbol{\zeta}_{\sdot k}} = \{\boldsymbol{\zeta}_{1k}, \ldots, \boldsymbol{\zeta}_{Nk}\}$, $\boldsymbol{\zeta}_{jk} \in \mathcal{Z}_k \subset \mathbb{R}^d$, nodal strength $Y_{it} = Y_{i1t}+ \ldots + Y_{ij-1t}+ Y_{ij+1t} +\ldots + Y_{iNt} $, strength's pgf $\varphi_{ijk}(x;\boldsymbol{\zeta}_{ik}, \boldsymbol{\zeta}_{jk}) = \mathbb{E}\left(x^{Y_{ijt}}| s_t=k, \boldsymbol{\zeta}_{ik},\boldsymbol{\zeta}_{jk} \right) = e^{\lambda_{ijk}(\boldsymbol{\zeta}_{ik}, \boldsymbol{\zeta}_{jk})(x-1)}$. The $\theta_{ijk}(x;\boldsymbol{\zeta}_{ik}) = \int_{\mathcal{Z}_k}\varphi_{ijk}(x;\boldsymbol{\zeta}_{ik}, \boldsymbol{\zeta}_{jk})f(\boldsymbol{\zeta}_{jk})d\boldsymbol{\zeta}_{jk}$ is the pgf of the weight of the edge between a node chosen at random and a node with latent information $\boldsymbol{\zeta}_{ik}$.

\subsection{Derivatives of an LS model with individual effects}

Consider an LS model with Poisson likelihood for the edges $Y_{ijt} \sim\mathcal{P}oi\left(\lambda_{ijk}
\right)
$ with intensity parameter $ \lambda_{ijk} = \exp\left\{\alpha_i+\alpha_j- \beta||\boldsymbol{\zeta}_{ik} - \boldsymbol{\zeta}_{jk}||^2\right\}$ (Assumption \ref{Ass:Poisson}). From the independence assumption in Assumption \ref{Ass:Indep}, and normal assumption for the latent features (Assumption \ref{Ass:Normal}), the \emph{m}-th derivative of the corresponding pgf can be written as:
\begin{align*}
&\left.\frac{\partial^{m} }{\partial x^{m}}\prod_{j \neq i}\theta_{ijk}(x, \boldsymbol{\zeta}_{ik})\right|_{x=1} = \left.\frac{\partial^{m}}{\partial x^{m}} \int_{\mathcal{Z}^{N-1}_k}\prod_{j \neq i}e^{(x-1)\lambda_{ijk}}f_N(\boldsymbol{\zeta}_{jk})d\boldsymbol{\zeta}_{jk} \right|_{x = 1} \\ &=  \left.\int_{\mathcal{Z}^{N-1}_k}\frac{\partial^{m}}{\partial x^{m}}e^{(x-1)\sum_{j \neq i}\lambda_{ijk}}\prod_{j \neq i}f_N(\boldsymbol{\zeta}_{jk}) d\boldsymbol{\zeta}_{jk} \right|_{x =1} \\ &= \left.\int_{\mathcal{Z}^{N-1}_k}\left(\sum_{j \neq i}\lambda_{ijk}\right)^{m}e^{(x-1) \sum_{j \neq i}\lambda_{ijk}}\prod_{j \neq i}f_N(\boldsymbol{\zeta}_{jk}) d\boldsymbol{\zeta}_{jk}\right|_{x =1} \\ &=
 \int_{\mathcal{Z}^{N-1}_k}\left(\sum_{j \neq i}\lambda_{ijk}\right)^{m}\prod_{j \neq i}f_N(\boldsymbol{\zeta}_{jk}) d\boldsymbol{\zeta}_{jk} \\ &=
\int_{\mathcal{Z}^{N-1}_k}\sum_{\underline{h}_i \in \mathcal{H}_i}  \binom{m}{\underline{h}_i} \prod_{j \neq i}\lambda^{h_j}_{ijk}f_N(\boldsymbol{\zeta}_{jk})d\boldsymbol{\zeta}_{jk} \\ &=
\sum_{\underline{h}_i \in \mathcal{H}_i}  \binom{m}{\underline{h}_i} \prod_{j\neq i}\int_{\mathcal{Z}}\lambda_{ijk}^{h_j}f_N(\boldsymbol{\zeta}_{jk})d\boldsymbol{\zeta}_{jk} \\ &=
\sum_{\underline{h}_i \in \mathcal{H}_i}  \binom{m}{\underline{h_i}}\prod_{j\in\mathcal{J}_{i}}e^{(\alpha_i+\alpha_j)h_j}\int_{\mathcal{Z}_h}e^{-\beta||\boldsymbol{\zeta}_{ik}-\boldsymbol{\zeta}_{jk}||^2 h_j}f_N(\boldsymbol{\zeta}_{jk})d\boldsymbol{\zeta}_{jk} \\ &= \sum_{\underline{h}_i \in \mathcal{H}_i}  \binom{m}{\underline{h}_i} \prod_{j\in\mathcal{J}_{i}}e^{(\alpha_i+\alpha_j)h_j}\int_{\mathcal{Z}_k}e^{-\beta||\boldsymbol{\zeta}_{ik}-\boldsymbol{\zeta}_{jk}||^2 h_j}f_N(\boldsymbol{\zeta}_{jk})\frac{(2 \pi)^{d / 2}}{(2 \pi)^{d / 2}} \frac{(2\beta h_j)^{d / 2}}{( 2\beta h_j)^{d / 2}}d\boldsymbol{\zeta}_{jk}\\ 
&= \sum_{\underline{h}_i \in \mathcal{H}_i}  \binom{m}{\underline{h}_i} \prod_{j\in\mathcal{J}_{i}}\frac{1}{(2\beta h_j)^{\frac{d}{2}}}(2\pi)^{\frac{d}{2}}e^{(\alpha_i+\alpha_j)h_j}f_N\left(\boldsymbol{\zeta}_{ik}; \boldsymbol{0},\left(\sigma_k^{2}+\frac{1}{2\beta h_j}\right) I_{d}\right),
\end{align*}
where $\underline{h}_{i} = \{h_1,\ldots,h_{i-1}, h_{i+1},\ldots, h_N \} $, $\mathcal{H}_{i} = \{h_j \in \mathbb{N} \quad j \neq i| \sum_{j \neq i}h_j = m\}$,  $\mathcal{J}_{i} = \{j|j \neq i, h_j > 0\}$, $\beta > 0$,
 and where the fifth equation follows from Proposition \ref{prop:Multinomial} and the last equation from Proposition \ref{prop:normalconv}. Thus we obtain the following:
\begin{align*}
    \left.\frac{\partial^{m} \widetilde{G}_k(x)}{\partial x^{m}}\right|_{x=1} &= \int_{\mathcal{Z}_k}f_N(\boldsymbol{\zeta}_{ik})\left(\sum_{\underline{h}_i \in \mathcal{H}_i}  \binom{m}{\underline{h}_i} \prod_{j\in\mathcal{J}_{i}}\frac{1}{(2\beta h_j)^{\frac{d}{2}}}(2\pi)^{\frac{d}{2}}e^{(\alpha_i+\alpha_j)h_j}f_N\left(\boldsymbol{\zeta}_{ik}; \boldsymbol{0},\tau_{kj} I_{d}\right) \right) d\boldsymbol{\zeta}_{ik},
\end{align*}
where $\tau_{kj}=\left(\sigma_k^{2}+\frac{1}{2\beta h_j}\right)$. 

Set $\alpha_i + \alpha_j = \alpha$ for each $i$ and $j$, then from Proposition \ref{prop:normalint}:
{\footnotesize
\begin{align*}
    &\left.\frac{\partial^{m} \widetilde{G}_k(x)}{\partial x^{m}}\right|_{x=1} = \int_{\mathcal{Z}_k}f_N(\boldsymbol{\zeta}_{ik})\left(\sum_{\underline{h}_i \in \mathcal{H}_i}  \binom{m}{\underline{h}_i}\prod_{j\in\mathcal{J}_i}\frac{1}{(2\beta h_j)^{\frac{d}{2}}}(2\pi)^{\frac{d}{2}}e^{\alpha h_j}f_N\left(\boldsymbol{\zeta}_{ik}; \boldsymbol{0},\tau_{kj} I_{d}\right) \right) d\boldsymbol{\zeta}_{ik}\\&=\left(\sum_{\underline{h}_i \in \mathcal{H}_i}  \binom{m}{\underline{h}_i}e^{\alpha m}\prod_{j\in\mathcal{J}_{i}}\frac{1}{(2\beta h_j)^{\frac{d}{2}}}(2\pi)^{\frac{d}{2}} \int_{\mathcal{Z}_k}f_N(\boldsymbol{\zeta}_{ik})\prod_{j\in\mathcal{J}_{i}}f_N\left(\boldsymbol{\zeta}_{ik}; \boldsymbol{0},\tau_{kj} I_{d}\right)d\boldsymbol{\zeta}_{ik} \right) \\ &= \sum_{\underline{h}_i \in \mathcal{H}_i}  \binom{m}{\underline{h}_i}e^{\alpha m}
    (\sigma_k^2)^{-\frac{d}{2}}  \left(\frac{1}{\sigma_k^2}+ \sum_{j\in\mathcal{J}_{i}}\tau_{kj}^{-1}\right)^{-\frac{d}{2}}\prod_{j\in\mathcal{J}_{i}}\left(\frac{1}{(2\beta h_j)^{\frac{d}{2}}}(2\pi)^{\frac{d}{2}}(2\pi)^{\frac{-d}{2}}(\tau_{kj})^{-\frac{d}{2}}\right) 
     \\&= \sum_{\underline{h}_i \in \mathcal{H}_i}  \binom{m}{\underline{h}_i}e^{\alpha m}(\sigma_k^2)^{-\frac{d}{2}}  \left(\frac{1}{\sigma_k^2}+ \sum_{j\in\mathcal{J}_{i}}\tau^{-1}_{kj}\right)^{-\frac{d}{2}} \prod_{j\in\mathcal{J}_{i}}\left(\frac{1}{(2\beta h_j)^{\frac{d}{2}}}\tau_{kj}^{-\frac{d}{2}}\right).
    \end{align*}
    }
\subsubsection{First Conditional Factorial  Moment}
\label{first_cond}

If we solve for $m = 1$, we obtain:
\begin{align*}
    \left.\frac{\partial \widetilde{G}_{k}(x)}{\partial x}\right|_{x=1} &=
    e^{\alpha}(\pi)^{\frac{d}{2}}\left(\sum_{\underline{h}_i \in \mathcal{H}_i}
\int_{\mathcal{Z}_k}f_N(\boldsymbol{\zeta}_{ik})
    f_N\left(\boldsymbol{\zeta}_{ik}; \boldsymbol{0},\left(\sigma_k^{2}+\frac{1}{2\beta}\right) I_{d}\right) d\boldsymbol{\zeta}_{ik}\right) \\ &=
    e^{\alpha}(\pi)^{\frac{d}{2}}(N-1)\left(2\sigma_k^2+ \frac{1}{2\beta}\right)^{-\frac{d}{2}}(2\pi\beta)^{-\frac{d}{2}}
    \end{align*}
    \begin{equation}
    \hspace{-98pt}
        =(N-1) e^{\alpha}\left(4 \sigma_k^{2}\beta+1\right)^{-\frac{d}{2}}.\label{eq:Gprime}
   \end{equation}

\subsubsection{Second Conditional Factorial  Moment}
If we solve for $m = 2$ we obtain:
\begin{align*}
&\left.\widetilde{G}_{k}^{\prime\prime}(x)\right|_{x=1} = (N-1) e^{2 \alpha}\left(\sigma_k^{2}+\frac{1}{4\beta}\right)  ^{-\frac{d}{2}}\left(\frac{1}{\sigma_k^{2} + \frac{1}{4\beta}}+\frac{1}{\sigma_k^{2}}\right)^{-\frac{d}{2}}\left(\sigma_k^{2}\right)^{-\frac{d}{2}}(4\beta)^{\frac{d}{2}}\\
&\quad + (2\beta)^{-d}(N-1)(N-2)e^{2 \alpha}\left(\sigma_k^{2}+\frac{1}{2\beta}\right)^{-d}\left(\frac{2}{\sigma_k^2 + \frac{1}{2\beta}}+\frac{1}{\sigma_k^{2}}\right)^{-\frac{d}{2}}\left(\sigma_k^{2}\right)^{-\frac{d}{2}}
\end{align*}
\begin{equation}
    =e^{2\alpha}(N-1)\left(8\sigma_k^2\beta+1\right)^{-\frac{d}{2}}+ (N-1)(N-2)e^{2\alpha}\left(2\sigma_k^2\beta+1\right)^{-\frac{d}{2}}\left(6\sigma_k^2\beta+1\right)^{-\frac{d}{2}}. \label{eq:Gprimeprime}
\end{equation}

\subsubsection{First Moment}
Employing the result in \ref{first_cond} along with the linearity property of differentiation, we can write the expected strength for our MS-LS model as: 
\begin{equation*}
\mathbb{E}(Y_t|S_{t-1} = l) = \left.G_l^{\prime}(x)\right|_{x=1} =\sum_{k =1}^Kq_{lk}\left.\widetilde{G}_{k}^{\prime}(x)\right|_{x=1} = (N-1) e^{\alpha}\sum_{k =1}^Kq_{lk}\left(4 \sigma_k^{2}\beta+1\right)^{-\frac{d}{2}}.
\end{equation*}

\subsubsection{Second Central Moment}

We notice that in general, for a given discrete random variable $X$ with pgf $G(x)$, the variance can be computed as:
\begin{equation*}\mathbb{V}ar(X) = \left.G^{\prime\prime}(x)\right|_{x=1} + \left.G^{\prime}(x)\right|_{x=1} - \left( \left.G^{\prime}(x)\right|_{x=1}\right)^2.\end{equation*}

We thus obtain the following expression for the variance of the strength distribution:
{\small
\begin{align*}
  &\mathbb{V}ar(Y_t| S_{t-1} = l) = 
 \sum_{k =1}^Kq_{lk}\left.\widetilde{G}_k^{\prime\prime}(x)\right|_{x=1} + \sum_{k =1}^Kq_{lk}\left.\widetilde{G}_k^{\prime}(x)\right|_{x=1} - \left(\sum_{k=1}^{K}q_{lk}\widetilde{G}_k^{\prime}(x)|_{x=1}\right)^2\\
  &= \sum_{k =1}^Kq_{lk}\left.\widetilde{G}_k^{\prime\prime}(x)\right|_{x=1} +  \sum_{k =1}^Kq_{lk}\left.\widetilde{G}_k^{\prime}(x)\right|_{x=1} - \left(\sum_{k =1}^Kq_{lk}\widetilde{G}_k^{\prime}(x)|_{x=1}\right)^2\\ &\quad+   \sum_{k =1}^Kq_{lk}\left.\widetilde{G}_k^{\prime 2}(x)\right|_{x=1}-  \sum_{k =1}^Kq_{lk}\left.\widetilde{G}_k^{\prime 2}(x)\right|_{x=1}\\
  &= \sum_{k =1}^Kq_{lk}\left(\left.\widetilde{G}_k^{\prime\prime}(x)\right|_{x=1} + \left.\widetilde{G}_k^{\prime}(x)\right|_{x=1} -  \left.\widetilde{G}_k^{\prime 2}(x)\right|_{x=1}\right)\\ &\quad + \sum_{k =1}^Kq_{lk}\left( \left.\widetilde{G}_k^{\prime}(x)\right|_{x=1}- \left.G_l^{\prime}(x)\right|_{x=1} \right)^2\\
   &= \sum_{k =1}^Kq_{lk}\mathbb{V}ar(Y_t|S_t =k)+\sum_{k =1}^Kq_{lk}\left( \left.\widetilde{G}_k^{\prime}(x)\right|_{x=1}- \left.G_l^{\prime}(x)\right|_{x=1} \right)^2.
  \end{align*}}

\subsubsection{Dispersion Index}\label{Sec:AppDisp}

We provide an analytical expression for the conditional dispersion index ($\mathfrak{D}$), which includes as a special case for a single regime the one derived in \citet{rastelli2016properties}. Let
\begin{equation*}\mathfrak{D}_k = 1 + \frac{\left.G^{\prime\prime}_k(x)\right|_{x=1}}{\left.G^{\prime}_k(x)\right|_{x=1}} - \left.G^{\prime}_k(x)\right|_{x=1} \end{equation*}
be the $k$th regime dispersion index. For our model, the dispersion index is the following:
{\small
\begin{align*}
  &\mathfrak{D}(Y_t|S_{t-1}= l) =  1+ \frac{  \sum_{k =1}^Kq_{lk}\left.\widetilde{G}_{k}^{\prime\prime}(x)\right|_{x=1}}{  \sum_{k =1}^Kq_{lk}\left.\widetilde{G}_{k}^{\prime}(x)\right|_{x=1}} - \sum_{k =1}^Kq_{lk}\widetilde{G}_{k}^{\prime}(x)|_{x=1}\\
 &=  1+ \frac{  \sum_{k =1}^Kq_{lk}\left.\widetilde{G}_{k}^{\prime\prime}(x)\right|_{x=1}}{  \sum_{k =1}^Kq_{lk}\left.\widetilde{G}_{k}^{\prime}(x)\right|_{x=1}} - \sum_{k =1}^Kq_{lk}\widetilde{G}_{k}^{\prime}(x)|_{x=1} +  \sum_{k =1}^Kq_{lk}\frac{\left.\widetilde{G}_{k}^{\prime\prime}(x)\right|_{x=1}}{\left.\widetilde{G}_{k}^{\prime}(x)\right|_{x=1}} - \sum_{k =1}^Kq_{lk}\frac{\left.\widetilde{G}_{k}^{\prime\prime}(x)\right|_{x=1}}{\left.\widetilde{G}_{k}^{\prime}(x)\right|_{x=1}}\\
 &=\sum_{k =1}^Kq_{lk}\left(1+ \frac{\left.\widetilde{G}_{k}^{\prime\prime}(x)\right|_{x=1}}{\left.\widetilde{G}_{k}^{\prime}(x)\right|_{x=1}} - \left.\widetilde{G}_{k}^{\prime}(x)\right|_{x=1}\right) + v \equiv \sum_{k =1}^Kq_{lk}\mathfrak{D}_k  + v,
\end{align*} \small}
where $v =   \frac{  \sum_{k =1}^Kq_{lk}\left.\widetilde{G}_{k}^{\prime\prime}(x)\right|_{x=1}}{  \sum_{k =1}^Kq_{lk}\left.\widetilde{G}_{k}^{\prime}(x)\right|_{x=1}} - \sum_{k =1}^Kq_{lk}\frac{\left.\widetilde{G}_{k}^{\prime\prime}(x)\right|_{x=1}}{\left.\widetilde{G}_{k}^{\prime}(x)\right|_{x=1}}$.\medskip

We can prove that $\mathfrak{D}(Y_t|S_{t-1}= l) > 1$. This can be done by showing that: 1) $\mathfrak{D}_k >1$ and 2) $v \geq 0$.

1) Proving that $\mathfrak{D}_k = \widetilde{G}_k'[\widetilde{G}_k''/\widetilde{G}_k'^2 - 1] + 1 > 1$ requires $[\widetilde{G}_k'' - \widetilde{G}_k'^2] > 0$, given that $\widetilde{G}_k'> 0$ for non--degenerate cases. Assuming $\beta = 1$ and substituting $\widetilde{G}_k'$ and $\widetilde{G}_k''$, after some standard algebraic manipulation we obtain:
\begin{align*}
[\widetilde{G}_k'' - \widetilde{G}_k'^2] &= \cancel{e^{2\alpha}}\cancel{(N-1)}(8\sigma^2_k +1)^{-\frac{d}{2}} +  \cancel{e^{2\alpha}}\cancel{(N-1)}(N-2)(12\sigma^4_k +8\sigma^2_k +1 )^{-\frac{d}{2}} \\ &\quad- \cancel{e^{2\alpha}}\cancel{(N-1)}(N-1)(16\sigma^4_k +8\sigma^2_k +1)^{-\frac{d}{2}}\\
&= [1/(p_k-16\sigma^4_k)^{d/2} - 1/(p_k-4\sigma^4_k)^{d/2}]\\ &\quad + (N-1)[1/(p_k-4\sigma^4_k)^{d/2} - 1/(p_k)^{d/2}] > 0,
\end{align*}
for $p_k = 16\sigma^4_k + 8\sigma^2_k + 1$ and where one can check that both terms are strictly positive given that $\sigma^2_k$ and $d$ are strictly positive.

2) We need to prove that $v = \sum_{k = 1}^K\widetilde{G}_k''/\sum_{k = 1}^K\widetilde{G}_k' - \sum_{k = 1}^K (\widetilde{G}''_k/\widetilde{G}_k') \geq 0$. Firstly, we note that

\begin{align*}
    \sum_{k = 1}^K q_k\frac{\widetilde{G}''_k}{\widetilde{G}_k'} &= e^{\alpha}\sum_{k = 1}^K q_k\left(\frac{4\sigma^2_k+1}{8\sigma^2_k +1}\right)^{d/2} + (N-2)e^{\alpha}\sum_{k = 1}^K q_k\left(\frac{4\sigma^2_k+1}{12\sigma^4_k + 8\sigma^2_k +1}\right)^{d/2}\\
    &=  e^{\alpha}\sum_{k = 1}^K q_k\left(\frac{\phi_k}{4\sigma^2_k + \phi_k}\right)^{d/2} + (N-2)e^{\alpha}\sum_{k = 1}^K q_k\left(\frac{\phi_k}{12\sigma^4_k + 4\sigma^2_k + \phi_k}\right)^{d/2},
\end{align*}

\begin{align*}
\sum_{k = 1}^K q_k\widetilde{G}''_k &= e^{2\alpha}(N-1)\sum_{k = 1}^K q_k(4\sigma^2_k + \phi_k)^{-d/2} + e^{2\alpha}(N-1)(N-2)\sum_{k = 1}^K q_k(12\sigma^4_k + 4\sigma^2_k + \phi_k)^{-d/2},
\end{align*}

\begin{align*}
 \sum_{k = 1}^K q_k\widetilde{G}'_k &= e^{\alpha}\sum_k q_k\left(\phi_k\right)^{-d/2},
\end{align*}
where $\phi_k = 4\sigma^2_k + 1$. This implies that:
\begin{align*}
v &= \left[\frac{\sum_{k = 1}^Kq_k(4\sigma^2_k + \phi_k)^{-d/2}}{\sum_{k = 1}^Kq_k\phi_k^{-d/2}} - \sum_{k = 1}^Kq_k \left(\frac{\phi_k}{4\sigma_k^2 + \phi_k}\right)^{d/2}\right] \\& \quad+ (N-2)\left[\frac{\sum_{k = 1}^Kq_k(12\sigma^4_k +4\sigma^2_k + \phi_k)^{(-d/2)}}{\sum_{k = 1}^Kq_k\phi_k^{-d/2}} -\sum_{k = 1}^Kq_k \left(\frac{\phi_k}{12\sigma^4_k + 4\sigma^2_k + \phi_k}\right)^{d/2}\right].
\end{align*}
It is sufficient to show that:
\begin{equation*}
\frac{\sum_{k = 1}^K q_k(4\sigma^2_k + \phi_k)^{-d/2} }{\sum_{k = 1}^K q_k\phi_k^{-d/2}} \geq \sum_{k = 1}^K q_k \left(\frac{\phi_k}{4\sigma^2_k + \phi_k}\right)^{d/2},
\end{equation*}
and that:
\begin{equation*}
\frac{\sum_{k = 1}^K q_k(12\sigma^4_k + 4\sigma^2_k + \phi_k)^{-d/2} }{\sum_{k = 1}^K q_k\phi_k^{-d/2}} \geq \sum_{k = 1}^K q_k\left(\frac{\phi_k}{12\sigma^4_k + 4\sigma^2_k + \phi_k}\right)^{d/2}.
\end{equation*}
The two inequalities can be verified with the use of Chebyshev's order inequality (\citealp{hardy1988inequalities}) bearing in mind that $q_k \in [0,1]$ and $\sigma^2_k > 0$. 

\begin{figure}[!htb]
  \centering
  
   \includegraphics[width= 0.75\textwidth]{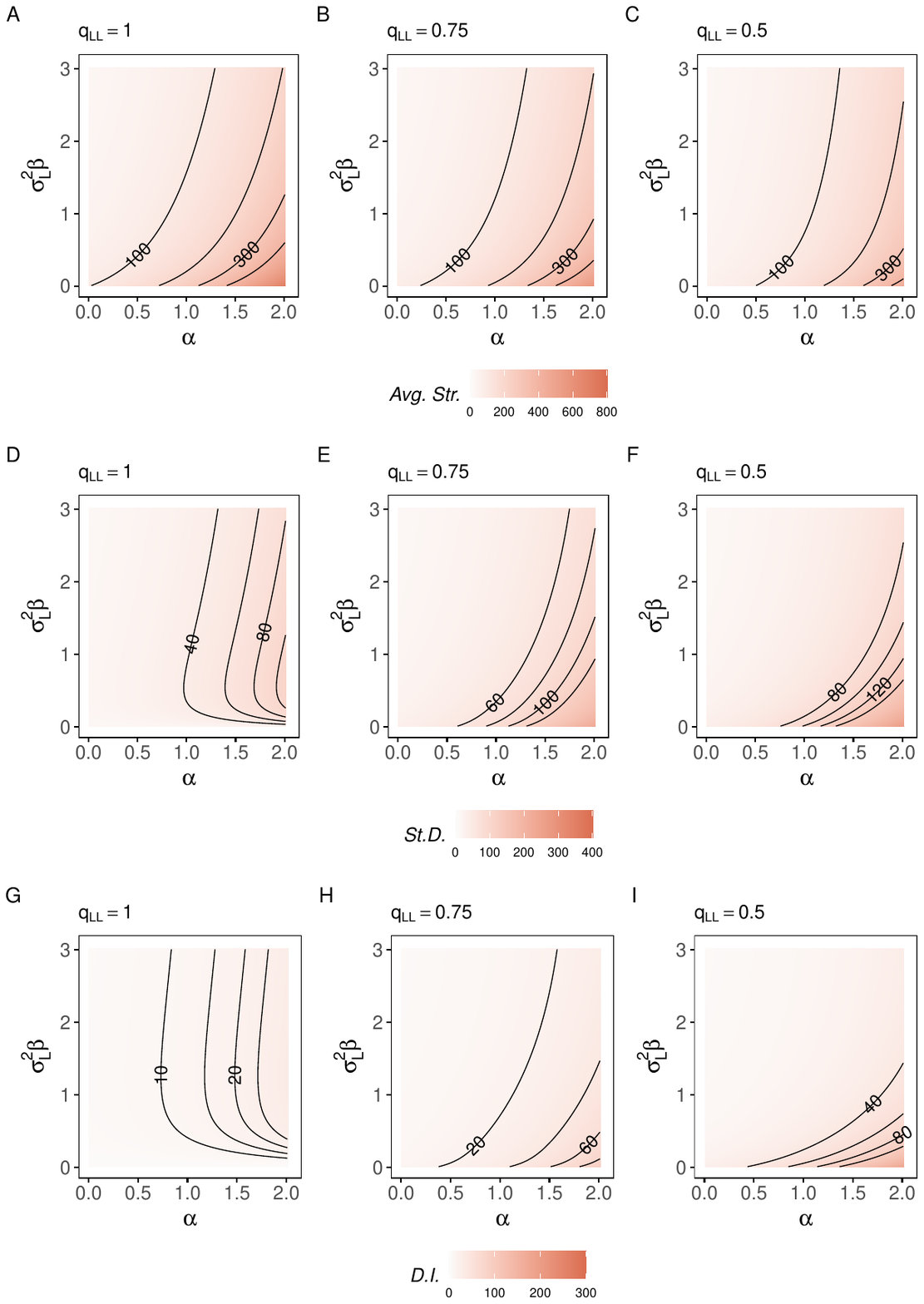}
 
 \caption
 {{\textbf{Strength Distribution of an MS Poisson LS model:} Contour plots of the expected value (top panels), standard deviation (middle panels), and dispersion index (bottom panels) of the strength distribution of a two-state MS Poisson LS model for different values of $\alpha$ and {$\sigma^2_{\mbox{\footnotesize L}}\beta$}, and for different values of the transition probabilities {$q_{\mbox{\footnotesize LL}}$} and {$q_{\mbox{\footnotesize LH}} = 1-q_{\mbox{\footnotesize LL}}$}. We assumed $d = 1$, $N = 100$, and {$\sigma^2_{\mbox{\footnotesize H}}\beta = 4$.}}}
 \label{fig:lvm_prop} 
 \end{figure}

\hfill \break
Figure \ref{fig:lvm_prop} displays the contour plots of the expected value (top panels), standard deviation (middle panels), and dispersion index (bottom panels) of the strength distribution of a two-state MS Poisson LS. Labels "L" and "H" denote low and high polarization states.

\begin{figure}[!htb]
  \centering
  
   \includegraphics[width= 0.75\textwidth]{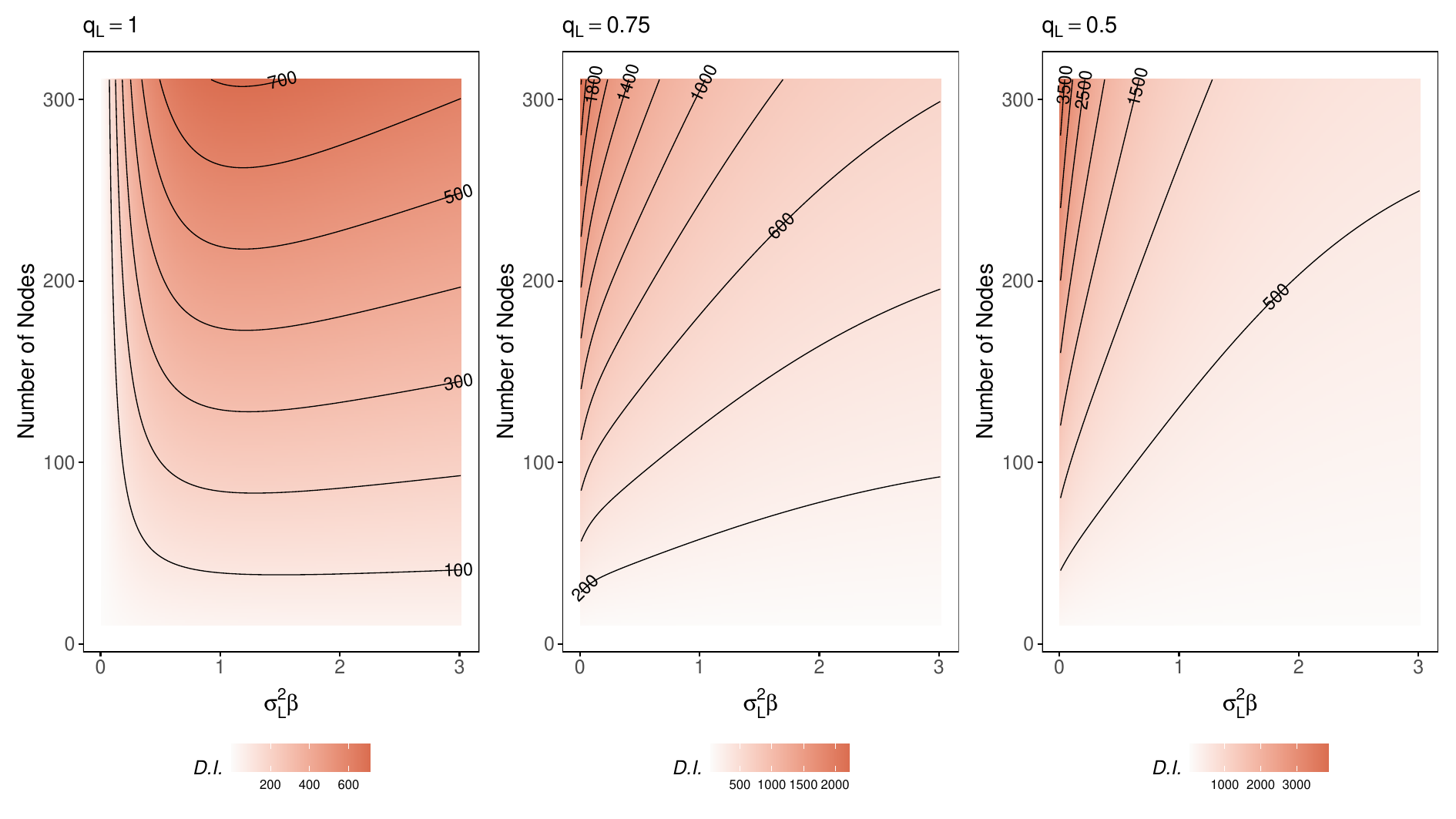}
 
 \caption
 {{\textbf{Strength Distribution of an MS Poisson LS model:} Contour plots of the dispersion index of the strength distribution of a two-state MS Poisson LS model for different values of $N$ and {$\sigma^2_{\mbox{\footnotesize L}}\beta$}, and for different values of the transition probabilities {$q_{\mbox{\footnotesize LL}}$} and {$q_{\mbox{\footnotesize LH}} = 1-q_{\mbox{\footnotesize LL}}$}. We assumed $d = 1$, $\alpha = 4$, and {$\sigma^2_{\mbox{\footnotesize H}}\beta = 4$.}}}
 \label{fig:di_nodes} 
 \end{figure}

\renewcommand\thefigure{B.\arabic{figure}}
\setcounter{figure}{0}
\renewcommand\theequation{B.\arabic{equation}}
\setcounter{equation}{0}
\renewcommand\thetable{B.\arabic{table}}
\setcounter{table}{0}

\section{Properties with Simulated Data}
\label{B:simulation}

As a robustness check, we study through simulation the statistical properties of a weighted network generated by the LS model. We draw the latent coordinates ${\mathbf{x}_i}|\sigma^2 \sim \mathcal{N}(0, \sigma^2 I_d)$ for $i = 1, \ldots, 100$, $d= 1$ and $K = 1$ and generate the networks weights ${Y_{ij}|\lambda_{ij}}\sim \mathcal{P}oi(\lambda_{ij})$ with log-intensity parameter $\log\lambda_{ij} = \alpha - \beta||{\mathbf{x}_i}-{\mathbf{x}_j}||^2$.\vspace{0.4cm}

As is common in network theory, we define the strength of node $i$ as $Y_i = \sum_{j \neq i}Y_{ij}$. We study the properties of the empirical distribution of $Y = \{Y_1, \ldots, Y_N\}$. In particular, we focus attention on the following statistics: the sample mean $\overline{Y}$, the sample standard deviation $SD(Y)$, the dispersion index $\mathfrak{D}(Y)$ and the clustering coefficient $CC(Y)$ as defined in \citet{barrat2004architecture}.

\begin{figure}[htp!]
    \centering
    \includegraphics[width = 1\textwidth]{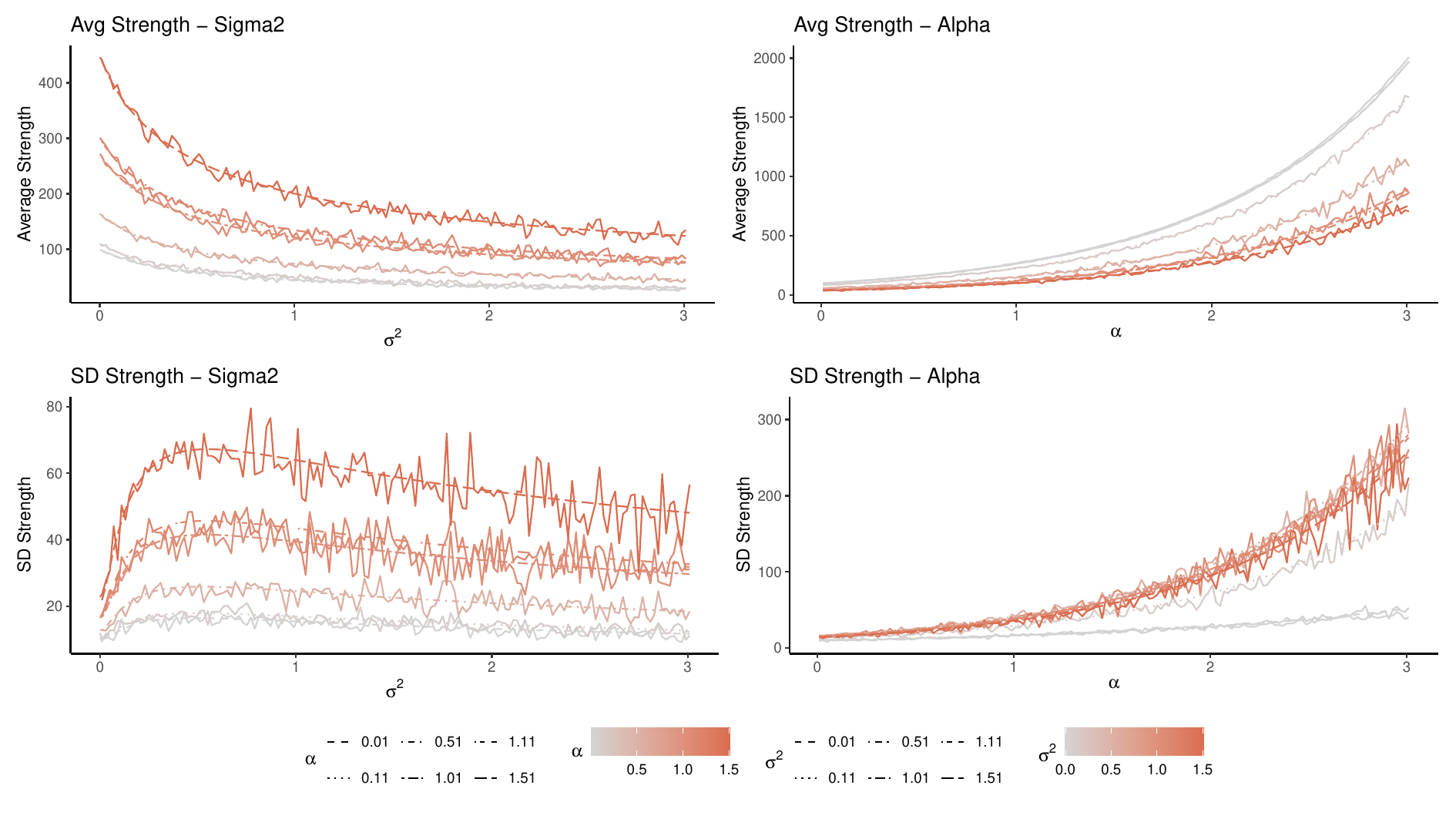}
    \caption{\textbf{Simulation Result - 1:} Sensitivity analysis for the expected strength and standard deviation of the strength as $\alpha$ and $\sigma^2$ vary. In this simulation, we assumed $\beta = 1$.  }
    \label{fig:simulation1}
\end{figure}

\begin{figure}[htp!]
    \centering
    \includegraphics[width = 1\textwidth]{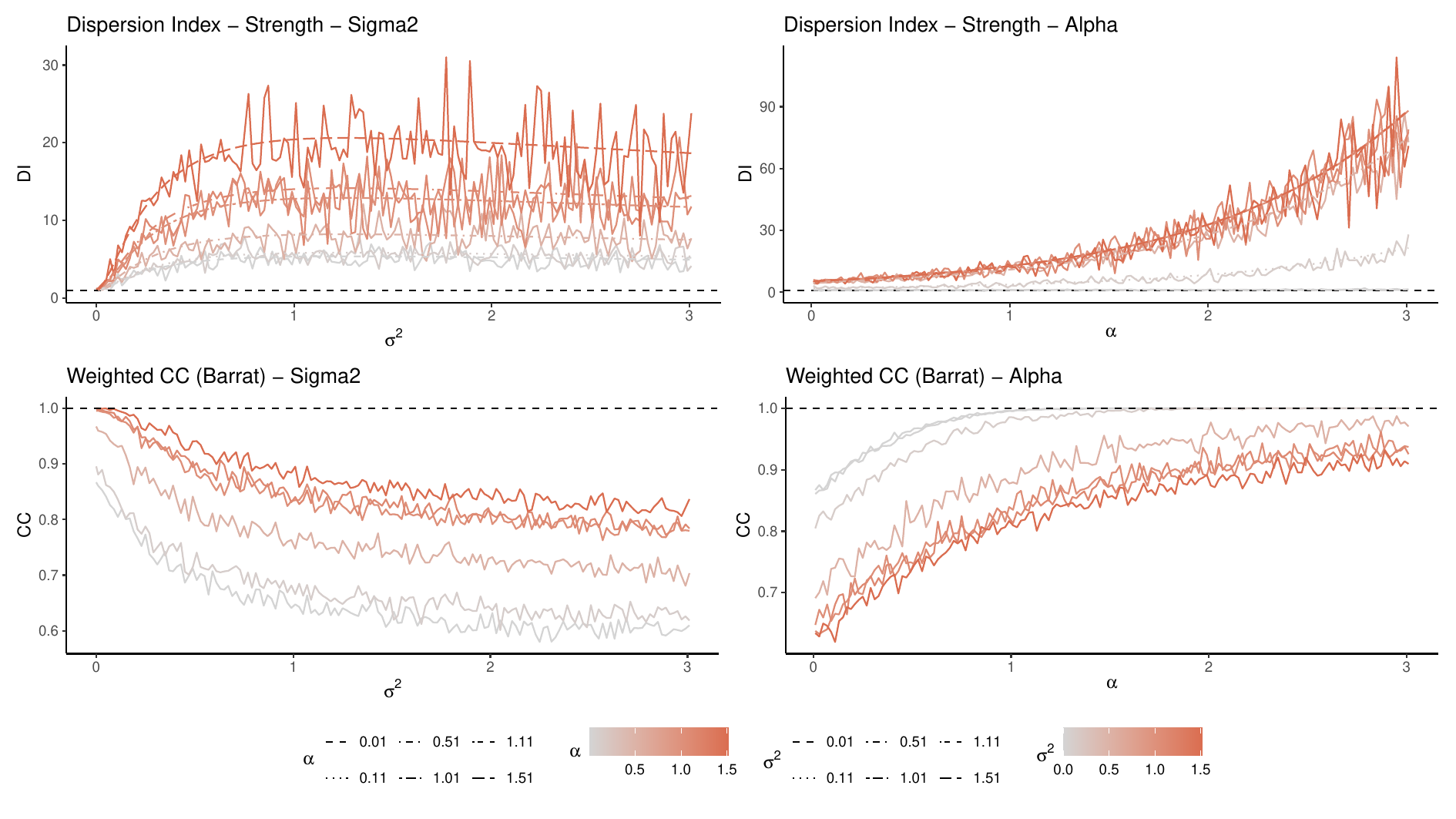}
    \caption{\textbf{Simulation Result - 2: } Sensitivity analysis for the dispersion index and clustering coefficient $CC$ (\citealp{barrat2004architecture}) as $\alpha$ and $\sigma^2$ vary. In this simulation, we assumed $\beta = 1$.} 
    \label{fig:simulation2}
\end{figure}

Figures \ref{fig:simulation1} and \ref{fig:simulation2} report the sensitivity analysis for the aforementioned statistics as $\alpha$ and $\sigma^2$ vary. In this simulation we assume $\beta = 1$. Notice how the theoretical quantities obtained in Appendix \ref{A:properties} (dashed lines) match with empirical quantities (solid lines).

\renewcommand\thefigure{C.\arabic{figure}}
\setcounter{figure}{0}
\renewcommand\theequation{C.\arabic{equation}}
\setcounter{equation}{0}
\renewcommand\thetable{C.\arabic{table}}
\setcounter{table}{0}

\section{Details of the prior and MCMC sampler}
\label{C:mcmc}

Let $\mathbf{Y}=(\mathbf{Y}_1,\ldots,\mathbf{Y}_T)$ and $\mathbf{L}=(\mathbf{L}_1,\ldots,\mathbf{L}_T)$ be the collections of the observables. The joint likelihood function $f( \mathbf{Y}, {\mathbf{L}}| \boldsymbol{\theta})$ can be written as:
\begin{equation*}
f(\mathbf{Y},{\mathbf{L}}|\boldsymbol{\theta})=\sum_{{\mathbf{S}}\in\{1,\ldots,K\}^{T}}\prod_{t=1}^{T}\prod_{i=1}^{N}\left(f_{B}({L_{it}}|{S_t},\boldsymbol{\theta})\prod_{j = i +1}^{N}f_{P}(Y_{ijt}|{S_t},\boldsymbol{\theta})\right)h({S_t}|{S_{t-1}}),
\end{equation*}
where $f_{P}(Y_{ijt}|{S_{t}},\boldsymbol{\theta})$ is the pmf of the Poisson distribution with dynamic intensity given in  (\ref{eq:pois}), $f_{B}({L_{it}}|{S_{t}}, \boldsymbol{\theta})$ the density of the Beta distribution given in (\ref{eq:betaLean}) and
\begin{equation*}
h({S_{t}}|{S_{t-1}})=\prod_{l=1}^{K}\prod_{k=1}^{K} q_{lk}^{\mathbb{I}({S_{t-1}}=l)\mathbb{I}({S_{t}}=k)},
\end{equation*}
the hidden Markov chain transition distribution. The joint posterior distribution is not tractable; thus, data augmentation was performed. In addition, given the cross-country independence assumption for both the likelihood and the joint prior, the Gibbs sampler iterates independent chains over the four countries in the application. We report below the prior specification and the full conditional distributions for the components of $\boldsymbol{\theta}$ together with a brief discussion of the sampling method.

\subsection[Prior Specification]{Prior Specification}\label{sec:AppPrior}

We take a Bayesian approach to inference and in the prior structure (\ref{prior}), we choose the following prior distributions (as explained in Section \ref{subsec:identification} we fix $\beta = 1$ in our empirical application):
\begin{align*}
\alpha_{i} &\sim\mathcal{N}(0, \sigma^2_\alpha ),\,
{\mathbf{Z}_{ik}|\sigma^2_k  \sim \mathcal{N}
(\boldsymbol{0}, \sigma^2_{k} I_{d})}, \sigma^2_{k} \sim \mathcal{IG}(a_{\sigma^2}, b_{\sigma^2}), \tag{1}\label{prior1}\\
\gamma_{0}  \sim \mathcal{N}(0, b_{\gamma_0} ),
\boldsymbol{\gamma}_{1}  &\sim \mathcal{N}
(\boldsymbol{0}, b_{\gamma_1}I_{d} ),\,
\phi \sim \mathcal{G}a(a_{\phi},b_{\phi}),\,
\boldsymbol{q}_{l} \sim \mathcal{D}ir(\omega_1, \ldots , \omega_{K}),\tag{2}\label{prior2}
\end{align*}
where $\mathcal{N}(\boldsymbol{\mu},\Sigma)$ denotes a multivariate normal distribution with mean vector $\boldsymbol{\mu}$ and variance-covariance matrix $\Sigma$, $\mathcal{G}a(a,b)$ denotes the gamma distribution with shape parameter $a$ and precision parameter $b$, with mean $a/b$, $\mathcal{IG}(a,b)$ denotes the inverse-gamma distribution with shape parameter $a$ and scale parameter $b$, with mean $b/(a-1)$ existing for $a>1$ and $\mathcal{D}ir(c_1, \ldots, c_K)$ denotes the Dirichlet distribution with parameters $c_1, \ldots, c_K$.

In our implementations of the model, we have little prior information at our disposal and we opt for the use of relatively vague priors to let the data speak: $\alpha_{i} \sim\mathcal{N}(0, 15^2 )$, {$\mathbf{Z}_{ik}|\sigma_k^2  \sim \mathcal{N}
(\mathbf{0}, \sigma^2_{k}I_d)$},  $\sigma^2_{k} \sim \mathcal{IG}(0.1, 0.1)$, $\gamma_{0}  \sim \mathcal{N}(0, 15^2 )$, $\boldsymbol{\gamma}_{1}  \sim \mathcal{N}(\mathbf{0}, 15^2I_d )$, $\phi \sim \mathcal{G}a(0.01,0.01)$ and $\boldsymbol{q}_{l} \sim \mathcal{D}ir( 2, \ldots, 2)$.
Our results are robust to substantial changes in these prior hyperparameters.

\subsection[Full conditional distribution of theta]{Full conditional distribution of $\alpha_{i}$}

\begin{align*}
& p(\alpha_{i}| \ldots ) \propto \pi(\alpha_{i})f(\mathbf{Y}, {\mathbf{L}}| \boldsymbol{\theta} ) \propto f_N(\alpha_{i}; 0,\sigma^2_\alpha)\prod_{t=1}^{T}\prod_{i = 1}^{N}\prod_{j = i +1}^{N}f_{P}(Y_{ijt}|\lambda_{ijt})\\
&\propto f_N(\alpha_{i}; 0,\sigma^2_\alpha)\prod_{t=1}^{T}\prod_{i = 1}^{N}\prod_{j = i +1}^{N}\lambda_{ijt}^{Y_{ijt}} e^{-\lambda_{ijt}}, 
\end{align*}
where $\log \lambda_{ijt} = \alpha_{i} + \alpha_{j} - \vert\vert {\boldsymbol{\zeta}_{i\sdot}}\boldsymbol{\xi}_{t}-{\boldsymbol{\zeta}_{j\sdot}}\boldsymbol{\xi}_{t}\vert\vert^2$. We sample from $p(\alpha_{i}| \ldots )$ via Adaptive RW-MH. 

\subsection[Full conditional distribution of phi]{Full conditional distribution of $\phi$}
\begin{align*}
    & p(\phi|\ldots) \propto \pi(\phi)f(\mathbf{Y}, {\mathbf{L}}| \boldsymbol{\theta} ) \\
    & \propto \phi^{a_{\phi}-1}e^{-b_{\phi}\phi}\prod_{t=1}^{T}\prod_{i=1}^{N}\frac{{L_{it}}^{a_{it} - 1}(1 -  {L_{it}})^{b_{it} -1 }}{B(a_{it}, b_{it})},
\end{align*}
where $a_{it} = \varphi(\gamma_{0}+ \boldsymbol{\gamma}_{1}'{\boldsymbol{\zeta}_{i\sdot}}\boldsymbol{\xi}_{t})\phi$, $b_{it} = \left(1- \varphi(\gamma_{0} +\boldsymbol{\gamma}_{1}'{\boldsymbol{\zeta}_{i\sdot}}\boldsymbol{\xi}_{t})\right)\phi $.  We sample from $p(\phi|\ldots) $ via RW-MH with a truncated Gaussian proposal.

\subsection[Full conditional distribution of gamma0 and gamma1]{Full conditional distribution of $\gamma_{0}$ and $\gamma_{1}$}

\begin{align*}
    &p(\gamma_{0}, \boldsymbol{\gamma}_{1}|\ldots) \propto \pi(\gamma_{0},  \boldsymbol{\gamma}_{1})f(\mathbf{Y}, {\mathbf{L}}|\boldsymbol{\theta} )\\
    &\propto f_N(\gamma_{0}; 0, b_{\gamma_0})f_N(\boldsymbol{\gamma}_1; \boldsymbol{0}, b_{\gamma_1}I_d)\prod_{t=1}^{T}\prod_{i=1}^{N}\frac{{L_{it}}^{a_{it} - 1}(1 -  {L_{it}})^{b_{it} -1 }}{B(a_{it}, b_{it})},
\end{align*}
where $a_{it} = \varphi(\gamma_{0} + \boldsymbol{\gamma}_{1}'{\boldsymbol{\zeta}_{i\sdot}}\boldsymbol{\xi}_{t})\phi$, $b_{it} = (1- \varphi(\gamma_{0} + \boldsymbol{\gamma}_{1}'{\boldsymbol{\zeta}_{i\sdot}}\boldsymbol{\xi}_{t}))\phi $.  We sample from $p(\gamma_{0}, \boldsymbol{\gamma}_{1}|\ldots)$ via RW-MH.

\subsection[Full conditional distribution of z]{Full conditional distribution of $\zeta_{ik}$}
\begin{align*}
&p(\boldsymbol{\zeta}_{ik}|\ldots) \propto \pi(\boldsymbol{\zeta}_{ik}|\sigma_{k}^2)f(\mathbf{Y}, {\mathbf{L}}| \boldsymbol{\theta})\\
&\propto f_N(\boldsymbol{\zeta}_{ik}; \mathbf{0}, \sigma^2_{k}I_d)\prod_{t \in \mathcal{T}_{k}}\frac{{L_{it}}^{a_{it} - 1}(1 -  {L_{it}})^{b_{it} -1 }}{B(a_{it}, b_{it})} \prod_{j \in \mathcal{S}^{-i} }\lambda_{ijt}^{Y_{ijt}} e^{-\lambda_{ijt}},
\end{align*}
where $\mathcal{T}_{k} = \{t : \xi_{kt} = 1\}$,  $\mathcal{S}^{-i} = \{j \in 1, \ldots, N : j \neq i \}$, $\log \lambda_{ijt} = \alpha_{i} + \alpha_{j} - \vert\vert {\boldsymbol{\zeta}_{i\sdot}}\boldsymbol{\xi}_{t}-{\boldsymbol{\zeta}_{j\sdot}}\boldsymbol{\xi}_{t}\vert\vert^2$, $a_{it} = \varphi(\gamma_{0} + \boldsymbol{\gamma}_{1}'{\boldsymbol{\zeta}_{i\sdot}}\boldsymbol{\xi}_{t}) \phi$, $b_{it} = (1- \varphi(\gamma_{0} + \boldsymbol{\gamma}_{1}'{\boldsymbol{\zeta}_{i\sdot}}\boldsymbol{\xi}_{t}))\phi$.  We sample from $p(\zeta_{ik}|\ldots)$ via Adaptive RW-MH (see Subsection \ref{subsec:adaptive}).

\subsection[Full conditional distribution of sigma]{Full conditional distribution of $\sigma^2_{k}$}
Denoting by ${\zeta}_{ivk}$ the $v$th element of $\boldsymbol{\zeta}_{ik}$ for $v=1,\dots,d$, we can write
\begin{align*}
&p(\sigma^2_{k}|{\boldsymbol{\zeta}_{\sdot k}}, \boldsymbol{\xi}) \propto \pi(\sigma^2_{k})\pi({\boldsymbol{\zeta}_{\sdot k}}|\sigma^2_{k} , \boldsymbol{\xi})
\\ &\propto (1 / \sigma^2_{k})^{a_{\sigma^2}+1} e^{-b_{\sigma^2} \frac{1}{\sigma^2_{k}}}\prod_{i = 1}^{N}\prod_{v = 1}^{d} \left(1/\sigma_{k}^{2}\right)^{\frac{1}{2}} e^{-\frac{1}{2\sigma^2_{k}}{\zeta}_{ivk}^2}
\\ &\propto (1 / \sigma^2_{k})^{a_{\sigma^2}+1} e^{-b_{\sigma^2} \frac{1}{\sigma^2_{k}}} \left(1/\sigma_{k}^{2}\right)^{\frac{Nd}{2}} e^{-\frac{1}{2\sigma^2_{k}}\sum_{i= 1}^{N}\sum_{v= 1}^{d}{\zeta}_{ivk}^2}\\
&\propto (1 / \sigma^2_{k})^{a_{\sigma^2} + \frac{Nd}{2} +1} e^{-\left(b_{\sigma^2}+ \frac{\sum_{i= 1}^{N}\sum_{v= 1}^{d}{\zeta}_{ivk}^2}{2} \right)\frac{1}{\sigma^2_{k}}}, 
\end{align*}
which is the density function of an $\mathcal{IG}(a^{*}_{\sigma^2 k}, b^{*}_{\sigma^2 k})$ distribution, where $a^{*}_{\sigma^2 k} = a_{\sigma^2} + \frac{Nd}{2}$ and $b^{*}_{\sigma^2 k} = b_{\sigma^2} + \frac{\sum_{i =1}^{N}\sum_{v =1}^{d}\zeta^2_{ivk}}{2}$.

\subsection[Full conditional distribution of pl]{Full conditional distribution of $\boldsymbol{q}_{l}$}
\begin{align*}
&p(\boldsymbol{q}_{l}|\boldsymbol{\xi}) \propto \left(\prod_{t=1}^{T}\prod_{k=1}^{K} q_{lk}^{\xi_{lt-1}\xi_{kt}}\right)\left(\prod_{k=1}^{K}q^{\omega_k - 1}_{lk}\right)
\\&\propto \prod_{k=1}^{K} q_{lk}^{\sum_{t=1}^{T}(\xi_{lt-1}\xi_{kt})+\omega_k - 1 }, 
\end{align*}
which is the density function of a Dirichlet distribution with parameters $\Bar{\omega}_{1}, \ldots, \Bar{\omega}_{K}$ where $\Bar{\omega}_{k} = \sum_{t=1}^{T}(\xi_{lt-1}\xi_{kt})+\omega_k - 1 $.

\subsection{Adaptive MCMC}
\label{subsec:adaptive}

Sampling from the full conditional of $\boldsymbol{\alpha}$ and $\boldsymbol{\zeta}$ is obtained via the Adaptive MH algorithm with global adaptive scaling proposed in \citet{andrieu2008tutorial}.  Adaptive MH generates samples from the distribution of $\boldsymbol{\theta}$ by iterating the following steps:
\begin{enumerate}
    \item Starting values for the parameter of interest $\boldsymbol{\theta}_0$ and for $\boldsymbol{\mu}_0$ and $\boldsymbol{\Sigma}_0$ are chosen. 
    \item For each iteration $h$, given $\boldsymbol{\theta}_{h}$, $\boldsymbol{\mu}_{h}$, $\boldsymbol{\Sigma}_{h}$ and $\delta_{h}$:
    \begin{enumerate}
        \item $\Tilde{\boldsymbol{\theta}}_{h} \sim f_N(\boldsymbol{\theta}_{h-1}, \delta_{h-1}\boldsymbol{\Sigma}_{h-1})$ is sampled and $\boldsymbol{\theta}_{h} = \Tilde{\boldsymbol{\theta}}_{h} $ with probability $\alpha(\boldsymbol{\theta}_{h-1}, \Tilde{\boldsymbol{\theta}}_{h})$, otherwise $\Tilde{\boldsymbol{\theta}}_{h}  = \boldsymbol{\theta}_{h-1}$;
        \item Update $\log(\delta_h) = \log(\delta_{h-1}) + \gamma_h[\alpha(\boldsymbol{\theta}_{h-1}, \Tilde{\boldsymbol{\theta}}_{h}) - \alpha^*]$;
        \item Update $\boldsymbol{\mu}_{h} = \boldsymbol{\mu}_{h-1}+\gamma_h(\boldsymbol{\theta}_h - \boldsymbol{\mu}_{h-1}) $;
        \item Update $\boldsymbol{\Sigma}_h = \boldsymbol{\Sigma}_{h-1} + \gamma_h[(\boldsymbol{\theta}_h - \boldsymbol{\mu}_{h-1})(\boldsymbol{\theta}_h - \boldsymbol{\mu}_{h-1})' - \boldsymbol{\Sigma}_{h-1}  ] $;
    \end{enumerate}
where $\gamma_h = \frac{1}{h^\psi}$ for $\psi \in (0,1)$ and $\alpha^*$ is a target acceptance rate, here chosen to be 25\%.
\end{enumerate}

\renewcommand\thefigure{D.\arabic{figure}}
\setcounter{figure}{0}
\renewcommand\theequation{D.\arabic{equation}}
\setcounter{equation}{0}
\renewcommand\thetable{D.\arabic{table}}
\setcounter{table}{0}

\renewcommand\thefigure{D.\arabic{figure}}
\setcounter{figure}{0}
\renewcommand\theequation{D.\arabic{equation}}
\setcounter{equation}{0}
\renewcommand\thetable{D.\arabic{table}}
\setcounter{table}{0}

\newpage
\section{MCMC Properties for the Simulated Data}
\label{E:traceplots}

 \begin{figure}[!htb]
  \centering
   \includegraphics[width=0.8\linewidth]{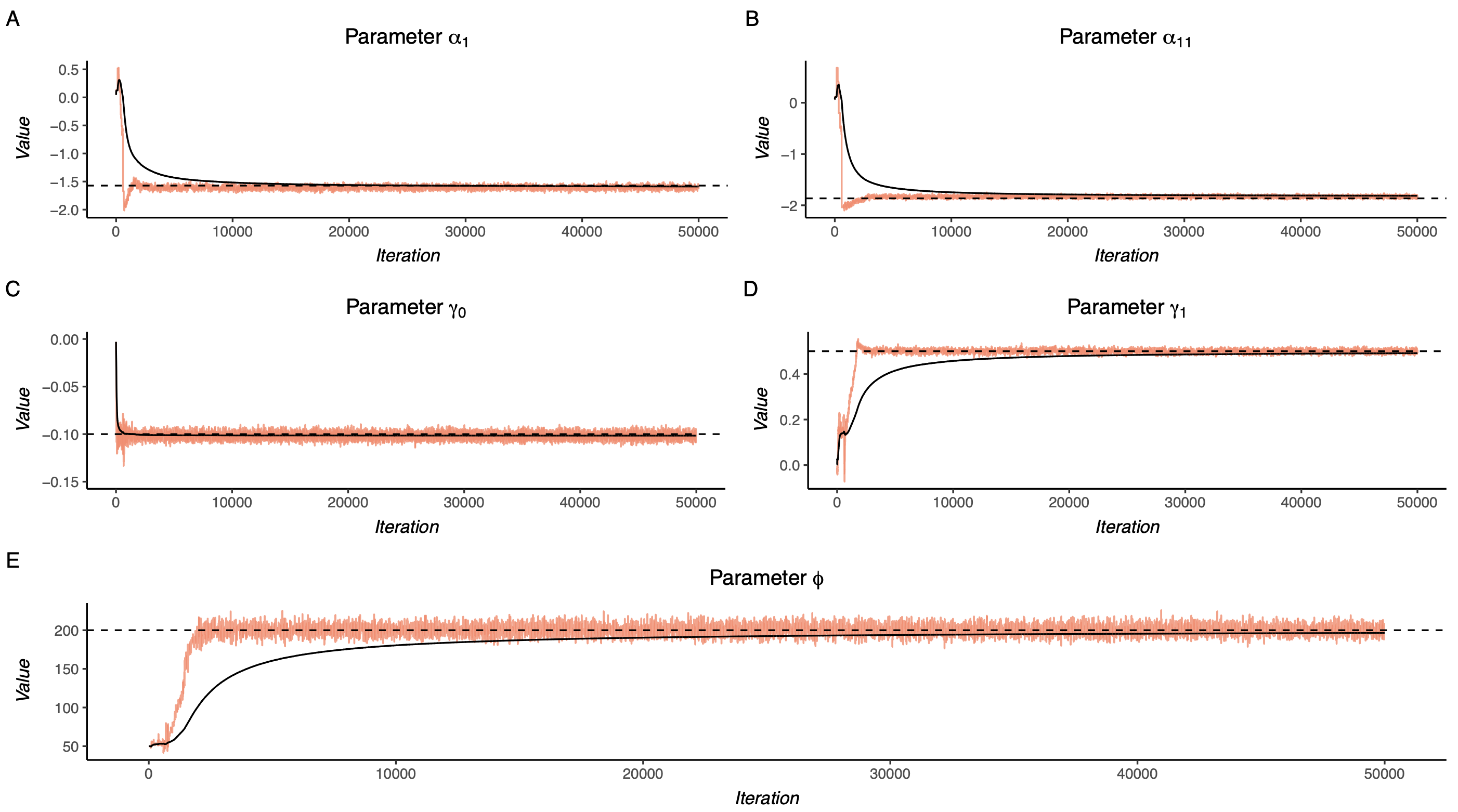}
 \caption{\textbf{Simulation Output:} Trace plots for the parameters $\alpha_1$  (Panel A),  $\alpha_{11}$  (Panel B), $\gamma_0$ (Panel C), $\gamma_1$ (Panel D) and $\phi$ (Panel E). The solid black line represents the cumulative average while the dashed black line represents the true parameter value used in the simulation. }
 \label{fig:simul_plane_1}
 \end{figure}

 \begin{figure}[!htb]
  \centering
   \includegraphics[width=0.8\linewidth]{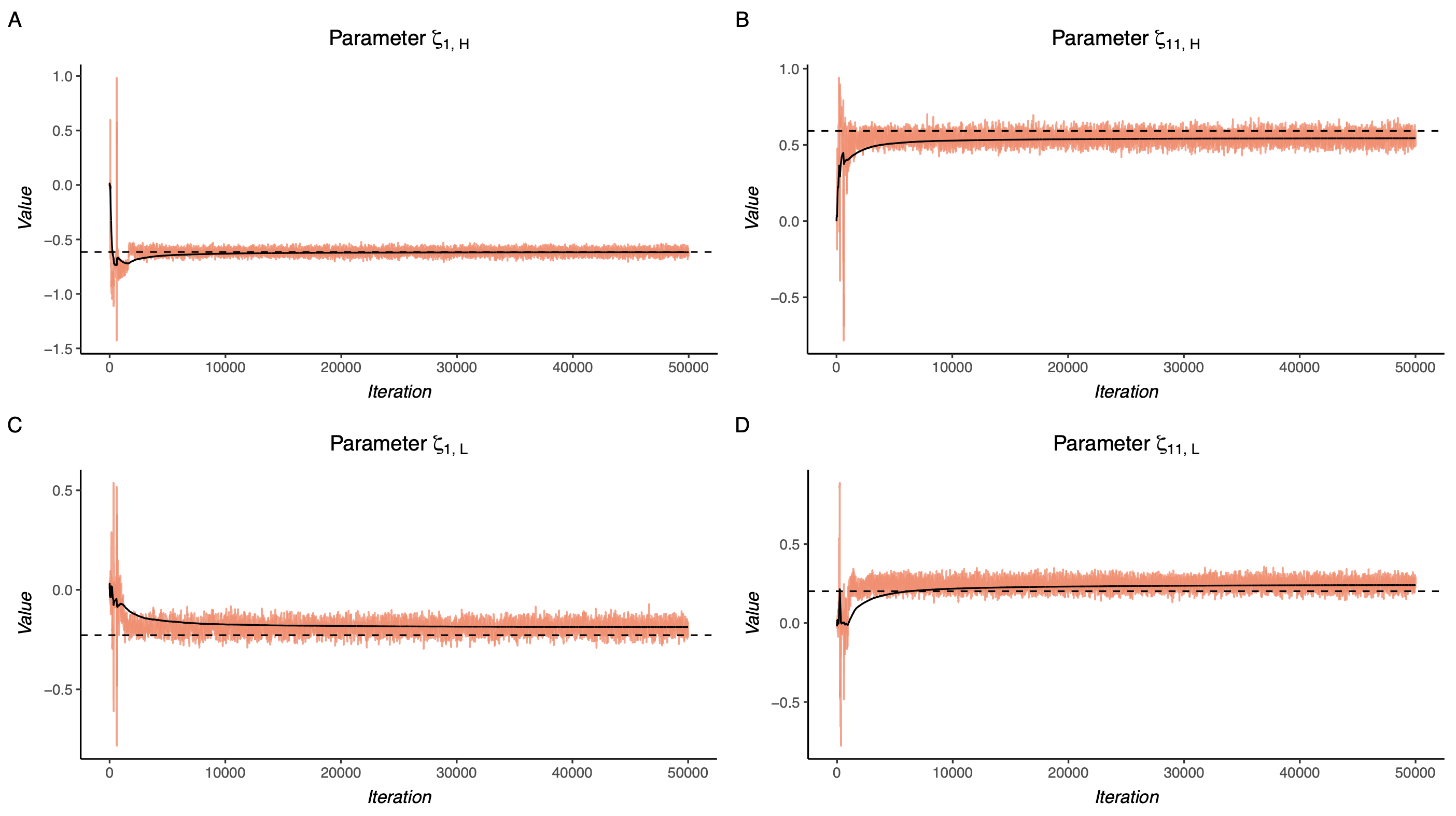}
 \caption{\textbf{Simulation Output:} Trace plots for the latent leaning for outlet $1$ --  $\zeta_{1,L}$ (Panel A), $\zeta_{1, H}$ (Panel B) -- an for outlet $11$ --  $\zeta_{11,L}$ (Panel C), $\zeta_{11, H}$ (Panel D). The solid black line represents the cumulative average while the dashed black line represents the true parameter value used in the simulation.}
 \label{fig:simul_plane_2}
 \end{figure}

 \begin{figure}[!htb]
  \centering
   \includegraphics[width=0.8\linewidth]{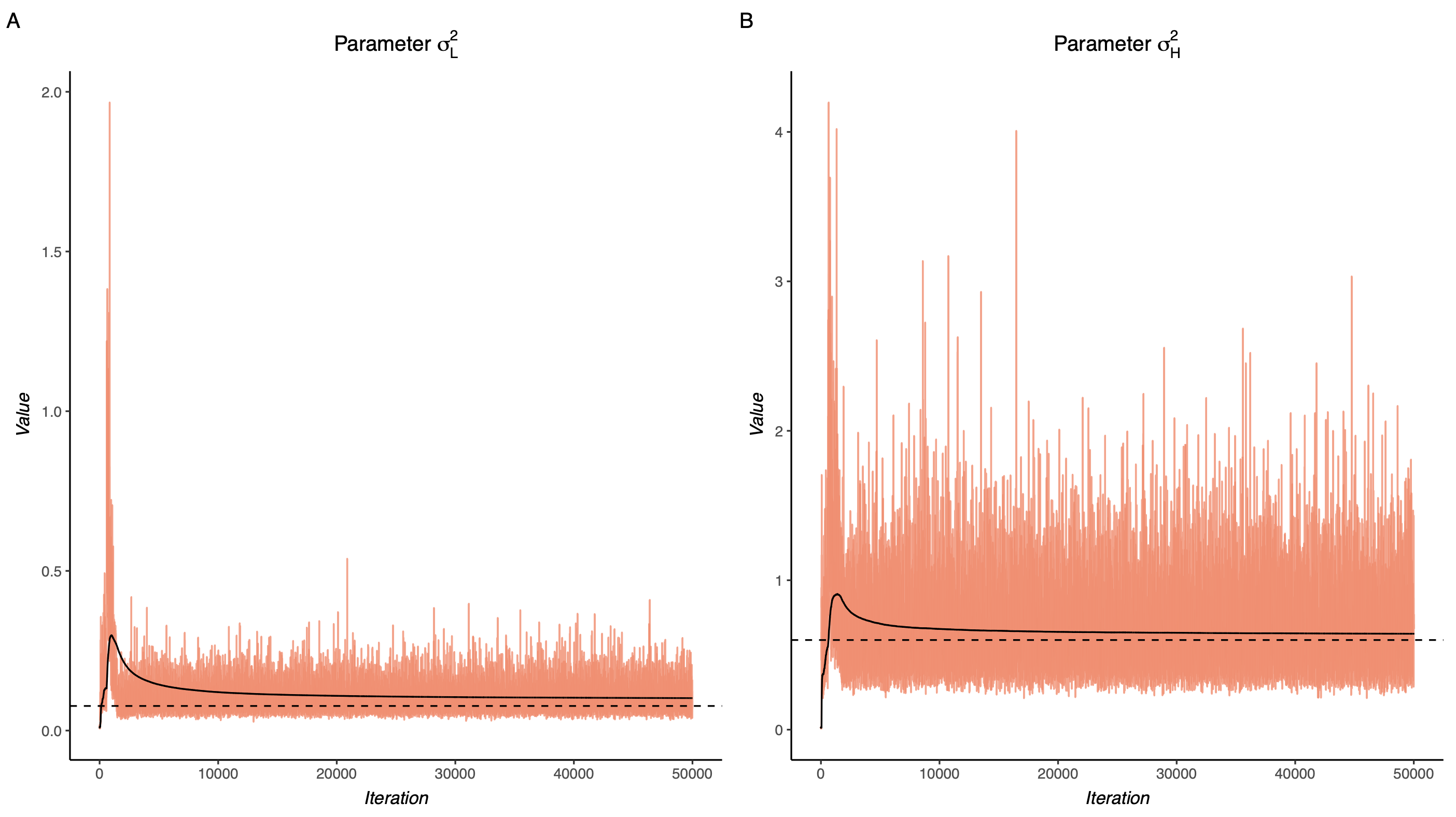}
 \caption{\textbf{Simulation Output:} Trace plots for the parameters $\sigma^2_L$ and $\sigma^2_H$. The solid black line represents the cumulative average while the dashed black line represents the sample variance of the latent coordinates values used in the simulation.}
 \label{fig:simul_plane_3}
 \end{figure}

\newpage

\begin{table}[htb!]
\caption{\textbf{Simulated data:} Auto-correlation (ACF) at lag 1, 10, 30, Acceptance Rate, Effective Sample Size over the number of draws (ESS) and Convergence Diagnostic p-value (CD) as defined in \citet{geweke1991evaluating} for the MCMC sampling of $\Bar{\alpha}$, $\gamma_0$, $\gamma_1$, $\phi$,  $\Bar{\zeta}_{L}$,  $\Bar{\zeta}_{H}$ respectively on the raw series, on the series collected after burn-in, and on the series applying burn-in and thinning. }
 \label{tab:diag}
 
\renewcommand{\arraystretch}{1.2}

\center\begin{tabular}{lcccccc}
\hline\hline
\multicolumn{7}{c}{Raw Series (50,000 obs.)}                                                             \\ \hline
Parameter & $\Bar{\alpha}_i$ & $\gamma_0 $ & $\gamma_1$ & $\phi$ & $\Bar{\zeta}_{L}$ & $\Bar{\zeta}_{H}$ \\ \hline
ACF(1)    & 0.996            & 0.788       & 0.997      & 0.980  & 0.924             & 0.956             \\
ACF(10)   & 0.972            & 0.190       & 0.978      & 0.925  & 0.713             & 0.803             \\
ACF(30)   & 0.930            & 0.009       & 0.947      & 0.911  & 0.487             & 0.641             \\
Acc.      & 25\%             & 26\%        & 26\%       & 21\%   & 25\%              & 25\%              \\
ESS       & 0\%              & 8\%         & 0\%        & 0\%    & 2\%               & 1\%               \\
CD p-val  & 0.050            & 0.175       & 0.000      & 0.000  & 0.109             & 0.142             \\ \hline
\multicolumn{7}{c}{With Burn-in (20,000 obs.)}                                                           \\ \hline
Parameter & $\Bar{\alpha}_i$ & $\gamma_0$  & $\gamma_1$ & $\phi$ & $\Bar{\zeta}_{L}$ & $\Bar{\zeta}_{H}$ \\ \hline
ACF(1)    & 0.891            & 0.730       & 0.898      & 0.750  & 0.849             & 0.766             \\
ACF(10)   & 0.529            & 0.034       & 0.398      & 0.067  & 0.374             & 0.149             \\
ACF(30)   & 0.318            & -0.017      & 0.145      & -0.011 & 0.174             & 0.048             \\
Acc.      & 25\%             & 25\%        & 25\%       & 22\%   & 25\%              & 25\%              \\
ESS       & 3\%              & 16\%        & 4\%        & 14\%   & 6\%               & 10\%              \\
CD p-val  & 0.213            & 0.399       & 0.040      & 0.146  & 0.300             & 0.256             \\ \hline
\multicolumn{7}{c}{With Burn-in and Thinning every 10 (2000 obs.)}                                       \\ \hline
Parameter & $\Bar{\alpha}_i$ & $\gamma_0 $ & $\gamma_1$ & $\phi$ & $\Bar{\zeta}_{L}$ & $\Bar{\zeta}_{H}$ \\ \hline
ACF(1)    & 0.527            & 0.017       & 0.401      & 0.040  & 0.367             & 0.149             \\
ACF(10)   & 0.135            & -0.005      & 0.054      & 0.028  & 0.057             & 0.011             \\
ACF(30)   & 0.022            & 0.022       & 0.003      & 0.004  & 0.005             & 0.008             \\
Acc.      & -                & -           & -          & -      & -                 & -                 \\
ESS       & 23\%             & 100\%       & 35\%       & 93\%   & 44\%              & 66\%              \\
CD p-val  & 0.209            & 0.402       & 0.033      & 0.013  & 0.275             & 0.259             \\ \hline\hline
\end{tabular}

\end{table}

\clearpage

\renewcommand\thefigure{E.\arabic{figure}}
\setcounter{figure}{0}
\renewcommand\theequation{E.\arabic{equation}}
\setcounter{equation}{0}
\renewcommand\thetable{E.\arabic{table}}
\setcounter{table}{0}

\section{Sensitivity Analysis}
\label{sens_analysis}

\begin{figure}[!htb]
  \centering
  \includegraphics[scale = 0.45]{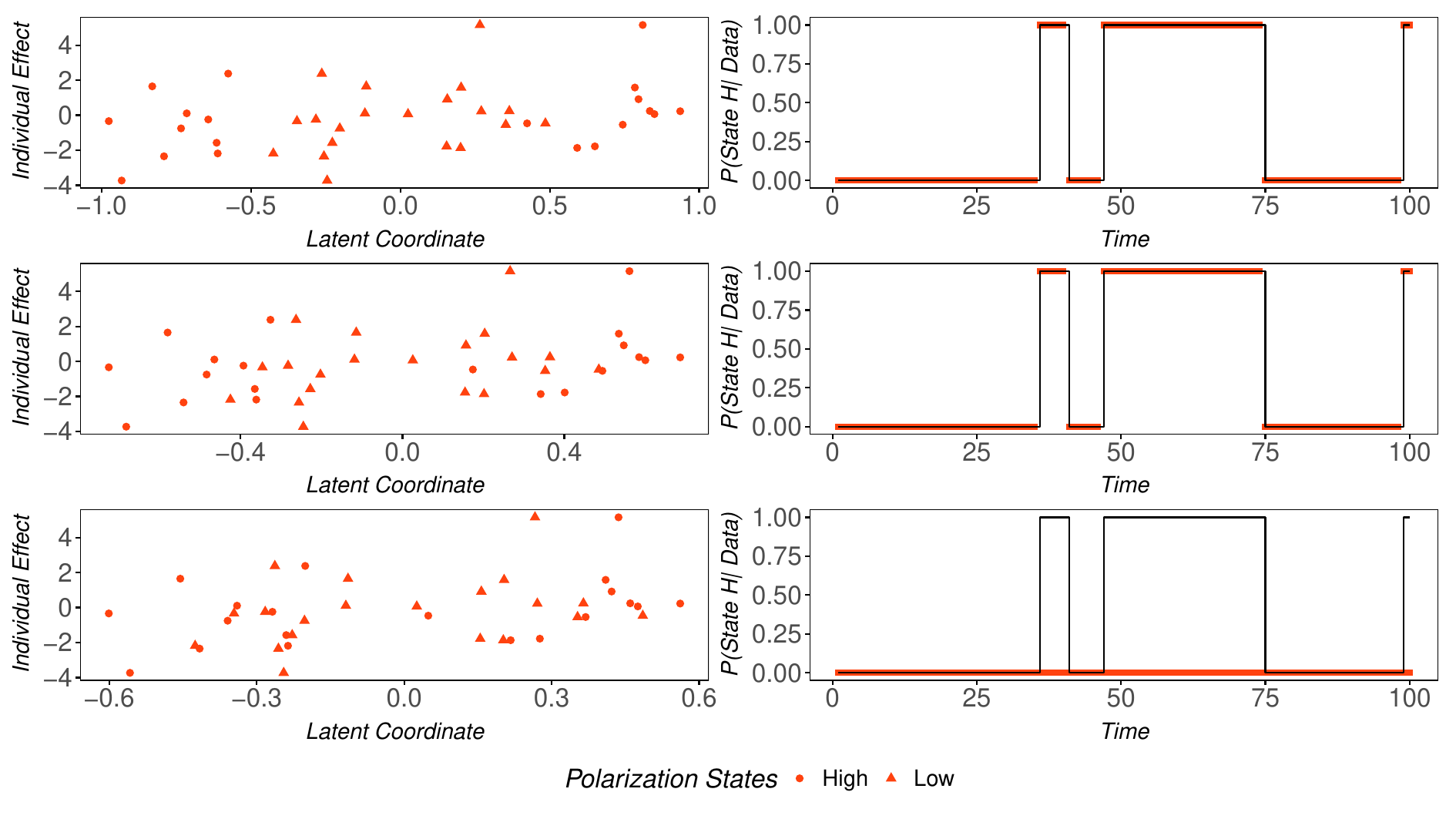}
  \caption{\textbf{Separation Sensitivity:} The plot reports the sensitivity analysis with respect to the separation of the latent coordinates in the two states. From top to bottom: high separation, intermediate separation, and low separation. The black solid line denotes the true states, while the thick red line denotes the posterior median.} 
  \label{simsens}
 \end{figure}

\begin{figure}[!htb]
  \centering
  \includegraphics[scale = 0.45]{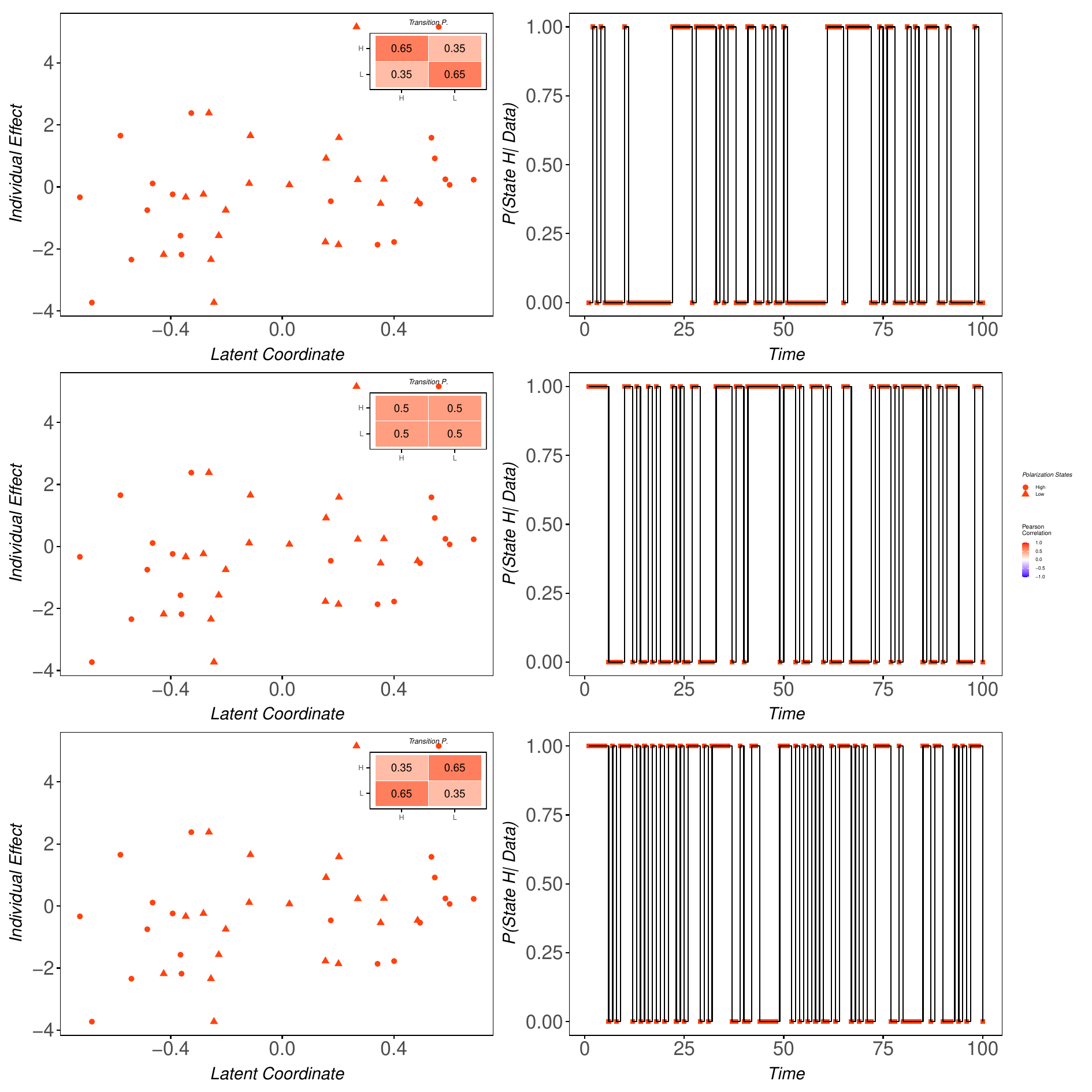}
  \caption{\textbf{Persistence Sensitivity:} The plot reports the sensitivity analysis with respect to the different degrees of persistence in the two states. The black solid line denotes the true states, while the thick red line denotes the posterior median.} 
  \label{persens}
 \end{figure}

Figure \ref{simsens} presents a sensitivity analysis of the state detection as the two regimes become increasingly indistinguishable in the latent space. We observe that increasing the similarity between the two regimes leads to the expected problems in accurately detecting the states. Figure \ref{persens} reports a sensitivity analysis of inference on the state as the transition between regimes gets less and less persistent. We don't notice any issue as the transition matrix elements vary for cases in which the two regimes are sufficiently separated.

\clearpage

\renewcommand\thefigure{F.\arabic{figure}}
\setcounter{figure}{0}
\renewcommand\theequation{F.\arabic{equation}}
\setcounter{equation}{0}
\renewcommand\thetable{F.\arabic{table}}
\setcounter{table}{0}

\section{Model Selection}
\label{subsec:model_sel}

We compute DIC to compare models as in \citet{gelman2014understanding}: $DIC(\mathcal{M}) = -2\overline{\log f}(\mathbf{Y},{\mathbf{L}}|\boldsymbol{\theta}, \mathcal{M} ) + p_D$, where $\overline{\log f}(\mathbf{Y},{\mathbf{L}}|\boldsymbol{\theta}, \mathcal{M}) = (1/H)\sum_{h = 1}^H\log f(\mathbf{Y},{\mathbf{L}}|\boldsymbol{\theta}^h, \mathcal{M} )$ is the sample average of the loglikelihood computed at each iteration $h$ of model $\mathcal{M}$ while $p_D = 2\mathbb{V}ar\left(\log f(\mathbf{Y},{\mathbf{L}}|\boldsymbol{\theta}^{1:H}, \mathcal{M} )\right)$. We also compute the Log Pointwise Predictive Density (lppd) as in \citet{gelman2014understanding}: $lppd(\mathcal{M}) =\sum_{i>j,t} \log \left(\frac{1}{H} \sum_{h=1}^H f_P\left(Y_{ijt} \mid \boldsymbol{\theta}^h, \mathcal{M}\right)\right)$.

\renewcommand\thefigure{G.\arabic{figure}}
\setcounter{figure}{0}
\renewcommand\theequation{G.\arabic{equation}}
\setcounter{equation}{0}
\renewcommand\thetable{G.\arabic{table}}
\setcounter{table}{0}

\section{Details of the Observed Media Slant Index}
\label{F:MediaSlantProxy}

To construct an observable proxy for media slant we rely on the methodology proposed by \citet{gentzkow2010drives} and extended to online textual data by \citet{garz2020partisan}.
Textual processing is carried out with the use of the R package $\texttt{quanteda}$.
The underlying intuition is that of computing the distance between the language used by news outlets in their posts and the language used by political parties. To do so we compose two corpora of textual data: the \emph{Parties Corpus} consisting of the textual content of the posts published by the major parties in the years 2015-2016 on their Facebook wall, the \emph{Outlets Corpus} consisting of the same information related to the posts published by the Italian news outlets considered in this work.

\begin{figure}[!htp]
  \centering
  \includegraphics[height=127pt,width= 0.9\textwidth]{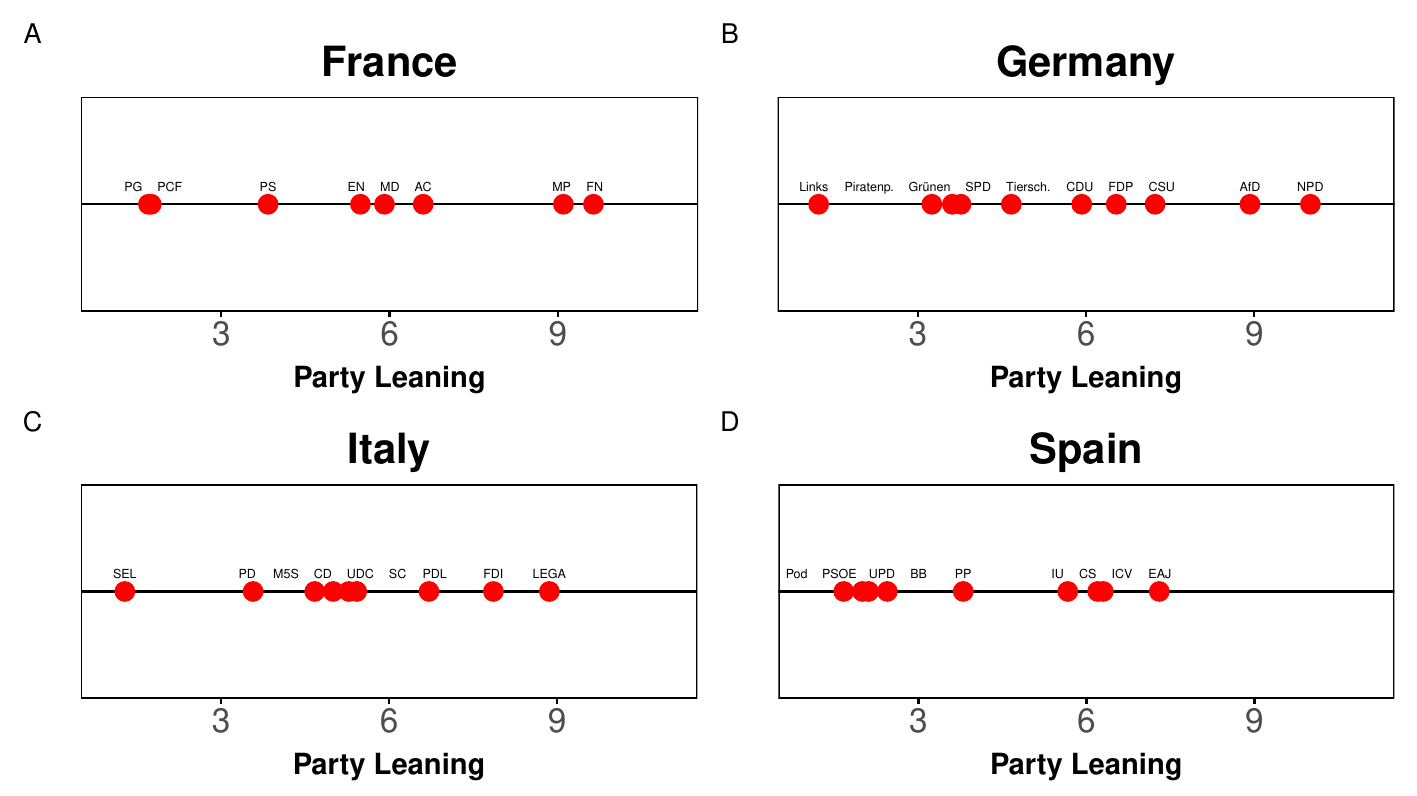}
\caption{\textbf{Ideology of Political Parties:} The index is computed according to the 2014 Chapel Hill Experts Survey \citep{polk2017explaining}.}
 \label{fig:PolLean}
\end{figure}

On both corpora, textual pre-processing is carried out (lower case transformation, punctuation removal, stopwords removal and \emph{n}-gram tokenization) and tokens not present in the Outlets corpus are filtered out from the Parties Corpus. By means of the TF-IDF score applied on the Parties Corpus, we retrieve the top 100 tokens with highest TF-IDF score for each party. We proceed assessing the cosine similarity between the vector of token occurrences $\mathbf{x}_{ot}$ obtained from the set of posts published by outlet $o$ at time $t$ and the set of token occurrences $\mathbf{y}_{p}$ characteristic of each party $p$:
\begin{equation*}
    sim_{p o t} = \frac{\sum_{k = 1}^K x_{k o t} y_{k p}}{\sqrt{\sum_{k = 1}^K x^2_{k o t}}\sqrt{\sum_{k = 1}^K  y^2_{k p} } }.
\end{equation*}
To take into account the fact that the style of posting adopted by some parties is closer to the one of news outlets and vice-versa, \citet{garz2020partisan} suggest regressing the similarity on a constant and both outlet and party fixed-effects to extract the residuals $\epsilon_{pot} $, which can be interpreted as a proxy of unexplained similarity. Finally the media slant for outlet $o$ at time $t$ is computed as: 
\begin{equation*}
    slant_{ot} = \sum_{p = 1}^{P} \epsilon_{pot} score_p,
\end{equation*}
where $score_p$ is the political leaning assigned to party $p$ by the 2014 Chapel Hill Experts Survey classification provided in \citet{polk2017explaining}, see Figure \ref{fig:PolLean}.

\begin{figure}[!htp]
  \centering
\includegraphics[height=127pt,width=0.9\textwidth]{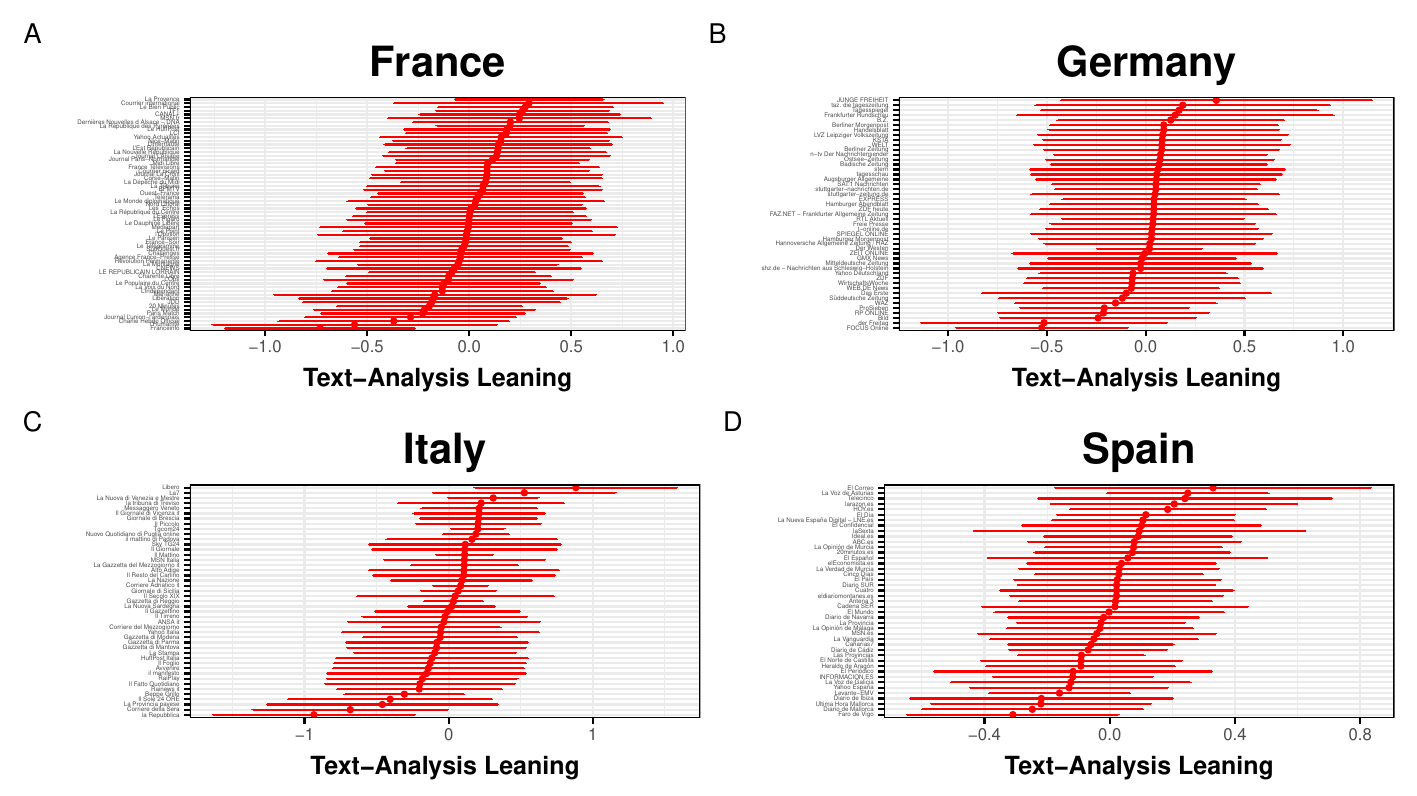}
\caption{\textbf{Media Slant of News Outlets:} The index is computed following the methodology proposed by \citet{gentzkow2010drives} and  \citet{garz2020partisan}. Full dots represent the average leaning and error bars represent the $\mu \pm \sigma$ interval over the time-lapse.}
 \label{fig:MediaBias}
\end{figure}

Figure \ref{fig:MediaBias} reports the average media slant for the available set of news outlets for France, Germany, Italy, and Spain, trough the whole time lapse.


\renewcommand\thefigure{H.\arabic{figure}}
\setcounter{figure}{0}
\renewcommand\theequation{H.\arabic{equation}}
\setcounter{equation}{0}
\renewcommand\thetable{H.\arabic{table}}
\setcounter{table}{0}

\section{Identifiers and List of Removed Outlets}
\label{G:Identifiers}

The table below lists the news outlets used as identifiers of the political orientation.

\begin{table}[htb!]

\caption{\textbf{Identifiers:} Set of outlets used to identify the political orientation of the news outlets. The column sign reports the assumed political orientation: either < 0 (left) or > 0 (right). The column PEW score reports the estimated leaning as per \citealp{mitchell2018western}.}
 \label{tab:identify}
 
\center
\begin{tabular}{ccccc}
\hline\hline

\multicolumn{1}{l}{News Outlet} & \multicolumn{1}{l}{Country} & \multicolumn{1}{l}{Assumed Media Bias} & \multicolumn{1}{l}{Sign} & PEW score (0-6) \\ \hline
l'Humanité                      & France                      & Left                                   & \textless{}0             & NA              \\
Bild                            & Germany                     & Center-Right                           & \textgreater{}0          & 3.1             \\
Libero                          & Italy                       & Center-Right                                  & \textgreater{}0          & 3.6             \\
ABC                             & Spain                       & Center-Right                                  & \textgreater{}0          & 3.3             \\ \hline\hline

\end{tabular}

\end{table}

 A few news outlets were removed in the dynamic analysis because of their prolonged inactivity, i.e.~15 days without any comment. They are {\small DE: \emph{Der Westen, GMX News, WEB.DE, News ZDF},
FR: \emph{Charlie Hebdo Officiel, France Télévisions, Franceinfo, LCI, Révolution Permanente}, IT: \emph{La Gazzetta del Mezzogiorno.it, MSN Italia}, SP: \emph{La Voz de Asturias, Yahoo España}}.

\renewcommand\thefigure{I.\arabic{figure}}
\setcounter{figure}{0}
\renewcommand\theequation{I.\arabic{equation}}
\setcounter{equation}{0}
\renewcommand\thetable{I.\arabic{table}}
\setcounter{table}{0}

\section{Models with different $d$ and $K$}
\label{G:high-order}

\begin{figure}[!htb]
  \centering
  \begin{tabular}{c}
\small {MS-LS model with $d = 1$, $K = 2$ ($\mathcal{M}_3$)}\\
   \includegraphics[width= 0.7\textwidth]{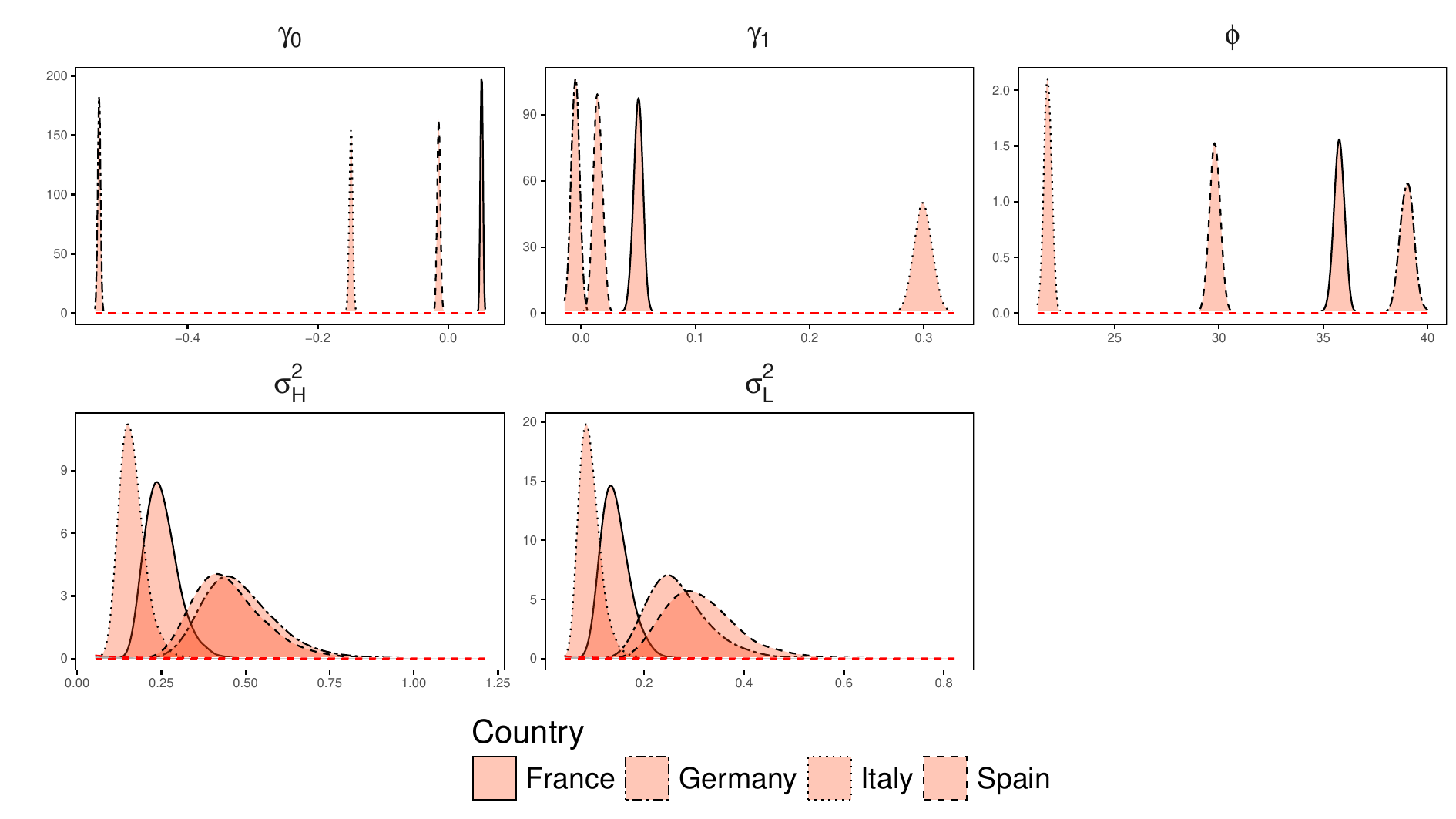}\\
\small {MS-LS model with $d = 2$, $K = 2$ ($\mathcal{M}_4$)}\\
    \includegraphics[width= 0.7\textwidth]{Figures/gamma_K2_D2/Figure8_v2_extended_gamma_K_2.pdf}\\
\small {MS-LS model with $d = 2$, $K = 3$ ($\mathcal{M}_5$)}\\
   \includegraphics[width= 0.7\textwidth]{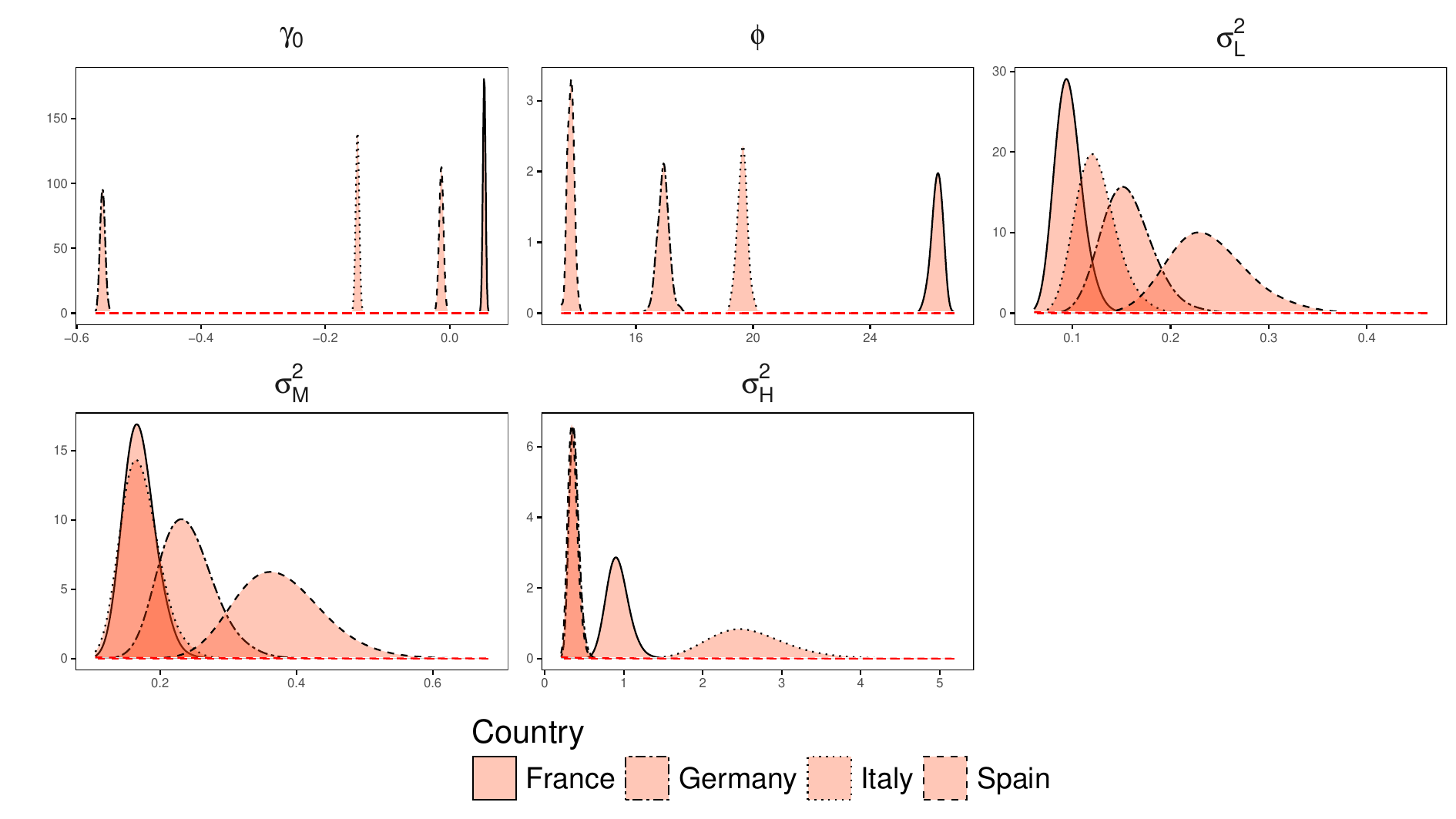}
\end{tabular}
\caption{\textbf{Marginal Posteriors - Dynamic Analysis} $\mathcal{M}_j$, $j=3,4,5$: Kernel posterior density estimates for the parameters $\gamma_{0}$, $\gamma_{1}$,  $\phi$, $\sigma^2_{L}$, $\sigma^2_{M}$ and $\sigma^2_{H}$ for France, Germany, Italy and Spain with prior pdfs indicated by red dashed lines (mostly indistinguishable from the horizontal axis).}
\label{fig:postDyn_sup}
 \end{figure}

\begin{figure}[htbp]
  \centering
   \includegraphics[width= 0.9\textwidth]{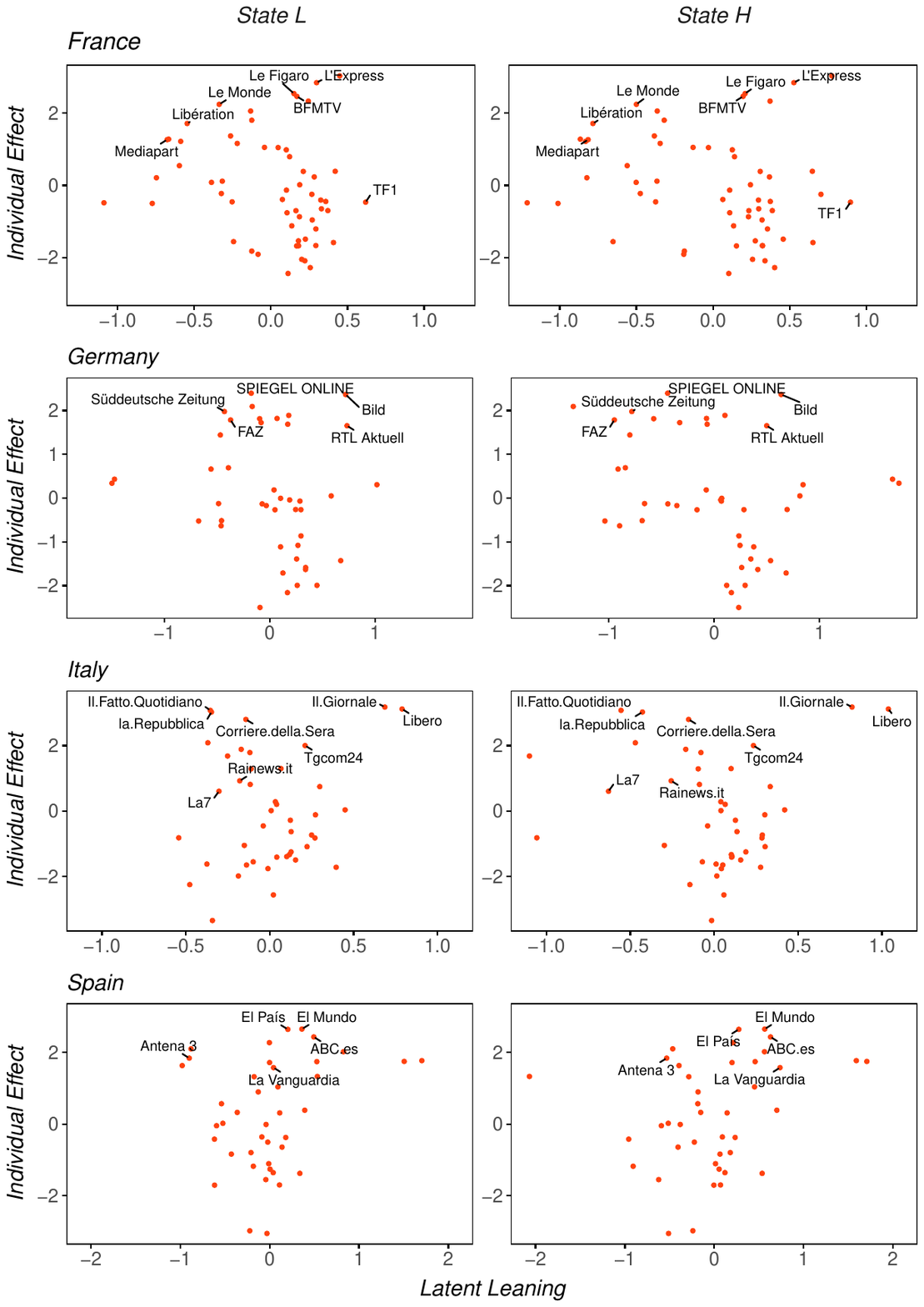}
 \caption{\textbf{Latent Positions - Dynamic Analysis $\mathcal{M}_3$:} Posterior means of individual effects and latent coordinates of the news outlets for France, Germany, Italy and Spain in State L and in State H.}
 \label{fig:dyn_latpos_sup}
 \end{figure}

\begin{figure}[htbp]
  \centering
   \includegraphics[width= 0.9\textwidth]{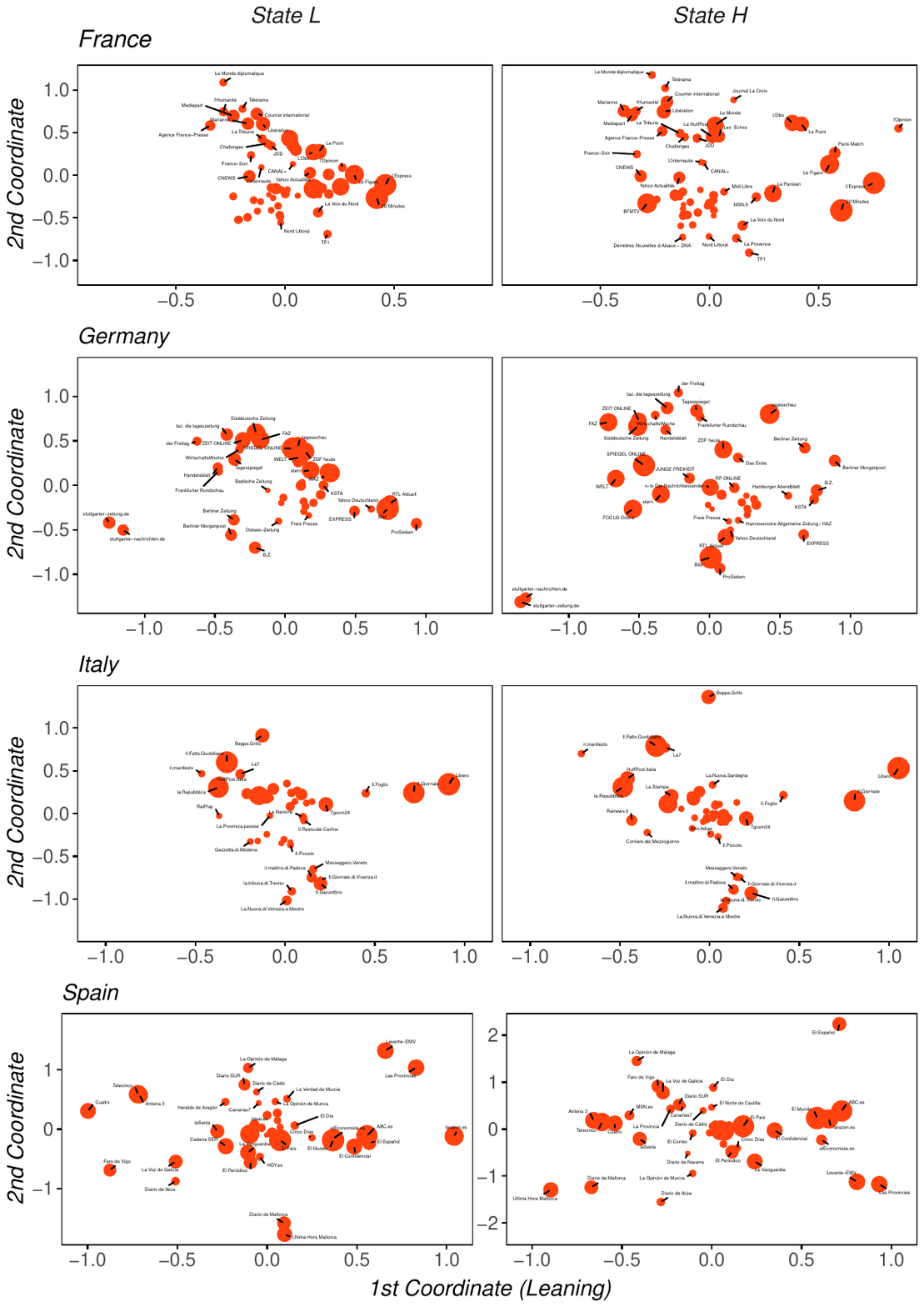}
 \caption{{\textbf{Latent Positions - Dynamic Analysis for $\mathcal{M}_4$:}} Posterior means of latent coordinates of the news outlets for France, Germany, Italy and Spain in State L, and H. The node size is proportional to the posterior mean of the individual effects.}
 \label{fig:dyn_latpos_m4}
 \end{figure}

\begin{figure}[htbp]
  \centering
   \includegraphics[width= 0.9\textwidth]{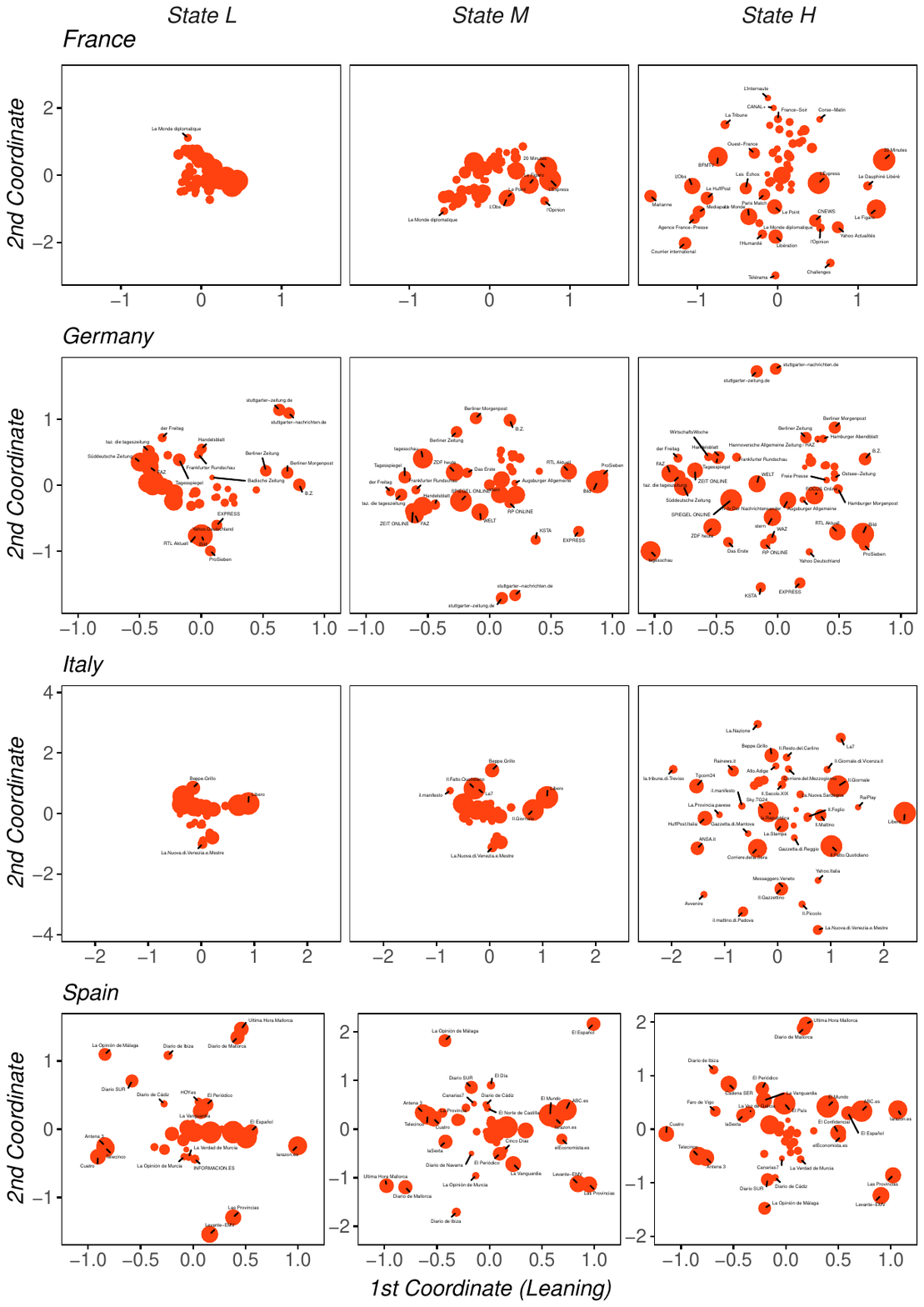}
 \caption{{\textbf{Latent Positions - Dynamic Analysis for $\mathcal{M}_5$:}} Posterior means of latent coordinates of the news outlets for France, Germany, Italy and Spain in State L, M and H. The node size is proportional to the posterior mean of the individual effects.}
 \label{fig:dyn_latpos_m5}
 \end{figure}

\begin{figure}[htbp]
  \centering
   \includegraphics[width= 0.9\textwidth]{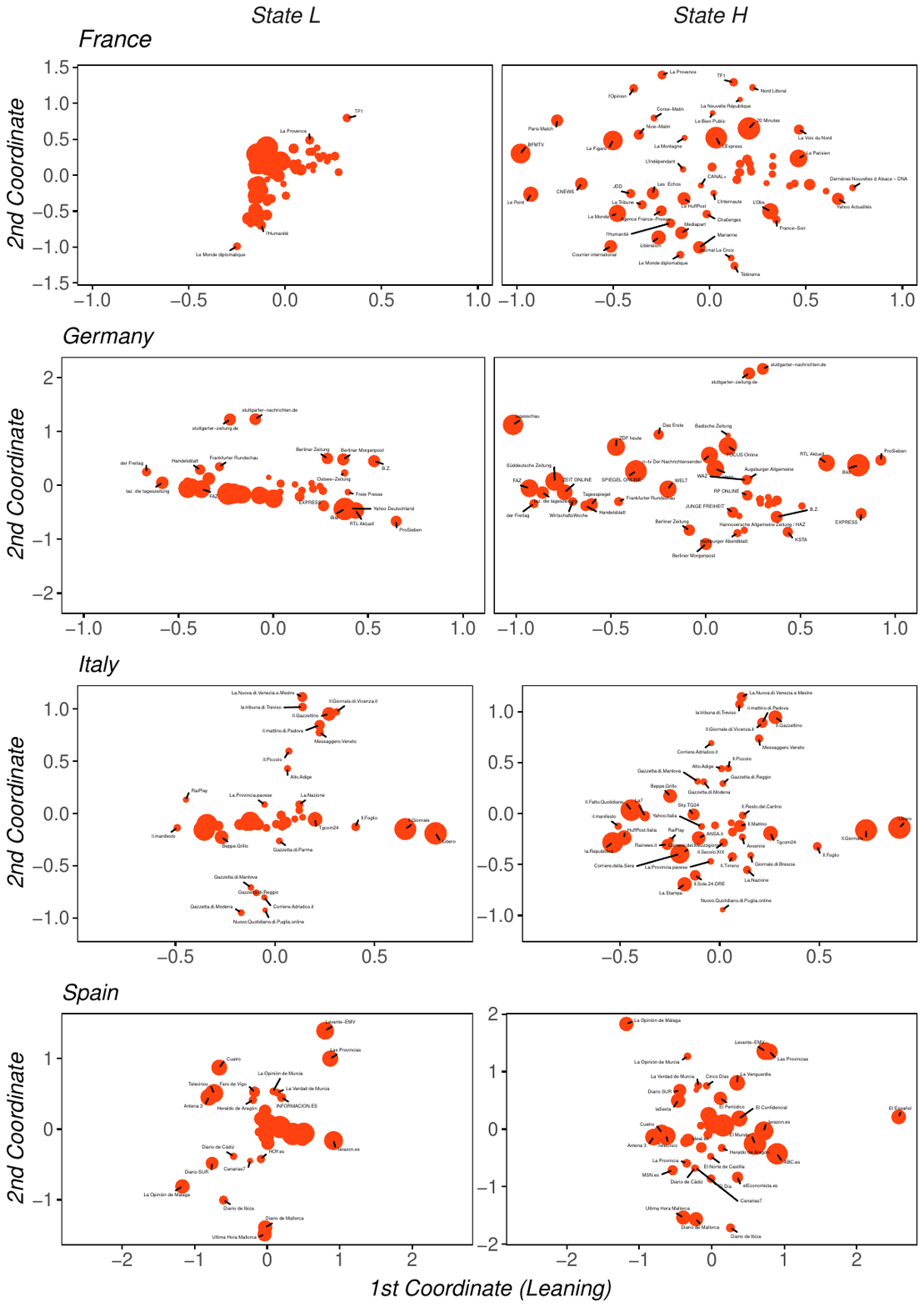}
 \caption{{\textbf{Latent Positions - Dynamic Analysis for $\mathcal{M}_6$:} Posterior means of latent coordinates of the news outlets for France, Germany, Italy and Spain in State of lowest polarization denoted as L, and highest polarization denoted as H. The node size is proportional to the posterior mean of the individual effects.}}
 \label{fig:dyn_latpos_m6}
 \end{figure}

\begin{figure}[!htb]
  \centering
  \begin{tabular}{c}
  \small {MS-LS model with $d = 1$, $K = 2$ ($\mathcal{M}_3$)}\\
   \includegraphics[height=155pt,width= \textwidth]{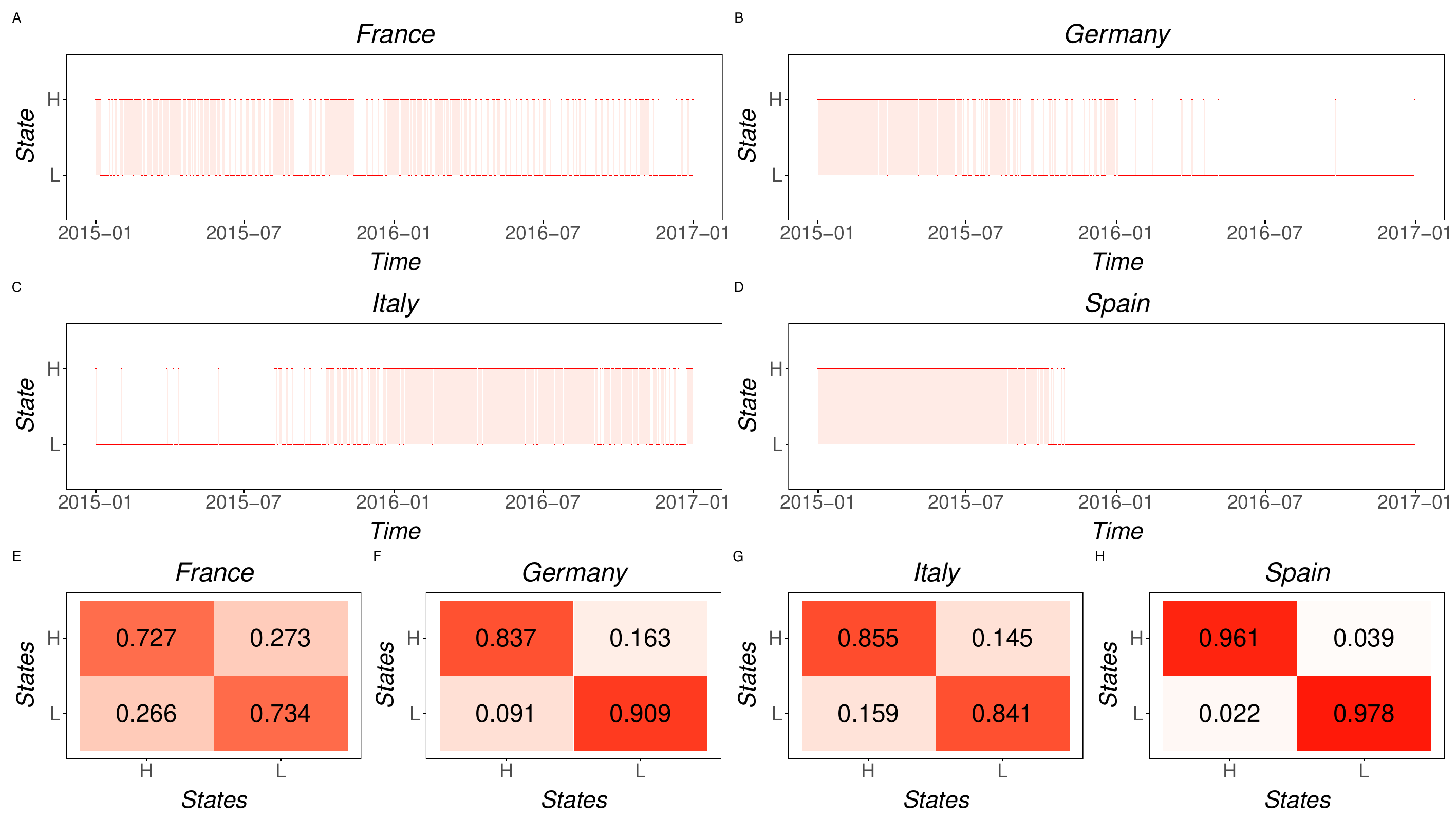}\\
     \small {MS-LS model with $d = 2$, $K = 2$ ($\mathcal{M}_4$)}\\
   \includegraphics[height=155pt,width= \textwidth]{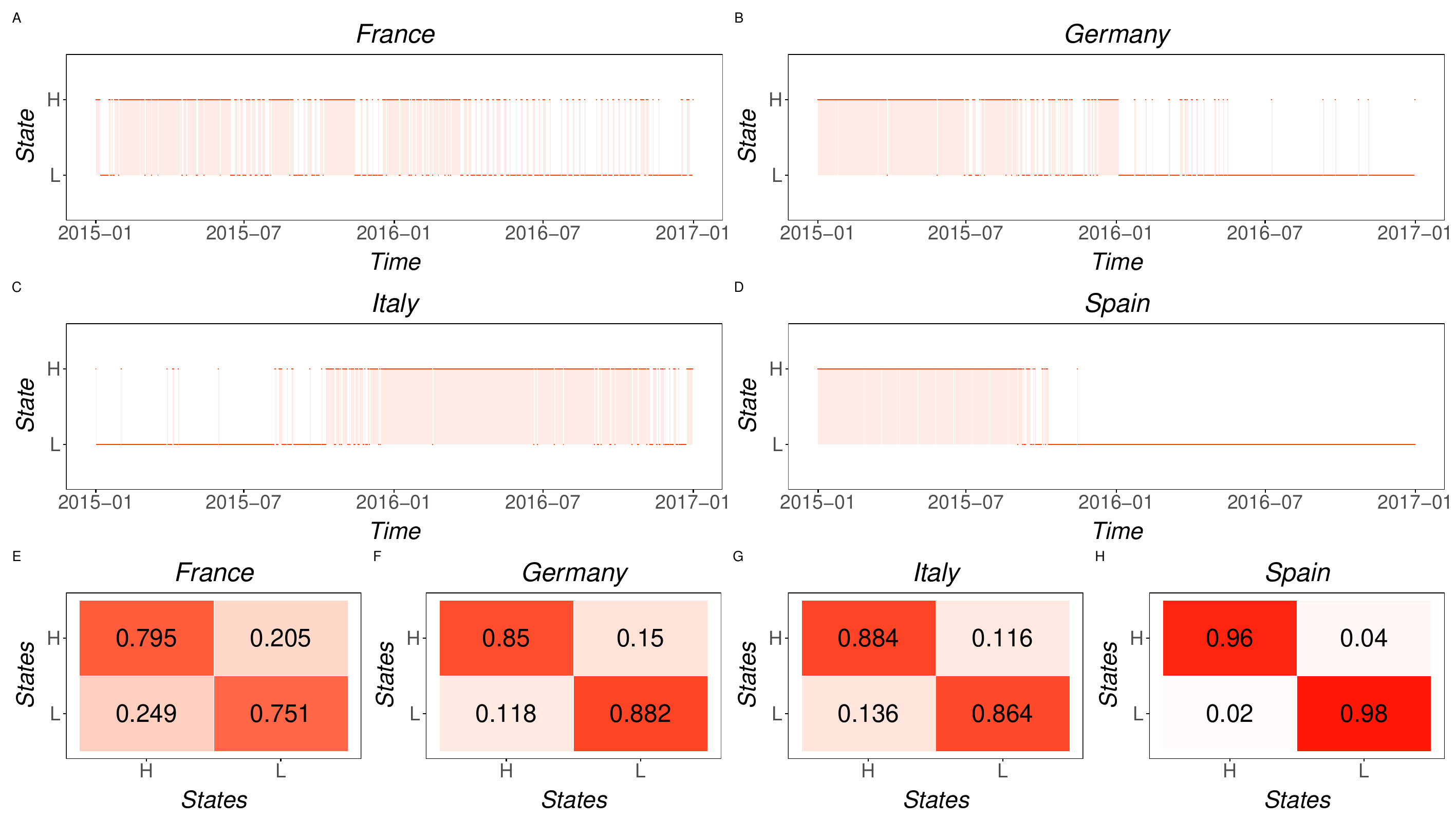}\\
  \small {MS-LS model with $d = 2$, $K = 3$ ($\mathcal{M}_5$)}\\
   \includegraphics[height=155pt,width= \textwidth]{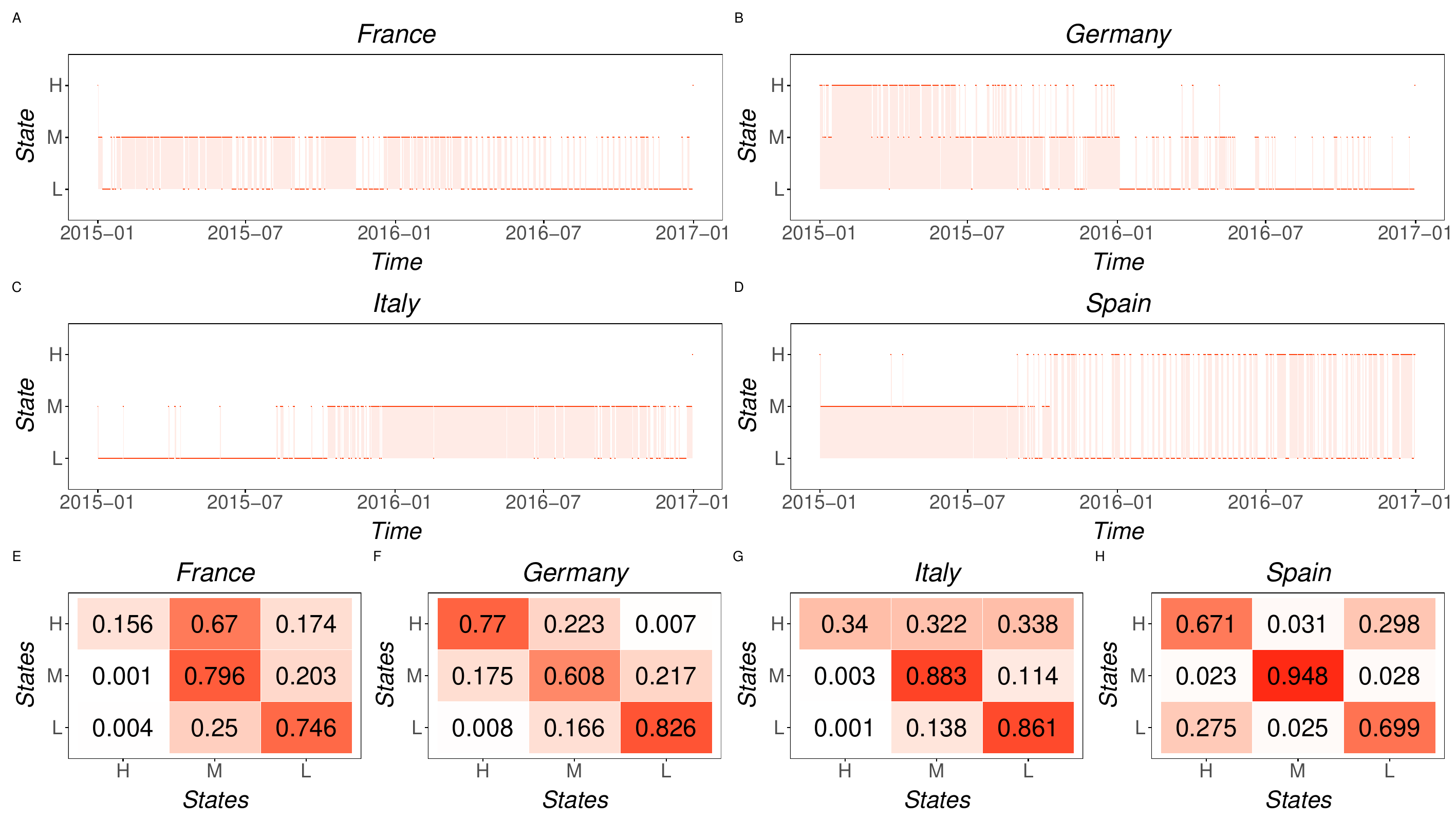}
   \end{tabular}
 \caption{\textbf{Latent States and Transition Probability Comparison:} $\mathcal{M}_j$, $j=3,4,5$  for  France, Germany, Italy and Spain. Posterior mean transition matrices with red shades proportional to the magnitude of the transition probability.}
 \label{fig:latent_states_res_sup}
 \end{figure}

\begin{figure}[!htb]
  \centering
   \includegraphics[width= 0.9\textwidth]{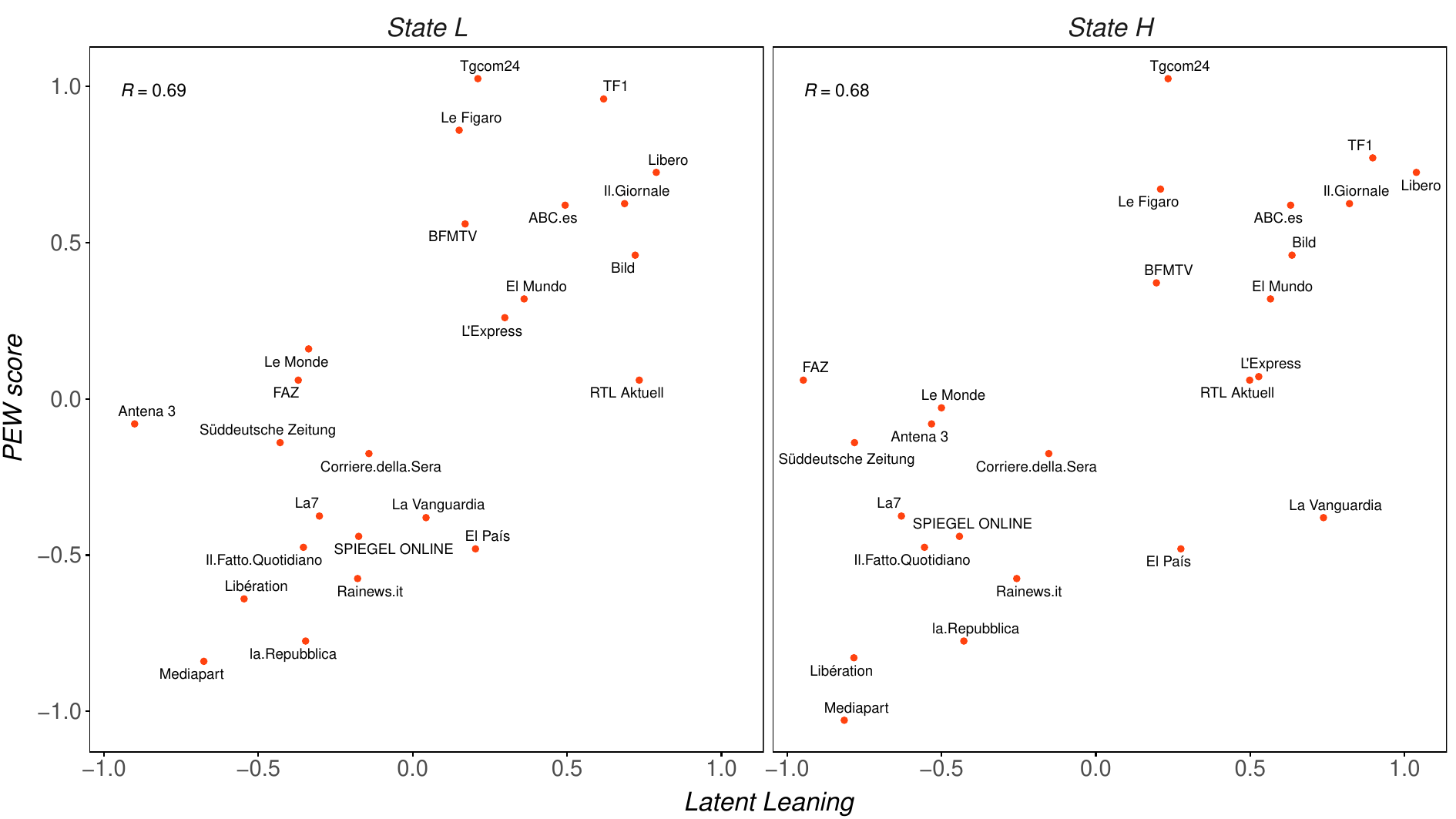}
 \caption{\textbf{PEW Index - Latent Leaning Dynamic Comparison for $\mathcal{M}_3$:} Scatter plot comparing the PEW survey results, available for 25 major national news outlets, with the estimated latent leaning variable in State L and H. The country-specific mean has been subtracted from the PEW Score to improve readability.} \label{fig:rob_dyn_lat_sup}
 \end{figure}

\begin{figure}[!htb]
  \centering
   \includegraphics[width= 0.9\textwidth]{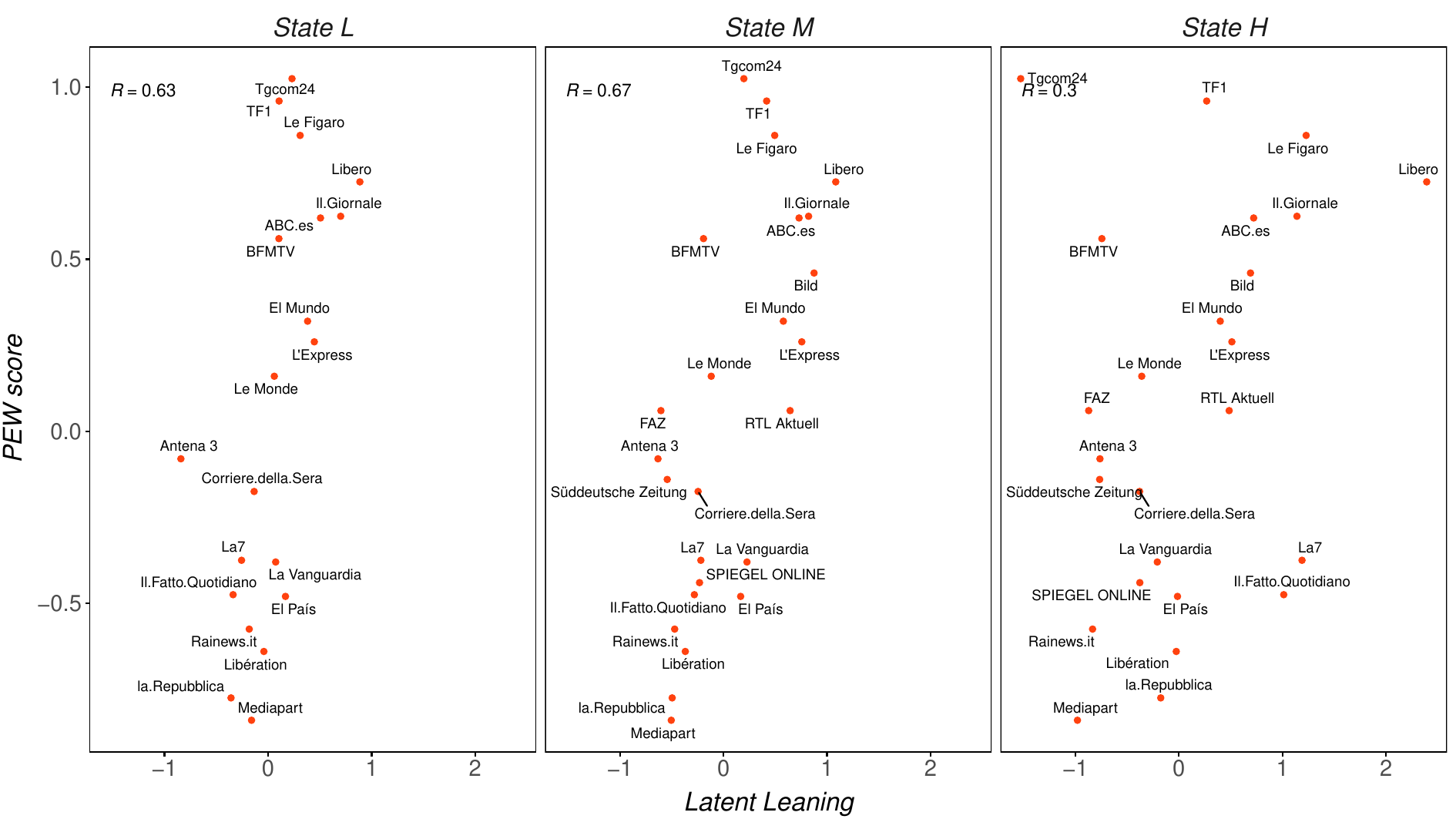}
 \caption{{\textbf{PEW Index - Latent Leaning Dynamic Comparison for $\mathcal{M}_5$:}} Scatter plot comparing the PEW survey results with the estimated latent leaning variable in State L, M and H. The country-specific mean has been subtracted from the PEW Score to improve readability.}
 \label{fig:rob_dyn_lat_m5}
 \end{figure}

\begin{figure}[!htb]
  \centering
  \begin{tabular}{c}
  \small {MS-LS model with $d = 1$, $K = 2$ ($\mathcal{M}_3$)}\\
   \includegraphics[height=135pt,width= \textwidth]{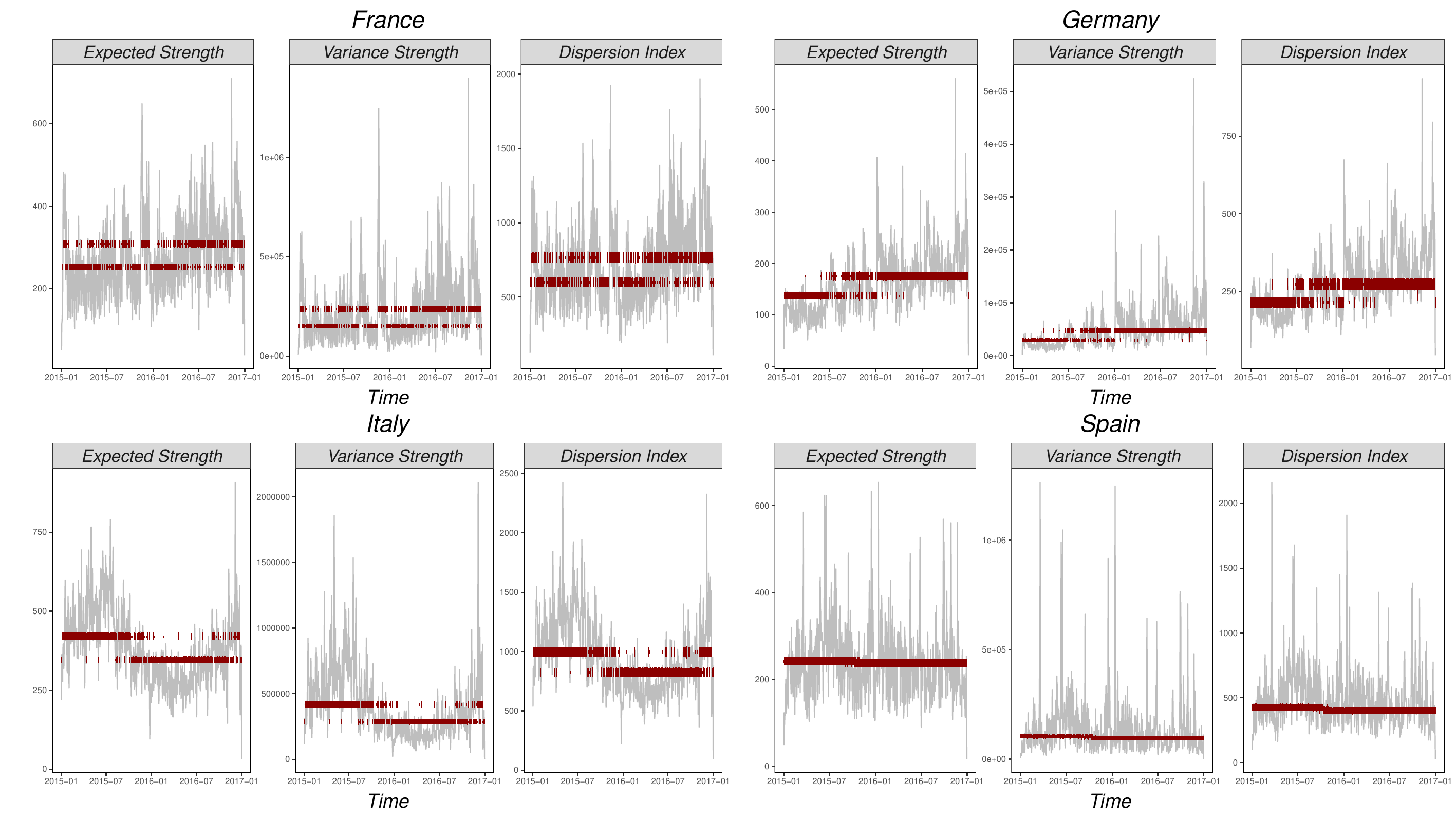}\\
     \small {MS-LS model with $d = 2$, $K = 2$ ($\mathcal{M}_4$)}\\
   \includegraphics[height=135pt,width= \textwidth]{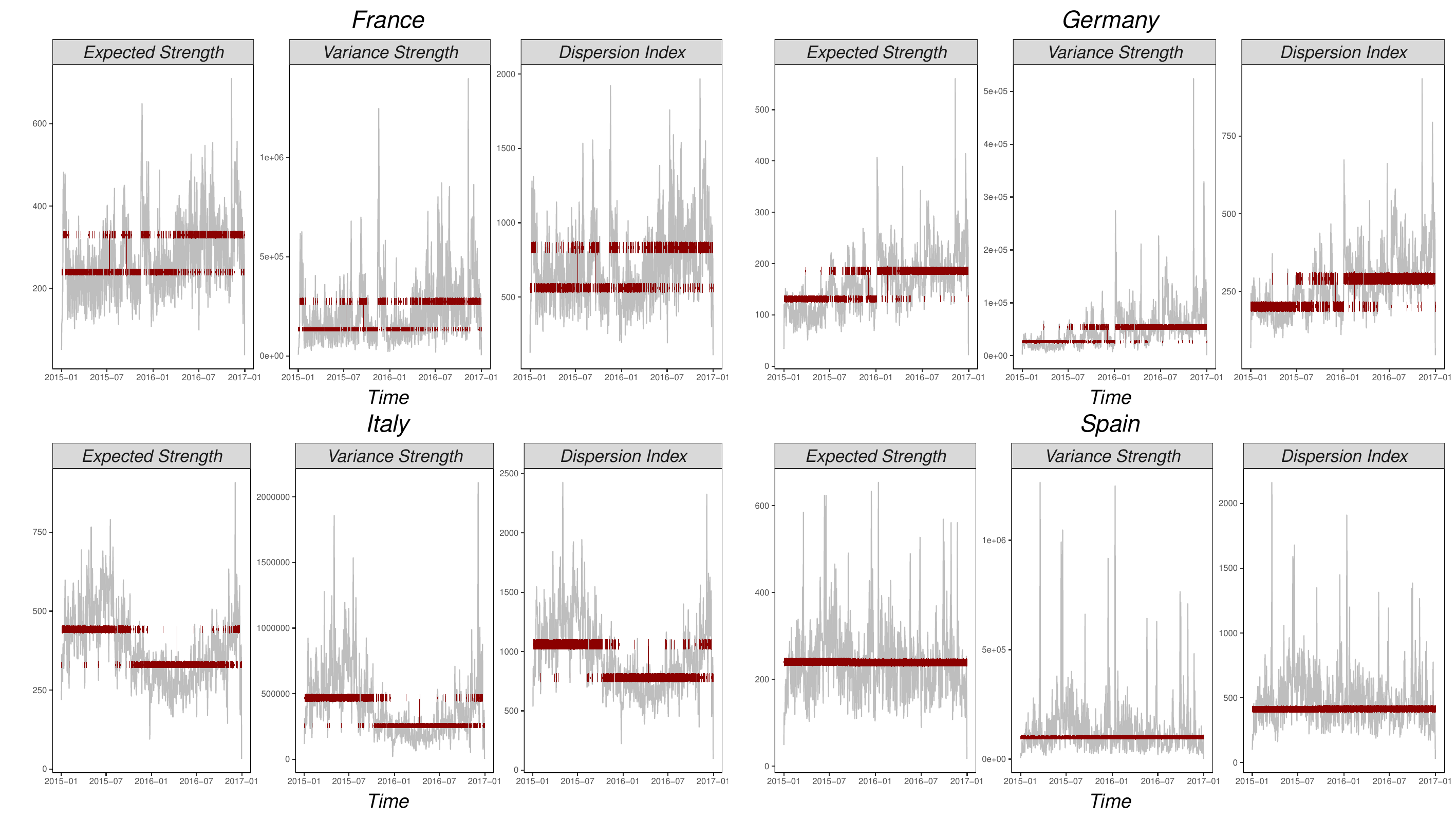}\\
  \small {MS-LS model with $d = 2$, $K = 3$ ($\mathcal{M}_5$)}\\
   \includegraphics[height=135pt,width= \textwidth]{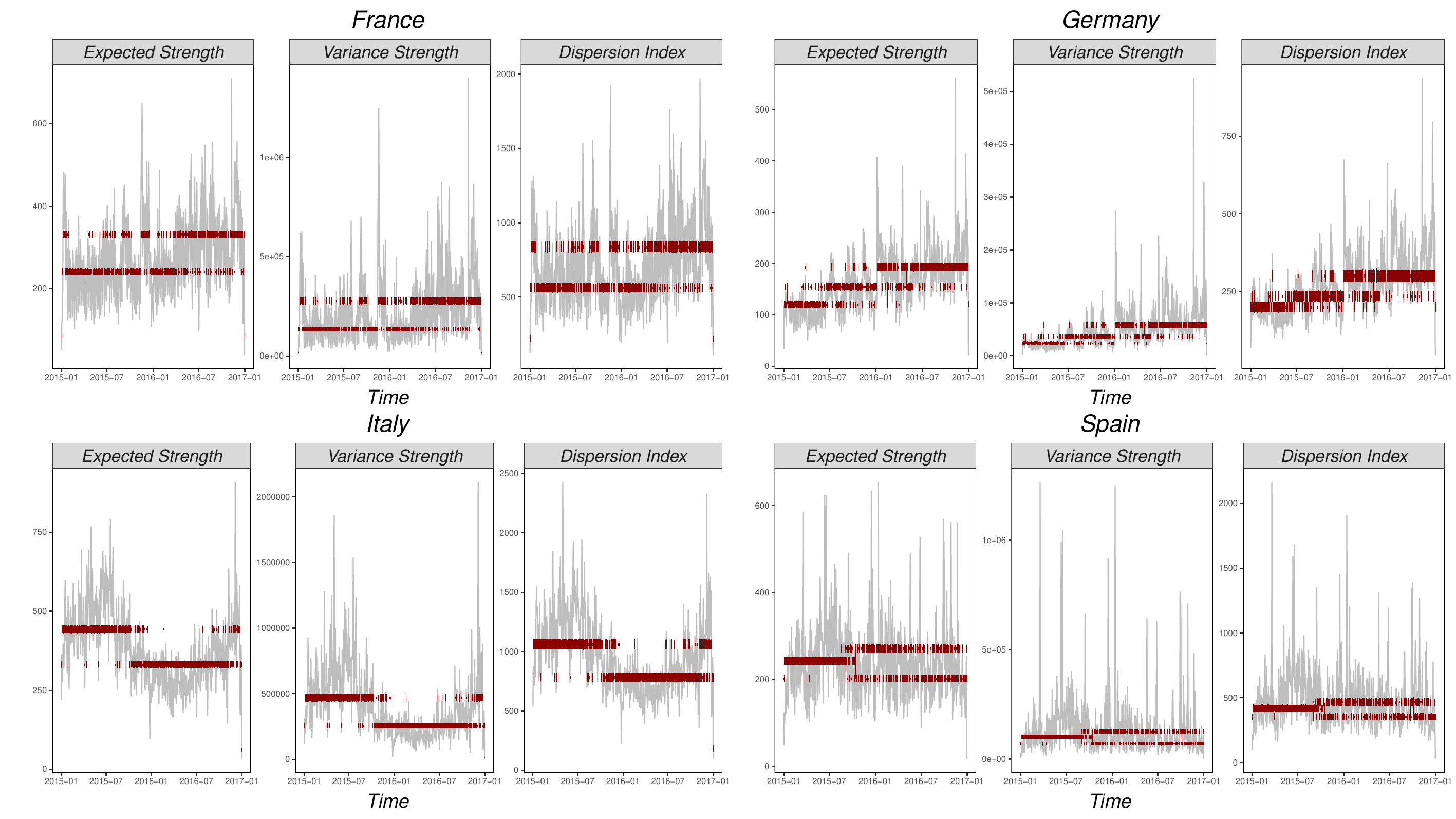}
   \end{tabular}
 \caption{{\textbf{Network Metrics over Time for ${\cal M}_j$ for $j= 3,4,5$:} Boxplots reporting the posterior 
distribution of the average expected strength, variance, and dispersion index in each regime. The underlying daily observations are displayed in the background for information. The presence of the boxplot indicates which regime each observation is assigned to.}}
 \label{fig:network_metrics_ts}
 \end{figure}

This section provides further results for four MS-LS model specifications:  $\mathcal{M}_3$ with $d = 1$ and $K = 2$, $\mathcal{M}_4$ with $d = 2$ and $K = 2$, $\mathcal{M}_5$ with $d = 2$ and $K = 3$, and $\mathcal{M}_6$ with $d = 2$ and $K = 5$. The following results are presented for France, Germany, Italy and Spain:
\begin{enumerate}[label=\roman*)]
\item Observed against predicted network characteristics (Table \ref{tab:net_char2}).
\item Parameter marginal posterior distributions (Figure \ref{fig:postDyn_sup}).
\item Individual effect and latent coordinate posterior means (scatter plots in Figures \ref{fig:dyn_latpos_sup}-\ref{fig:dyn_latpos_m6}).
\item Latent state and transition probability estimates (Figures \ref{fig:latent_states_res_sup}).
\item PEW survey index versus the fitted leaning variable for each state (scatter plots in Figures \ref{fig:rob_dyn_lat_sup} and \ref{fig:rob_dyn_lat_m5}). 
\item Network Metrics over Time for ${\cal M}_j$ for $j= 3,4,5$ (Figure \ref{fig:network_metrics_ts}).
\end{enumerate}

 \begin{table}[h!]
\caption{\textbf{Network Posterior Predictive Check:} The table compares the observed network metrics in terms of expected strength, strength's standard deviation, and dispersion index (first panel) against the posterior predictive network characteristics for the standard Poisson random graph model (second panel), the Poisson random graph model with individual effects and observed leaning distances as a covariate (third panel), and the dynamic latent space network models $\mathcal{M}_3$, $\mathcal{M}_4$, $\mathcal{M}_5$  and $\mathcal{M}_6$  (bottom panels). The observed and posterior predictive network metrics are averaged over time. The 95\% credible intervals are in parenthesis.} 
 \label{tab:net_char2}
 
\centering

\resizebox{0.8\textwidth}{!}{ 
\begin{tabular}{lcccc} 
\hline\hline
\multicolumn{1}{c}{}      & \multicolumn{4}{c}{Average Empirical Network Metrics}                                                                                                                                                                                                                                                      \\ 
\hline
                          & France                                                                   & Germany                                                                  & Italy                                                                    & Spain                                                                     \\ 
\hline
Expected Strength         & 280.83                                                                   & 161.82                                                                   & 380.98                                                                   & 239.32                                                                    \\
S.D. Strength             & 450.53                                                                   & 209.39                                                                   & 600.58                                                                   & 340.76                                                                    \\
Dispersion Index          & 814.20                                                                   & 306.11                                                                   & 1037.82                                                                  & 562.56                                                                    \\ 
\hline
\multicolumn{1}{c}{}      & \multicolumn{4}{c}{Random Graph with  $\lambda_{ijt} = \exp\{\alpha\}$ ($\text{RG}_1$)}                                                                                                                                                                                                                                                  \\ 
\hline
\vcell{Expected Strength} & \vcell{\begin{tabular}[b]{@{}c@{}}280.82\\(280.55, 281.10)\end{tabular}} & \vcell{\begin{tabular}[b]{@{}c@{}}161.82\\(161.56, 162.04)\end{tabular}} & \vcell{\begin{tabular}[b]{@{}c@{}}380.99\\(380.62, 381.37)\end{tabular}} & \vcell{\begin{tabular}[b]{@{}c@{}}239.30\\(239.00, 239.59)\end{tabular}}  \\[-\rowheight]
\printcelltop             & \printcellmiddle                                                         & \printcellmiddle                                                         & \printcellmiddle                                                         & \printcellmiddle                                                          \\
\vcell{S.D. Strength}     & \vcell{\begin{tabular}[b]{@{}c@{}}16.53\\(16.43, 16.62)\end{tabular}}    & \vcell{\begin{tabular}[b]{@{}c@{}}12.49\\(12.40, 12.57)\end{tabular}}    & \vcell{\begin{tabular}[b]{@{}c@{}}19.17\\(19.04, 19.29)\end{tabular}}    & \vcell{\begin{tabular}[b]{@{}c@{}}15.18\\(15.07, 15.28)\end{tabular}}     \\[-\rowheight]
\printcelltop             & \printcellmiddle                                                         & \printcellmiddle                                                         & \printcellmiddle                                                         & \printcellmiddle                                                          \\
\vcell{Dispersion Index}  & \vcell{\begin{tabular}[b]{@{}c@{}}0.98\\(0.97, 0.99)\end{tabular}}       & \vcell{\begin{tabular}[b]{@{}c@{}}0.98\\(0.96, 0.99)\end{tabular}}       & \vcell{\begin{tabular}[b]{@{}c@{}}0.98\\(0.96, 0.99)\end{tabular}}       & \vcell{\begin{tabular}[b]{@{}c@{}}0.98\\(0.96, 0.99)\end{tabular}}        \\[-\rowheight]
\printcelltop             & \printcellmiddle                                                         & \printcellmiddle                                                         & \printcellmiddle                                                         & \printcellmiddle                                                          \\ 
\hline
\multicolumn{1}{c}{}      & \multicolumn{4}{c}{Random Graph with {$\lambda_{ijt} = \exp\{\alpha_i + \alpha_j - \beta||L_{it}-L_{jt}||^2\}$ ($\text{RG}_2$)} }                                                                                                                                                                                                   \\ 
\hline
\vcell{Expected Strength} & \vcell{\begin{tabular}[b]{@{}c@{}}276.64\\(276.36, 276.92)\end{tabular}} & \vcell{\begin{tabular}[b]{@{}c@{}}162.73\\(162.47, 163.00)\end{tabular}} & \vcell{\begin{tabular}[b]{@{}c@{}}375.68\\(375.26, 376.09)\end{tabular}} & \vcell{\begin{tabular}[b]{@{}c@{}}239.50\\(239.19, 239.80)\end{tabular}}  \\[-\rowheight]
\printcelltop             & \printcellmiddle                                                         & \printcellmiddle                                                         & \printcellmiddle                                                         & \printcellmiddle                                                          \\
\vcell{S.D. Strength}     & \vcell{\begin{tabular}[b]{@{}c@{}}431.33\\(430.84, 431.85)\end{tabular}} & \vcell{\begin{tabular}[b]{@{}c@{}}201.97\\(201.60, 202.30)\end{tabular}} & \vcell{\begin{tabular}[b]{@{}c@{}}577.62\\(576.89, 578.30)\end{tabular}} & \vcell{\begin{tabular}[b]{@{}c@{}}312.12\\(311.67, 312.54)\end{tabular}}  \\[-\rowheight]
\printcelltop             & \printcellmiddle                                                         & \printcellmiddle                                                         & \printcellmiddle                                                         & \printcellmiddle                                                          \\
\vcell{Dispersion Index}  & \vcell{\begin{tabular}[b]{@{}c@{}}672.57\\(671.57, 673.75)\end{tabular}} & \vcell{\begin{tabular}[b]{@{}c@{}}250.68\\(250.10, 251.23)\end{tabular}} & \vcell{\begin{tabular}[b]{@{}c@{}}888.15\\(886.72, 889.50)\end{tabular}} & \vcell{\begin{tabular}[b]{@{}c@{}}406.79\\(405.96, 407.55)\end{tabular}}  \\[-\rowheight]
\printcelltop             & \printcellmiddle                                                         & \printcellmiddle                                                         & \printcellmiddle                                                         & \printcellmiddle                                                          \\ 
 \hline
\multicolumn{1}{c}{}      & \multicolumn{4}{c}{Dynamic MS-LS Model with $d = 1$, $K = 2$ ($\mathcal{M}_{3}$)}                                                                                                                                                                                                                                                                     \\ 
\hline
\vcell{Expected Strength} & \vcell{\begin{tabular}[b]{@{}c@{}}280.84\\(280.60, 281.09)\end{tabular}} & \vcell{\begin{tabular}[b]{@{}c@{}}161.81\\(161.57, 162.03)\end{tabular}} & \vcell{\begin{tabular}[b]{@{}c@{}}380.99\\(380.64, 381.36)\end{tabular}} & \vcell{\begin{tabular}[b]{@{}c@{}}239.32\\(239.00, 239.64)\end{tabular}}  \\[-\rowheight]
\printcelltop             & \printcellmiddle                                                         & \printcellmiddle                                                         & \printcellmiddle                                                         & \printcellmiddle                                                          \\
\vcell{S.D. Strength}     & \vcell{\begin{tabular}[b]{@{}c@{}}437.96\\(437.52, 438.44)\end{tabular}} & \vcell{\begin{tabular}[b]{@{}c@{}}201.90\\(201.59, 202.24)\end{tabular}} & \vcell{\begin{tabular}[b]{@{}c@{}}587.77\\(587.12, 588.43)\end{tabular}} & \vcell{\begin{tabular}[b]{@{}c@{}}313.64\\(313.16, 314.13)\end{tabular}}  \\[-\rowheight]
\printcelltop             & \printcellmiddle                                                         & \printcellmiddle                                                         & \printcellmiddle                                                         & \printcellmiddle                                                          \\
\vcell{Dispersion Index}  & \vcell{\begin{tabular}[b]{@{}c@{}}683.13\\(682.10, 684.22)\end{tabular}} & \vcell{\begin{tabular}[b]{@{}c@{}}251.94\\(251.33, 252.54)\end{tabular}} & \vcell{\begin{tabular}[b]{@{}c@{}}906.83\\(905.48, 908.02)\end{tabular}} & \vcell{\begin{tabular}[b]{@{}c@{}}411.12\\(410.20, 412.04)\end{tabular}}  \\[-\rowheight]
\printcelltop             & \printcellmiddle                                                         & \printcellmiddle                                                         & \printcellmiddle                                                         & \printcellmiddle                                                          \\\hline
\multicolumn{1}{c}{}      & \multicolumn{4}{c}{Dynamic MS-LS Model with $d = 2$, $K = 2$ ($\mathcal{M}_{4}$)}                                                  \\ 
\hline
\vcell{{Expected Strength}} & \vcell{\begin{tabular}[b]{@{}c@{}}280.83\\(280.49 , 281.16)\end{tabular}} & \vcell{\begin{tabular}[b]{@{}c@{}}161.82\\(161.51, 162.15)\end{tabular}} & \vcell{\begin{tabular}[b]{@{}c@{}}380.98\\(380.57, 381.41)\end{tabular}} & \vcell{\begin{tabular}[b]{@{}c@{}}239.32\\(238.97, 239.68)\end{tabular}}  \\[-\rowheight]
\printcelltop             & \printcellmiddle                                                         & \printcellmiddle                                                         & \printcellmiddle                                                         & \printcellmiddle                                                          \\
\vcell{{S.D. Strength}}     & \vcell{\begin{tabular}[b]{@{}c@{}}438.13\\(437.54, 438.76)\end{tabular}} & \vcell{\begin{tabular}[b]{@{}c@{}}201.83\\(201.39, 202.32)\end{tabular}} & \vcell{\begin{tabular}[b]{@{}c@{}} 587.83\\(587.11, 588.6)\end{tabular}} & \vcell{\begin{tabular}[b]{@{}c@{}} 314.69\\(314.16, 315.28)\end{tabular}}  \\[-\rowheight]
\printcelltop             & \printcellmiddle                                                         & \printcellmiddle                                                         & \printcellmiddle                                                         & \printcellmiddle                                                          \\
\vcell{{Dispersion Index}}  & \vcell{\begin{tabular}[b]{@{}c@{}}683.81\\(682.54, 685.19)\end{tabular}} & \vcell{\begin{tabular}[b]{@{}c@{}}251.76\\(250.99, 252.59)\end{tabular}} & \vcell{\begin{tabular}[b]{@{}c@{}}907.04\\(905.53 , 908.61)\end{tabular}} & \vcell{\begin{tabular}[b]{@{}c@{}}413.85\\(412.84, 414.91)\end{tabular}}  \\[-\rowheight]
\printcelltop             & \printcellmiddle                                                         & \printcellmiddle                                                         & \printcellmiddle                                                         & \printcellmiddle                                                          \\
\hline
\multicolumn{1}{c}{}      & \multicolumn{4}{c}{Dynamic MS-LS Model with $d = 2$, $K = 3$ ($\mathcal{M}_{5}$)}                                                                                                                                                                                                                                                                             \\ 
\hline
\vcell{{Expected Strength}} & \vcell{\begin{tabular}[b]{@{}c@{}}280.83\\(280.51 , 281.14)\end{tabular}} & \vcell{\begin{tabular}[b]{@{}c@{}}161.82\\(161.54, 162.1)\end{tabular}} & \vcell{\begin{tabular}[b]{@{}c@{}}380.97\\(380.54, 381.4)\end{tabular}} & \vcell{\begin{tabular}[b]{@{}c@{}}239.30\\(238.92, 239.64)\end{tabular}}  \\[-\rowheight]
\printcelltop             & \printcellmiddle                                                         & \printcellmiddle                                                         & \printcellmiddle                                                         & \printcellmiddle                                                          \\
\vcell{{S.D. Strength}}     & \vcell{\begin{tabular}[b]{@{}c@{}}438.14\\(437.56, 438.74)\end{tabular}} & \vcell{\begin{tabular}[b]{@{}c@{}}202.27\\(201.82, 202.71)\end{tabular}} & \vcell{\begin{tabular}[b]{@{}c@{}}587.82\\(587.07, 588.54)\end{tabular}} & \vcell{\begin{tabular}[b]{@{}c@{}} 314.93\\(314.36, 315.49)\end{tabular}}  \\[-\rowheight]
\printcelltop             & \printcellmiddle                                                         & \printcellmiddle                                                         & \printcellmiddle                                                         & \printcellmiddle                                                          \\
\vcell{{Dispersion Index}}  & \vcell{\begin{tabular}[b]{@{}c@{}}683.88\\(682.58, 685.18)\end{tabular}} & \vcell{\begin{tabular}[b]{@{}c@{}}252.92\\(252.19, 253.66)\end{tabular}} & \vcell{\begin{tabular}[b]{@{}c@{}}907.03\\(905.52 , 908.56)\end{tabular}} & \vcell{\begin{tabular}[b]{@{}c@{}}414.5\\(413.49, 415.54)\end{tabular}} \\[-\rowheight]
\printcelltop             & \printcellmiddle                                                         & \printcellmiddle                                                         & \printcellmiddle                                                         & \printcellmiddle                                                          \\
\hline
\multicolumn{1}{c}{}      & \multicolumn{4}{c}{Dynamic MS-LS Model with $d = 2$, $K = 5$ ($\mathcal{M}_{6}$)}                                                                                \\ 
\hline
\vcell{{Expected Strength}} & \vcell{\begin{tabular}[b]{@{}c@{}}280.84\\(280.52 , 281.13)\end{tabular}} & \vcell{\begin{tabular}[b]{@{}c@{}}161.82\\(161.51, 162.11)\end{tabular}} & \vcell{\begin{tabular}[b]{@{}c@{}}380.98\\(380.58, 381.41)\end{tabular}} & \vcell{\begin{tabular}[b]{@{}c@{}}239.32\\(238.98, 239.67)\end{tabular}}  \\[-\rowheight]
\printcelltop             & \printcellmiddle                                                         & \printcellmiddle                                                         & \printcellmiddle                                                         & \printcellmiddle                                                          \\
\vcell{{S.D. Strength}}     & \vcell{\begin{tabular}[b]{@{}c@{}}439.07\\(438.49 , 439.66)\end{tabular}} & \vcell{\begin{tabular}[b]{@{}c@{}}202.61\\(202.16, 203.04)\end{tabular}} & \vcell{\begin{tabular}[b]{@{}c@{}}588.78\\(588.03, 589.53)\end{tabular}} & \vcell{\begin{tabular}[b]{@{}c@{}} 315.82\\(315.27, 316.39)\end{tabular}}  \\[-\rowheight]
\printcelltop             & \printcellmiddle                                                         & \printcellmiddle                                                         & \printcellmiddle                                                         & \printcellmiddle                                                          \\
\vcell{{Dispersion Index}}  & \vcell{\begin{tabular}[b]{@{}c@{}}687.98\\(686.72, 689.21)\end{tabular}} & \vcell{\begin{tabular}[b]{@{}c@{}}253.79\\(253.04, 254.5)\end{tabular}} & \vcell{\begin{tabular}[b]{@{}c@{}}910.24\\(908.75 , 911.72)\end{tabular}} & \vcell{\begin{tabular}[b]{@{}c@{}}416.87\\(415.82, 417.93)\end{tabular}} \\[-\rowheight]
\printcelltop             & \printcellmiddle                                                         & \printcellmiddle                                                         & \printcellmiddle                                                         & \printcellmiddle                                                          \\
\hline\hline
\end{tabular}}
\end{table}

\clearpage

\renewcommand\thefigure{J.\arabic{figure}}
\setcounter{figure}{0}
\renewcommand\theequation{J.\arabic{equation}}
\setcounter{equation}{0}
\renewcommand\thetable{J.\arabic{table}}
\setcounter{table}{0}

\section{Controlling for Exposure}
\label{G:Control}

Here, we discuss the inclusion of an additional variable, $TotCom_{t}$,
accounting for the exposure to the total number of comments at time $t$ in a MS-LS model with $d =1$ and $K = 2$. This is to compare polarisation across periods with different engagement levels. To maintain parameter interpretation, we add to the log--intensity the de--meaned logarithm of the total number of comments, $\log TotCom_{t}$:

\begin{equation*}
\log \lambda_{ijt} = \alpha_{i} + \alpha_{j} +\delta \log TotCom_{t} - \beta \vert\vert x_{it}- x_{jt}\vert\vert^2.
\end{equation*}

\begin{figure}[!htb]
  \centering
   \includegraphics[width= 0.9\textwidth]{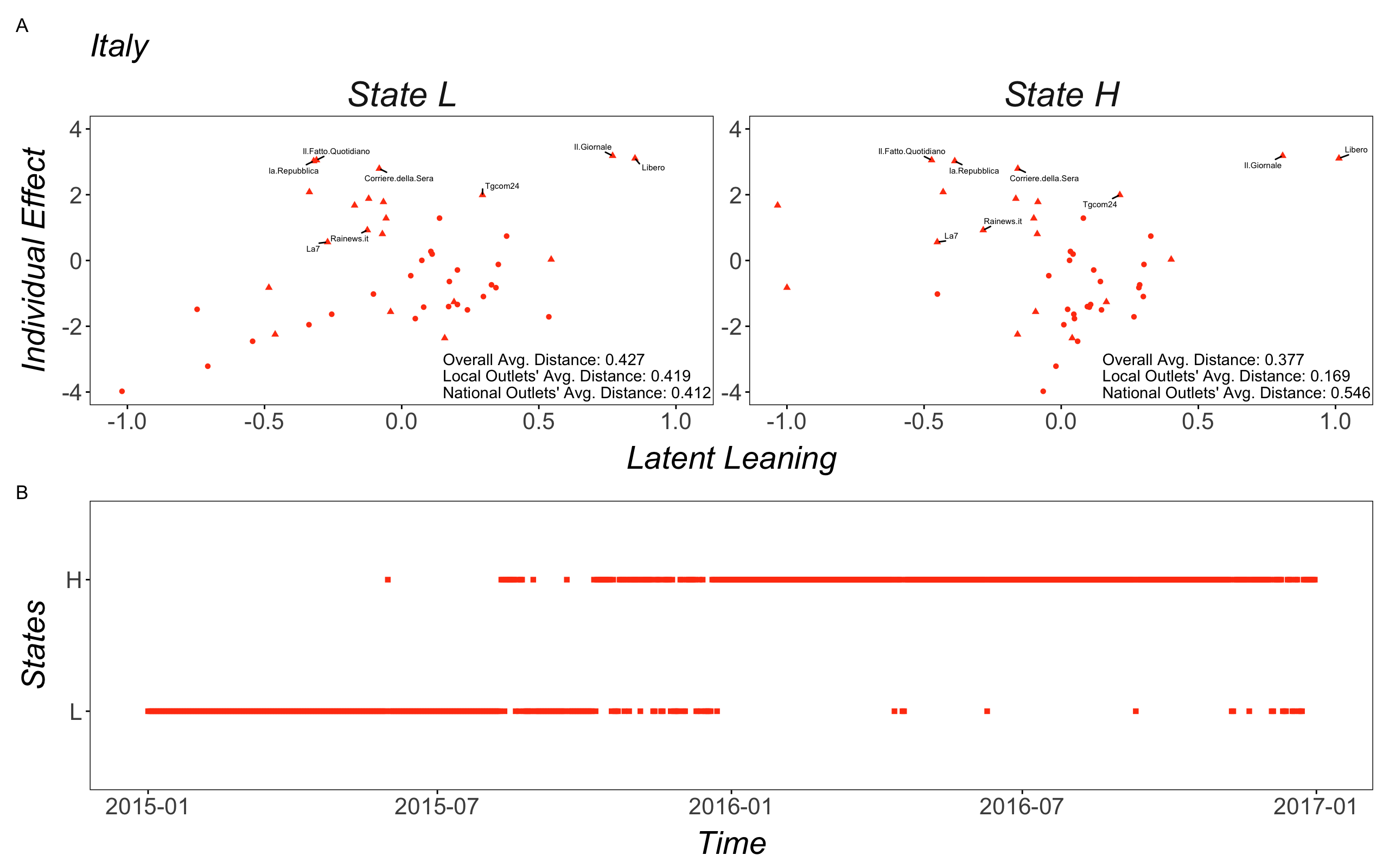}
 \caption{\textbf{Posterior Mean Latent Space and States with Exposure Control:} Panel A reports the latent space of Italian news outlets in the two states. State identification is achieved considering the average distance computed across national news outlets (triangles). Panel B reports the latent states through time. The states are coherent with those found in Sec. \ref{sec:application}. }
 
 \label{fig:control1}
 \end{figure}

\begin{figure}[!htb]
  \centering
   \includegraphics[width= 0.9\textwidth]{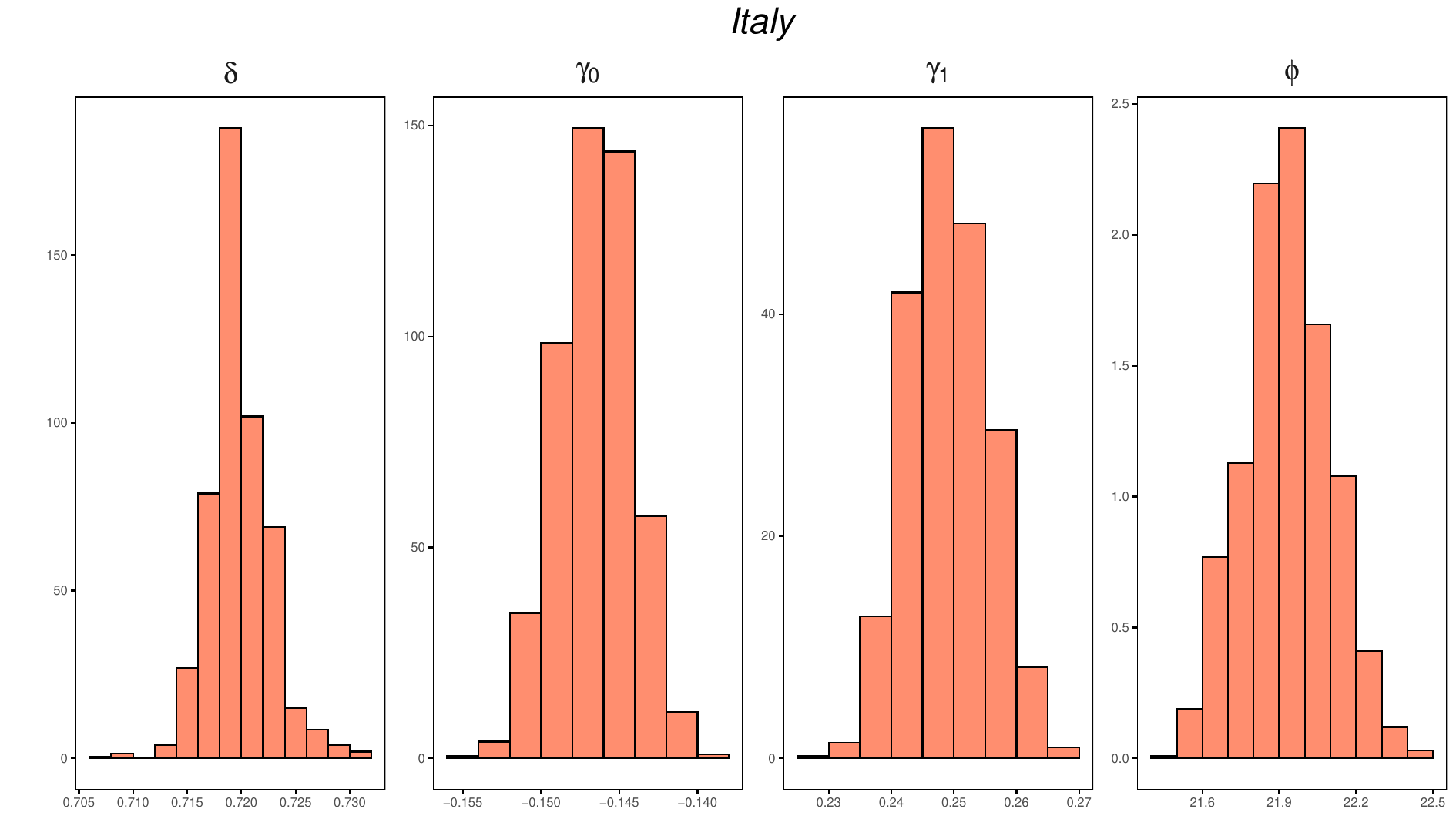}
 \caption{\textbf{Marginal Posterior Distributions with Exposure Control:}  Histograms for the MCMC draws for the parameters $\delta$, $\gamma_{0}$, $\gamma_{1}$ and $\phi$ for Italy. }
 
 \label{fig:control2}
 \end{figure}

We run our modified LS model for the Italian data. Figure \ref{fig:control1} reports the posterior estimates of the latent space and states through time when we include this control. Panel A reports the latent space of Italian news outlets in the two states. State identification is not trivial anymore. We notice heterogeneous behavior between local and national outlets. We achieve state identification by considering the average distance computed across national news outlets (triangles), as they involve a larger number of commenters. Panel B reports the latent states through time. The states are coherent with those found in Sec. \ref{sec:application}. Figure \ref{fig:control2} reports the marginal posteriors draws for the parameters $\delta$, $\gamma_{0}$, $\gamma_{1}$ and $\phi$. The sign of the parameter $\delta$ is as expected. Higher comments overall in the network implies also a higher number of comments in common between the pairs of pages.

 \newpage

\renewcommand\thefigure{K.\arabic{figure}}
\setcounter{figure}{0}
\renewcommand\theequation{K.\arabic{equation}}
\setcounter{equation}{0}
\renewcommand\thetable{K.\arabic{table}}
\setcounter{table}{0}

\section{Alternative dynamic Specifications }
\label{alternative_specs}

Additional dynamic features can be incorporated into the model if, in addition to inference on polarization features, one wants to track the data from day to day. 
While additional dynamic features can provide a better fitting of the daily data features, they can seriously hamper the inference of polarization regimes since the latent space will use the same signal for inference. This issue can be more severe for large $K$. For example, one can assume $$\log \lambda_{ijt} = \alpha_{i} + \alpha_{j} + f_t  - \beta||\mathbf{x}_{it}- \mathbf{x}_{jt}||^2,$$ where the latent dynamic factor $f_t = f_{t-1} + \varepsilon_t$, $\varepsilon_t \sim \mathcal{N}(0, \sigma_\epsilon^2)$ captures a common dynamic component.

Alternatively, one could opt for a less flexible but more parsimonious and computationally cheaper autoregressive structure $$\log \lambda_{ijt} =  \delta \log(\Bar{y}_{t-1}) + (1-\delta)(\alpha_{i} + \alpha_{j}   - \beta||\mathbf{x}_{it}- \mathbf{x}_{jt}||^2),$$ where $\Bar{y}_{t-1}=\sum_{ij}2y_{ijt}/(n(n-1))$
is the average strength at time $t-1$ and $\delta$ is an autoregressive coefficient with prior distribution $\delta \sim \mathcal{B}(a_\delta, b_\delta)$. The posterior approximation is obtained by modifying the MCMC sampler presented in Section 3 to account for the new dynamic features.

\begin{figure}[h!]
    \centering
    \begin{tabular}{c}
    {\small (a)  MS-LS. In-sample fitting (blue)}\\
    \includegraphics[width=0.7\linewidth]{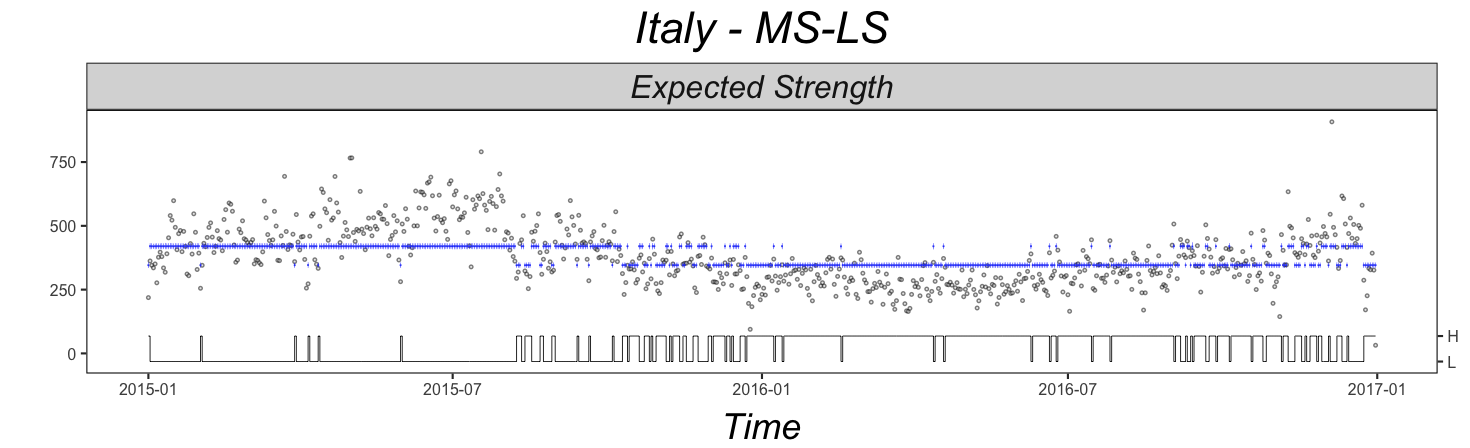}\\    
    {\small (b) Factor MS-LS. In-sample fitting (blue);  Factor and AR fitting contribution (green).}\\
    \includegraphics[width=0.7\linewidth]{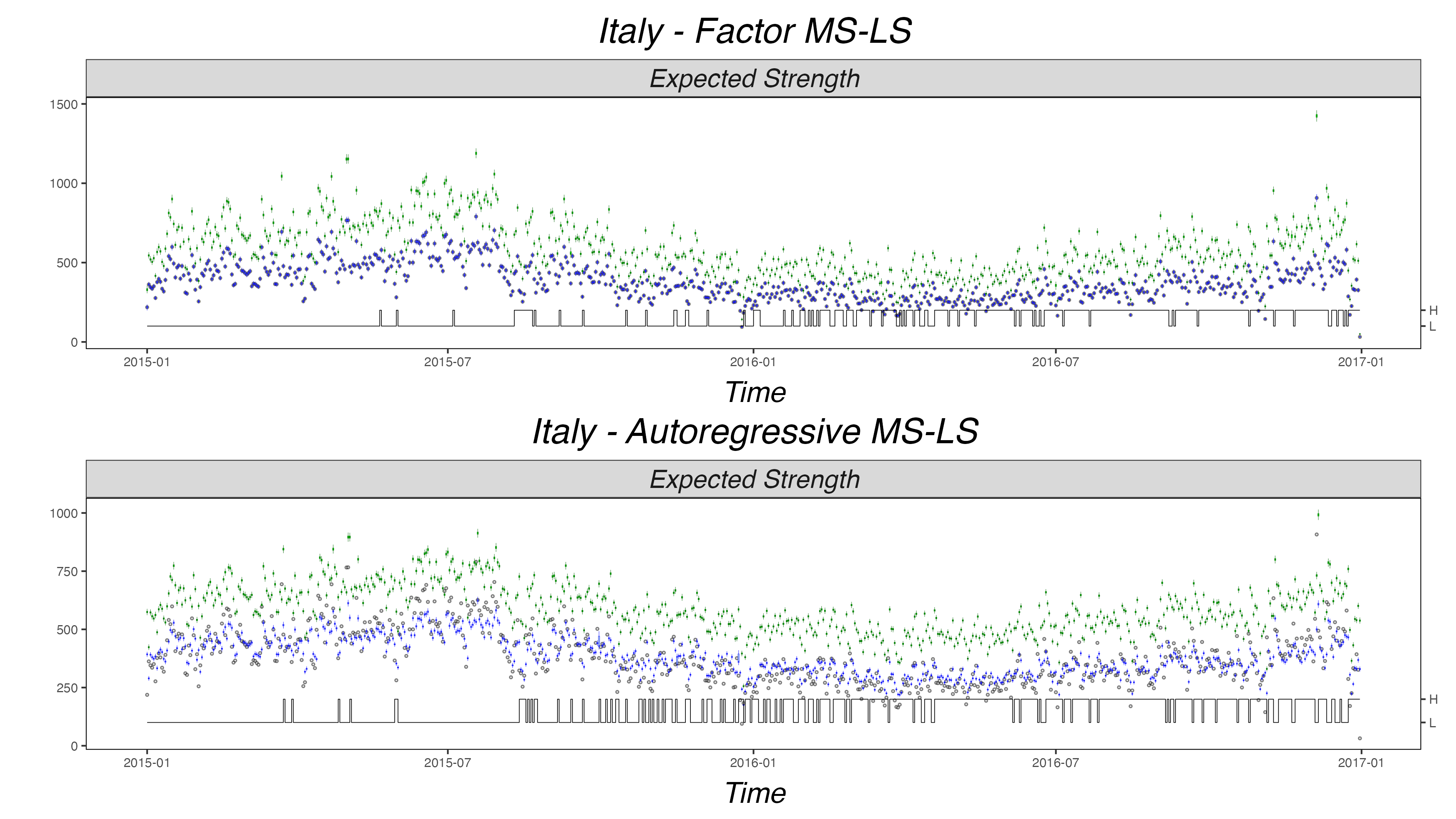}\\
    {\small (c) Factor MS-LS. In-sample fitting (blue);  MS process fitting contribution (red).}\\
        \includegraphics[width=0.7\linewidth]{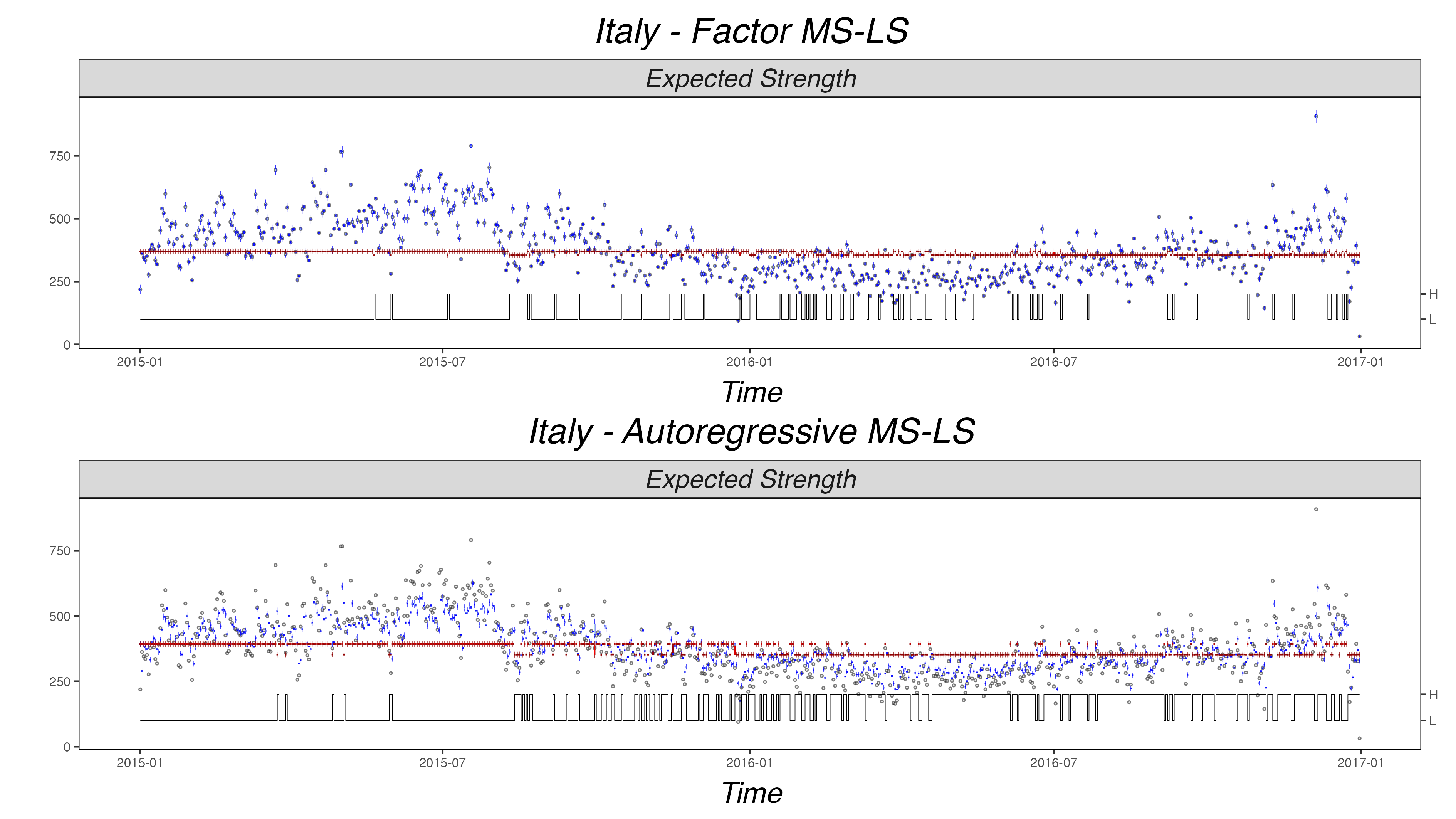}
\end{tabular}
    \caption{ The MS-LS model with $d = 1$, $K = 2$ (Panel a) against Factor and Autoregressive MS-LS with $d = 1$, $K = 2$ (Panels b and c). The blue boxplots report the fit of the model to daily observations of the Expected Strength. The step function displays the regime transitions for the two models  (with $K=2$). The green boxplots report the factor's contribution (or autoregressive component) to the fit. The red boxplots report the contribution of the MS component to the fit.
    }
    \label{fig:alternatives}
    \end{figure}

\begin{figure}[h!]
    \centering
    \begin{tabular}{cc}
        \includegraphics[height=120pt,width=0.45\textwidth]{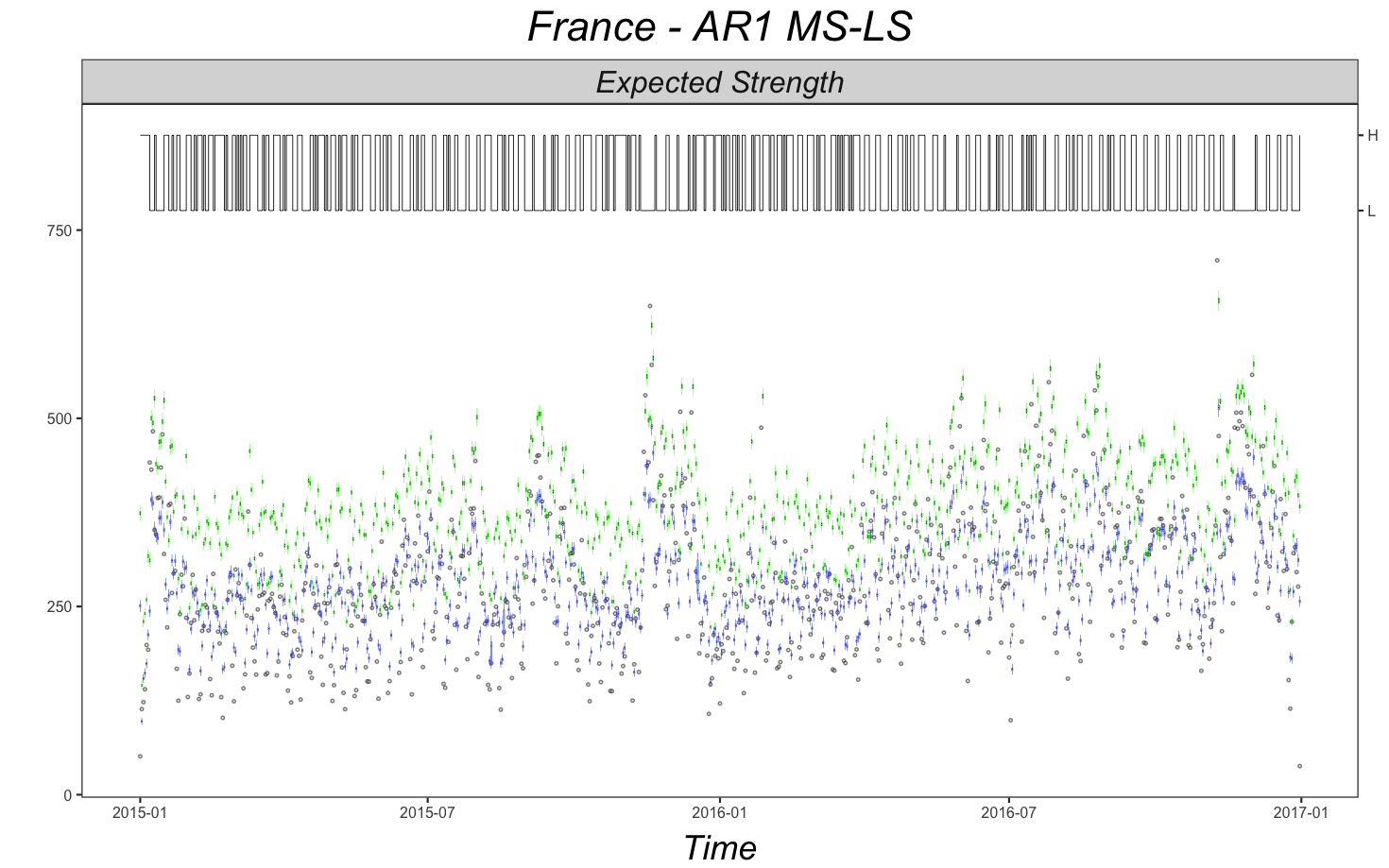} & 
        \includegraphics[height=120pt,width=0.45\textwidth]{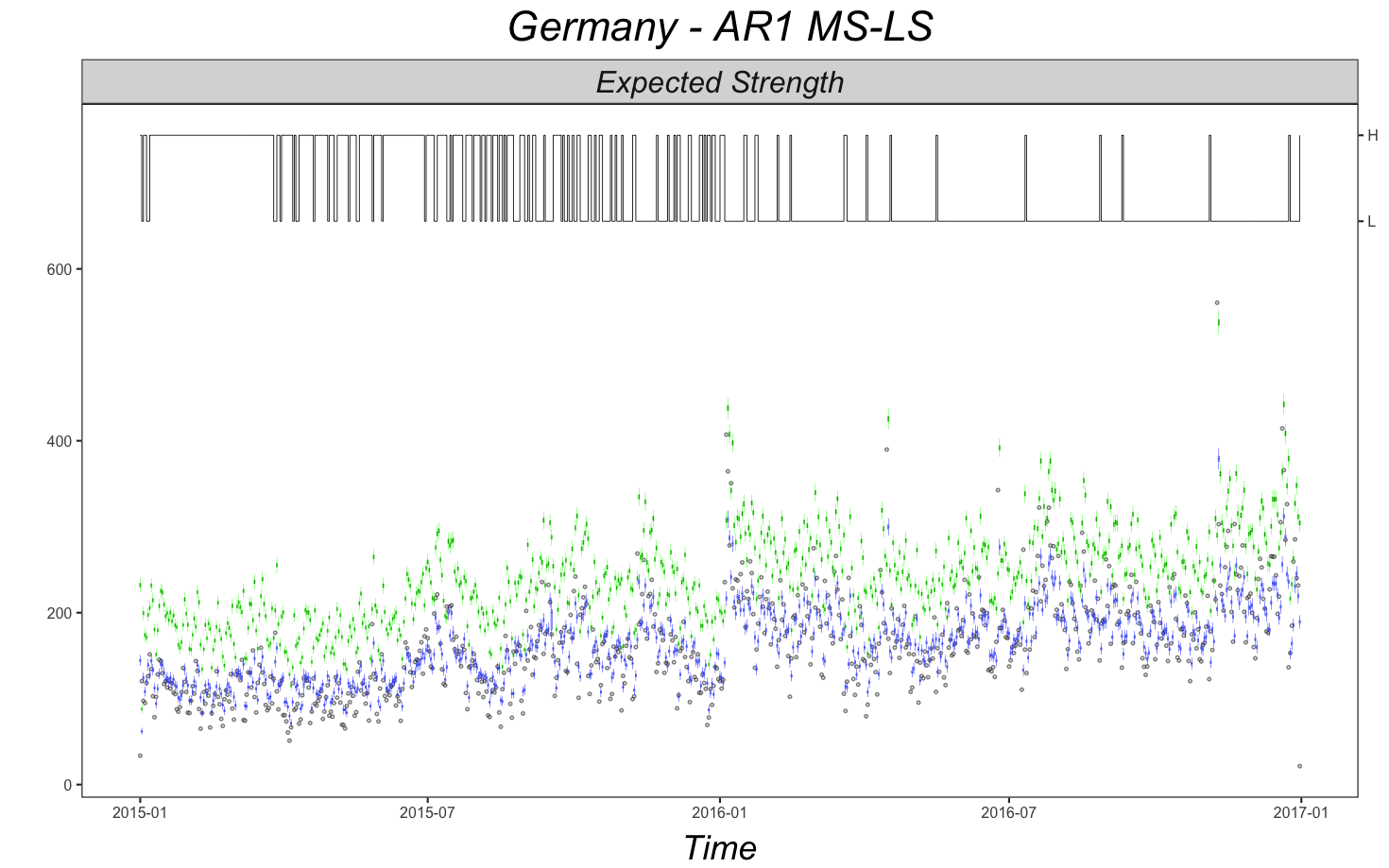}\\
        \includegraphics[height=120pt,width=0.45\textwidth]{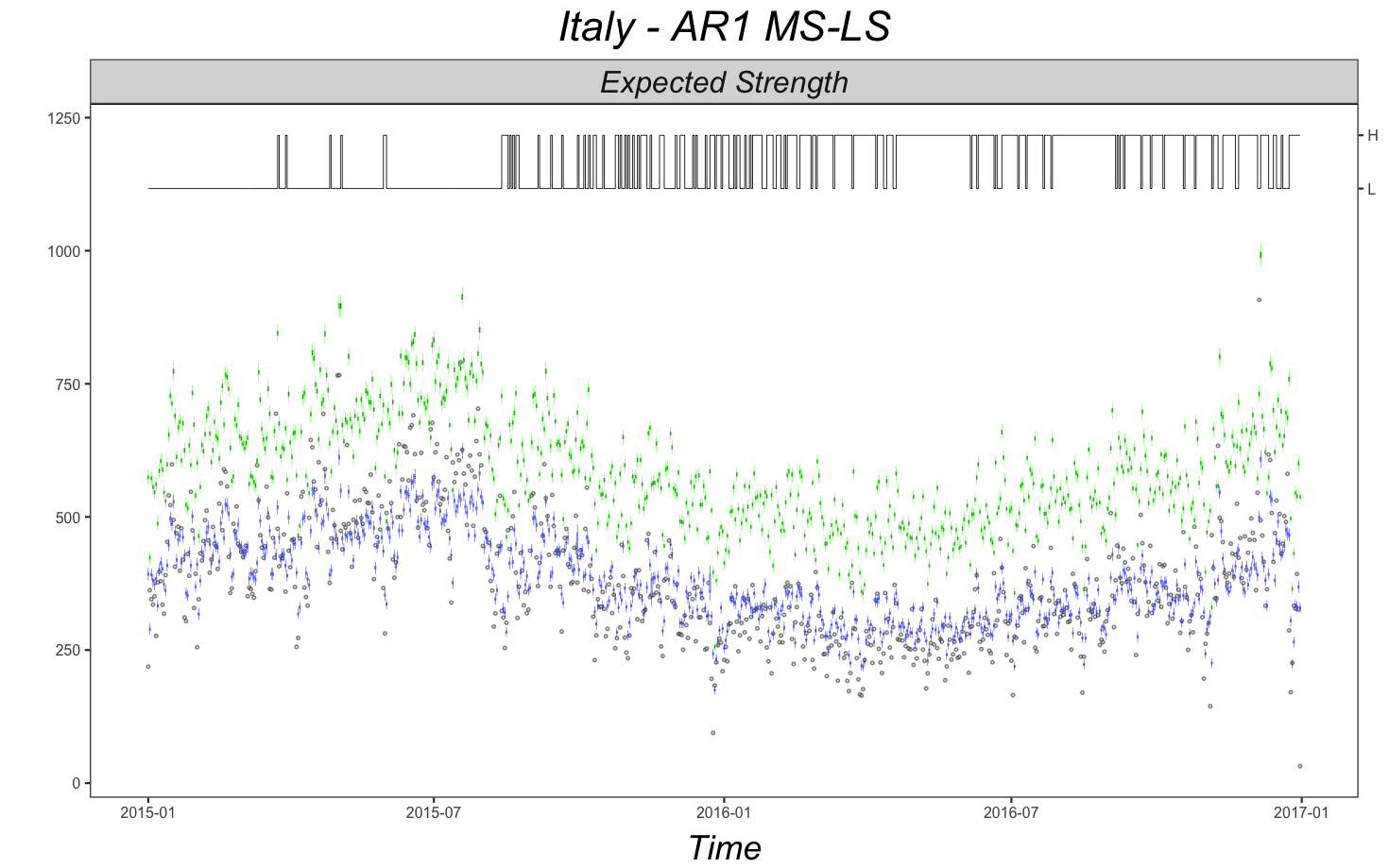} & 
        \includegraphics[height=120pt,width=0.45\textwidth]{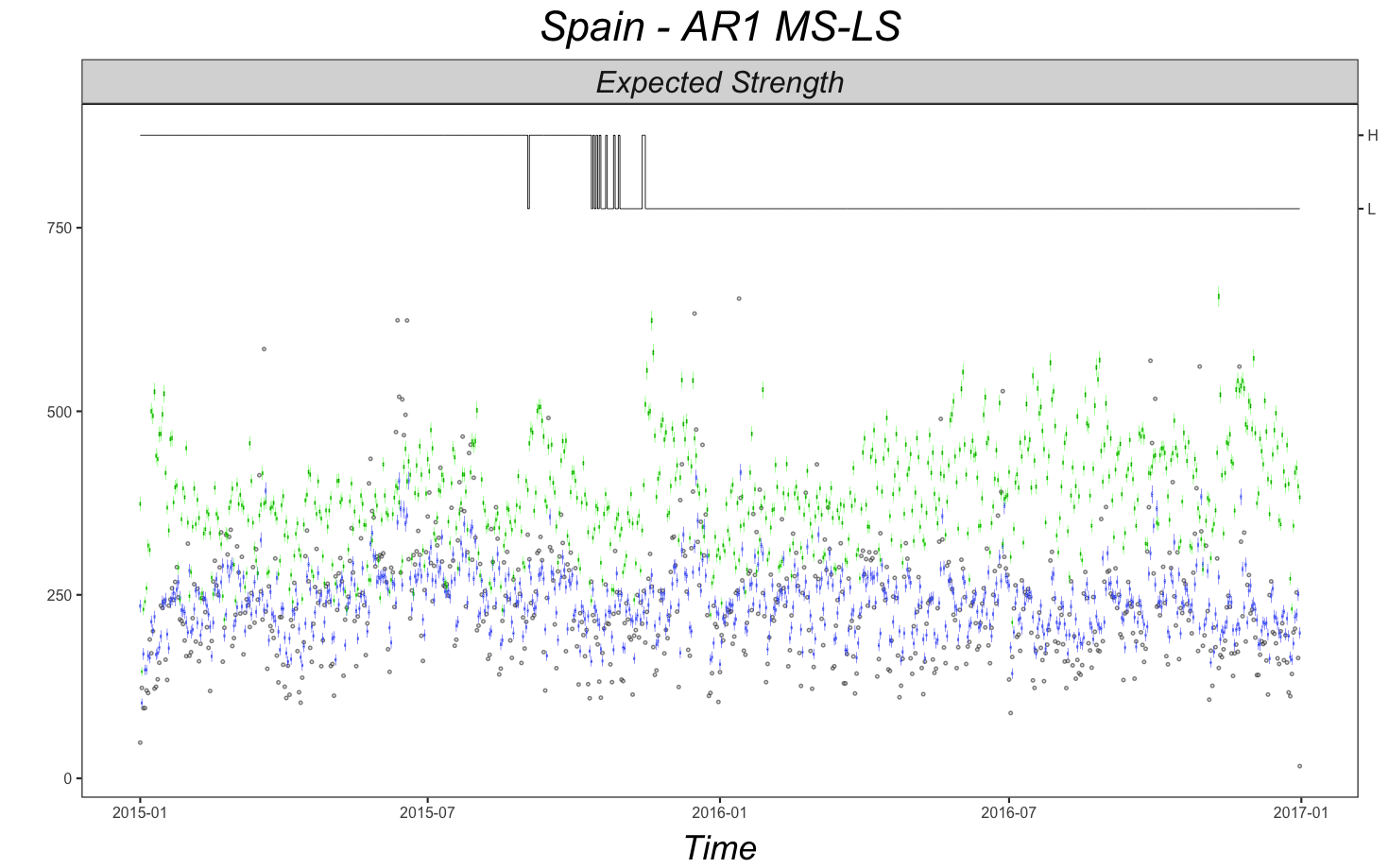}
    \end{tabular}
    \caption{AR1 MS-LS with $d = 1$, $K = 2$. The blue boxplots report the fit of the model to daily observations of the Expected Strength. The green boxplots report the fit of the model without the latent coordinates contribution. The step function displays the polarization regime transitions  (with $K=2$).}
    \label{fig:four_countries1}
\end{figure}

\begin{figure}[h!]
    \centering
    \begin{tabular}{cc}        \includegraphics[height=120pt,width=0.45\textwidth]{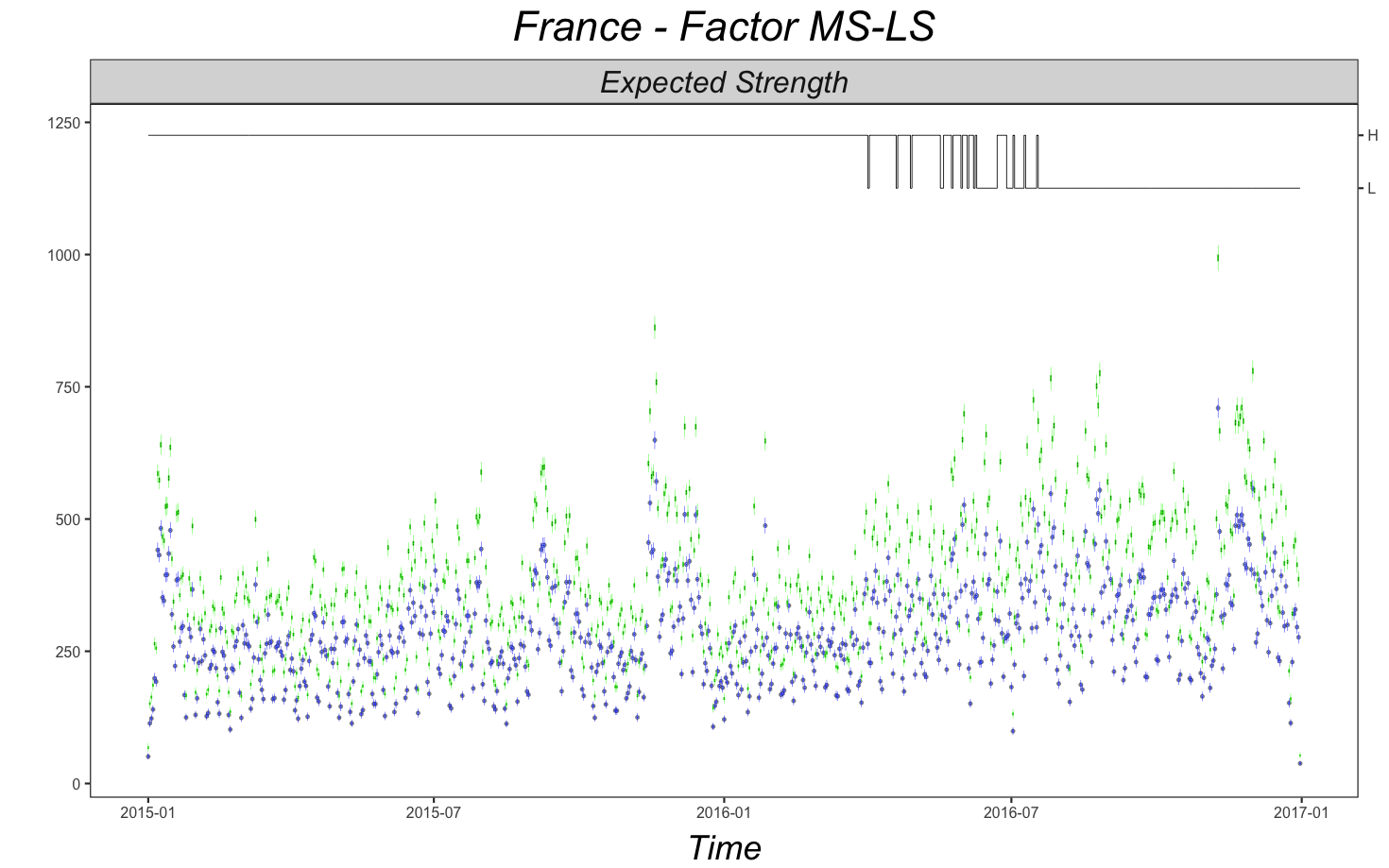} & 
        \includegraphics[height=120pt,width=0.45\textwidth]{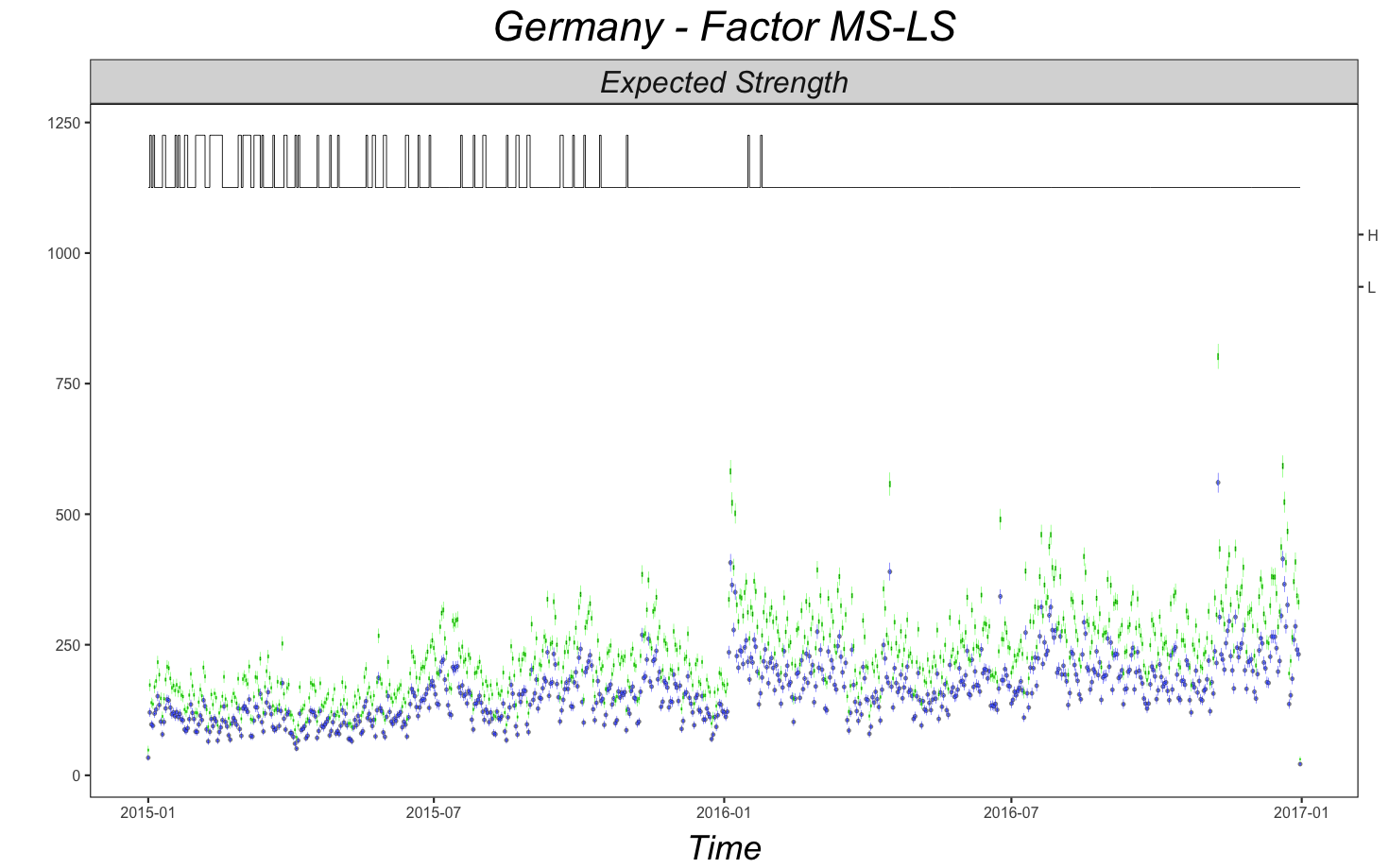}\\
        \includegraphics[height=120pt,width=0.45\textwidth]{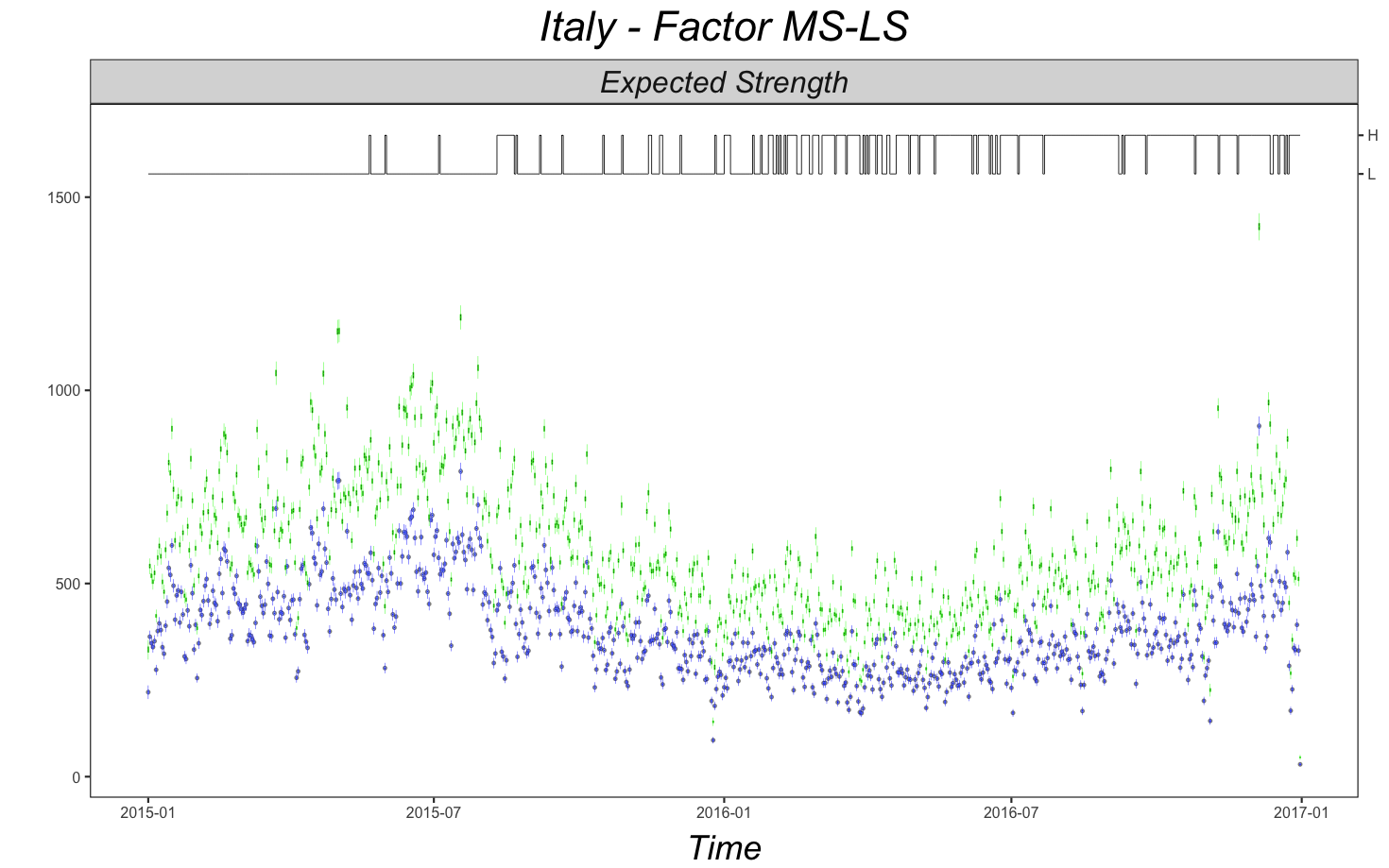} & 
        \includegraphics[height=120pt,width=0.45\textwidth]{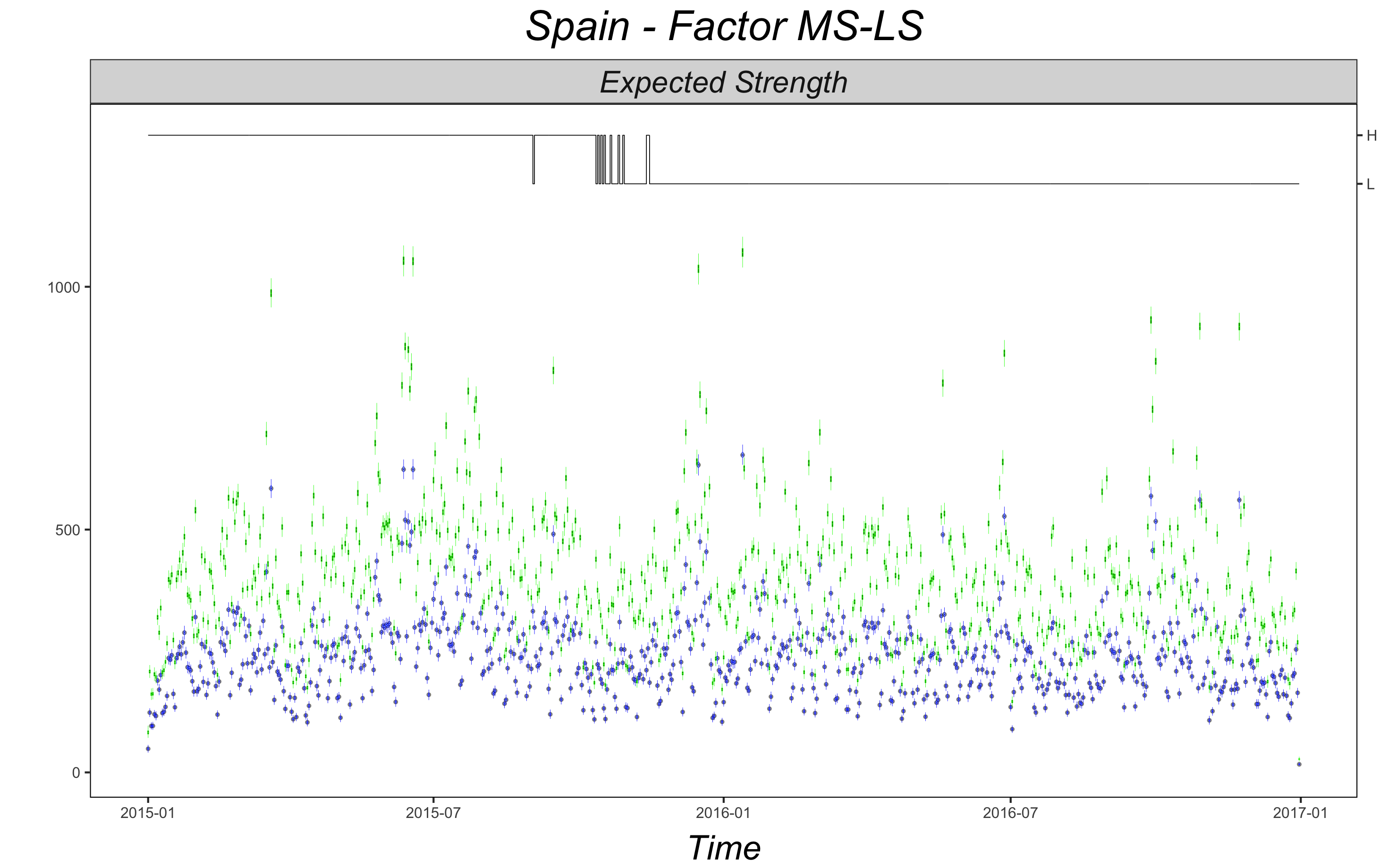}
    \end{tabular}
    \caption{Factor MS-LS with $d = 1$, $K = 2$. The blue boxplots report the fit of the model to daily observations of the Expected Strength. The green boxplots report the fit of the model without the latent coordinates contribution. The step function displays the polarization regime transitions  (with $K=2$).}
    \label{fig:four_countries2}
\end{figure}

We report in Figure \ref{fig:alternatives} the fit of these two alternative specifications for Italy.
While these models come closer to fitting the daily oscillations (blue boxplots in Panel (b) of Figure \ref{fig:alternatives} and also Figures \ref{fig:four_countries1}-\ref{fig:four_countries2} for the other countries), especially the factor model, they do not manage to properly differentiate latent positioning across the polarization regimes, which is the main purpose of our analysis. Despite the regimes being roughly coherent with those reported in panel (a) (for the case $d = 1$ and $K = 2$), the two-regime models with enhanced dynamic specification do not manage to match the regime variation to the data variability (see red boxplots in Panel (c) of Figure \ref{fig:alternatives}). In contrast, the pure MS-LS models recommended in the paper are able to detect well-separated polarization regimes in the latent space (as shown in Figure \ref{fig:latent_states_res} of the manuscript for $d=2$ and $K=5$.

\vfill
\clearpage

\renewcommand\thefigure{L.\arabic{figure}}
\setcounter{figure}{0}
\renewcommand\theequation{L.\arabic{equation}}
\setcounter{equation}{0}
\renewcommand\thetable{L.\arabic{table}}
\setcounter{table}{0}

\section{Further Results for the Facebook Data}
\label{H:furtherplots}

\subsection{Fast Greedy Clustering}
\begin{figure}[!htb]
  \centering
  \includegraphics[scale = 0.45]{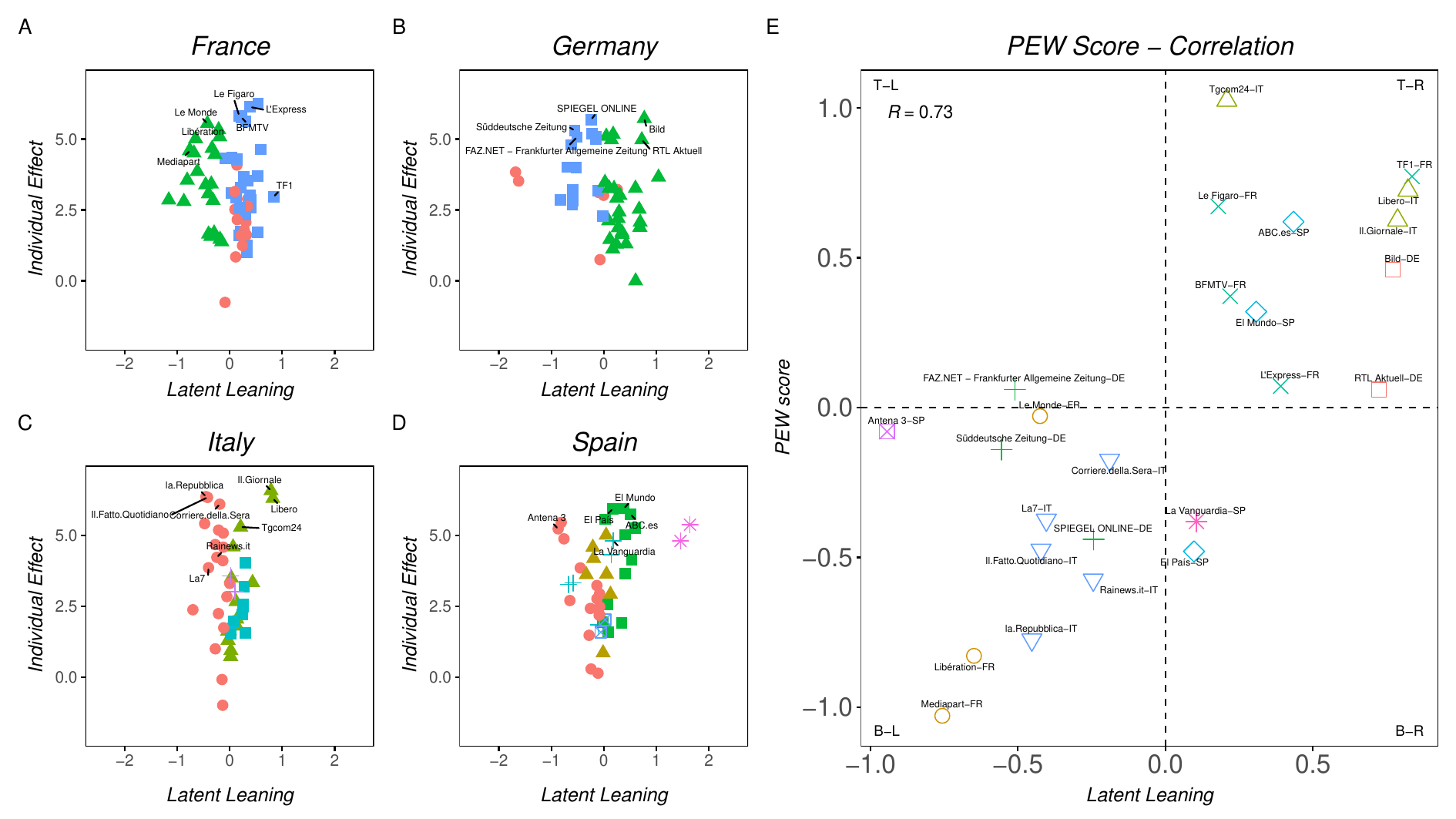}
   \caption{\textbf{PEW Index - Latent Leaning (Static Model) Comparison with Fast Greedy Algorithm:} Panels A, B, C, and D present the posterior mean of the latent coordinates of our news outlets, while Panel E displays the scatter plot comparing the PEW survey results with our estimated latent leaning variable (country-specific means have been subtracted from the PEW scores to improve readability). Colors and shapes denote the different clusters detected by the Fast Greedy algorithm (\citealp{clauset2004finding}).}
  \label{fastgreedy}
 \end{figure}

Figure \ref{fastgreedy} compares the results of our static analysis and the clustering provided by the Fast Greedy algorithm (\citealp{clauset2004finding}). Colors and shapes denote the different clusters detected by the clustering algorithm. It is worth noticing that the partition preserves the left-right distinction (see Panel E), although clustering alone is not enough to discriminate between left and right-leaning news outlets.

\subsection{Comparison with SBM}\label{sec:sbm}
We provide a comparison with a time-varying version of the stochastic block model implemented in the \texttt{sbm} package (\url{https://github.com/GrossSBM/sbm}). The algorithm provides information about the optimal number of clusters. Running the algorithm on our data allows us to detect from 5 to 12 clusters of news outlets for each point in time. In \texttt{sbm}, estimation is performed using Variational Inference, and model comparison can be conducted using the Evidence Lower Bound (ELBO), an output from the algorithm. Although we acknowledge that a direct comparison between the Log Pointwise Predictive Density ($llpd$) criterion we provide and ELBO may not be entirely appropriate, we present it in Table \ref{tab:comparison}. Despite this limitation, our model yields an $llpd$ higher than the ELBO of the dynamic stochastic block model for France, Germany, and Italy. Only in the case of Spain are the results in favor of SBM.

\begin{table}[!htb]
\caption{\textbf{Model Comparison:} Log pointwise predictive density ($lppd$) as in \cite{gelman2014understanding} for the models introduced here against the evidence lower bound provided in \texttt{sbm}.} 
 \label{tab:comparison}
 
\renewcommand{\arraystretch}{1.1}

  \centering

\resizebox{0.7\textwidth}{!}{  

\begin{tabular}{ccccc}
\hline \hline
\multicolumn{1}{l}{}                                                                                           & \multicolumn{4}{c}{Model Comparison}                                                                \\ \hline
Model                                                                                                          & France                & Germany               & Italy                 & Spain                 \\ \hline

{$\mathcal{M}_6$ ($lppd\times 10^{-6}$)}                                                                                                & -$\mathbf{ 2.0827 } $        &   -$\mathbf{1.1107}$       &  -$\mathbf{1.4641}$    &   -$2.0633$     \\

{$SBM$ ($ELBO\times 10^{-6}$)}                                                                                                & -$2.14423 $       &   -$1.1698$       &  -$1.7894$    &   -$\mathbf{1.6557}$     \\
 \hline \hline
\end{tabular}

}
\end{table}

\begin{figure}[h!]
  \centering
   \includegraphics[width= 0.7\textwidth]{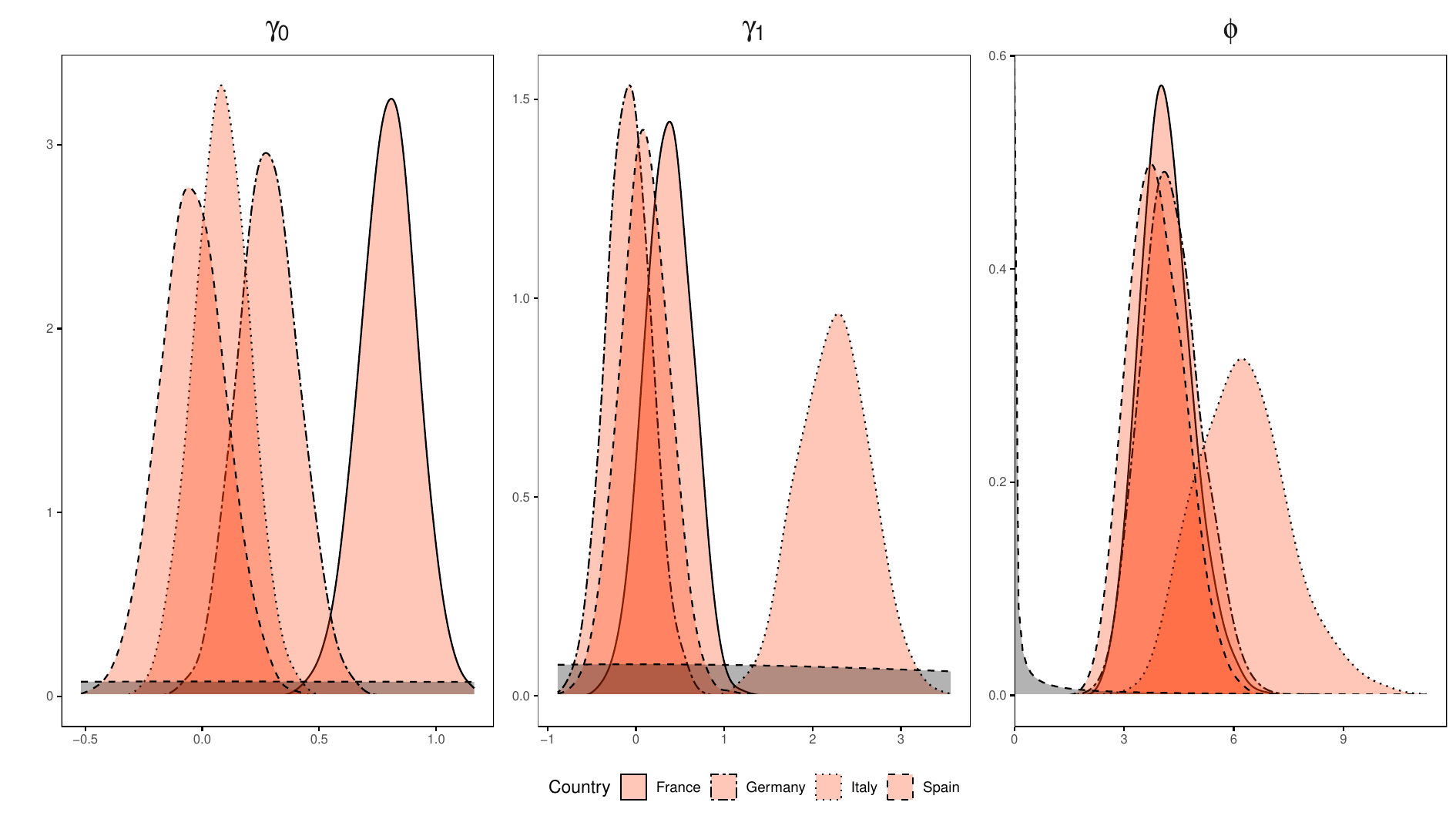}
 
 \caption{\textbf{Marginal Posteriors - Static Analysis:} Kernel density estimates of the posterior for the parameters $\gamma_{0}$, $\gamma_{1}$ and $\phi$ for France, Germany, Italy and Spain with prior pdfs denoted by dashed lines and gray shading.}
 
 \label{fig:boxplots_static}
 \end{figure}

\subsection{Further Results}
Figure \ref{fig:boxplots_static} presents the posterior distributions of the parameters of the static LS model for the Facebook data.

\section{Data and Scripts Repository}
\label{J:repository}
Data and Scripts are stored in the following Repository:  \url{https://github.com/BayesianEcon/Dyn-MS-LS-Media}. 
Refer to the README file in the repository for a complete description of each file.
A CodeOcean capsule is also available at the following link: \url{https://codeocean.com/capsule/9380600/tree/v1}.

 \subsection{Data}
\label{J1:repository_data}

The data entirely covers 729 days from "2015-01-01" to "2016-12-31". The number of news outlets included in each dataset is the following: Germany 47 news outlets, France 62, Italy 45 and Spain 43.

\begin{description}

\item[Network Dataset:] Data set used to illustrate the MS-LS Network Model in Section 4. The data set is an edge-list representation of the media networks where columns "i" and "j" refer to the nodes, column "t" refers to the day and column "w" refers to the number of unique Facebook commenters in common between "i" and "j" at time "t". The static version of the Network Dataset of each country is contained within the file ("Data\_Env\_single\_(country).RData"), while the dynamic version is in the file ("DataEnv\_(country)\_all.RData").

\item[Slant-Index Dataset:] Data set used to illustrate the MS-LS Network Model in Section 4. Refer to Appendix \ref{F:MediaSlantProxy} and the README file for a description of the index construction procedure. The data set represents a nodal feature of the media networks where column "i" refers to the nodes, column "t" refers to the day and column "leaning" refers to the Slant Index.

The static version of the Slant-Index Dataset of each country is contained within the file ("Data\_Env\_single\_(country).RData"), while the dynamic version is included in the file ("DataEnv\_(country)\_all.RData").

\end{description}

\subsection{Scripts}
\label{J2:repository_script}
We report here a brief description of the main scripts used to estimate the Bayesian MS-LS network model on the datasets studied in the main paper (Sections 3 and 4) and the supplementary material. Our MCMC algorithm is entirely implemented in C++, enabling shorter execution time than interpreted languages like R or Python. However, we still rely on R for data manipulation and plotting. The smooth integration of the two languages has been made possible using the \texttt{Rcpp} package, which offers a convenient interface for invoking C++ scripts within R. Refer to the README.txt file in the repository for a complete description of each script.

\begin{description}

\item[{\small Simulation\_02\_results.R:}] \hfill \\
Estimates the MS-LS model on the simulated network dataset.\\ Running time $\sim$ 12 mins (50,000 iterations, Apple M2, 8 GB Memory) 

\item[{\small Static\_01\_Results\_(country).R:}] \hfill \\
Estimates the MS-LS model on the static network dataset.\\ Running time $\sim$ 45 mins. (15,000 iterations, Apple M2, 8 GB Memory) 

\item[{{\small Dynamic\_01\_Results\_(country).R and Dynamic\_01\_Results\_(country)\_extended.R:}}] \hfill \\
Estimates the MS-LS model on the dynamic network dataset. \\ Running time > 20 hrs. (35,000 iterations, Apple M2, 8 GB Memory) 

\item[{{\small MS\_LS\_FE.cpp and MS\_LS\_FE\_extended.cpp:}}] \hfill \\
The scripts contain the functions to generate MCMC draws for the dynamic Bayesian MS-LS network model.

\end{description}

\end{document}